\documentclass[aps,rmp,%
twocolumn%
,amsmath,amssymb,superscriptaddress]{revtex4-2}
\usepackage{braket}
\usepackage{siunitx}
\usepackage[dvipsnames]{xcolor}
\usepackage{graphicx}
\usepackage{xcolor}
\graphicspath{{./},{./figures/}}
\usepackage{titlesec}
\usepackage{tabularx}
\usepackage{soul} % for strike-outs
\usepackage{dsfont}
\usepackage{multirow}
\usepackage[toc]{glossaries}
\usepackage[colorlinks=true,
                linkcolor=blue,
	        citecolor=blue,
	        urlcolor=blue]{hyperref}
\usepackage[normalem]{ulem}

\newacronym{DFT}{DFT}{density-functional theory}
\newacronym{DFPT}{DFPT}{density-functional perturbation theory}
\newacronym{GWPT}{GWPT}{GW perturbation theory}
\newacronym{KS}{KS}{Kohn--Sham}
\newacronym{MBPT}{MBPT}{many-body perturbation theory}
\newacronym{WFPT}{WFPT}{Wannier function perturbation theory}
\newacronym{DMFT}{DMFT}{dynamical mean-field theory}
\newacronym{BvK}{BvK}{Born--von K\'{a}rm\'{a}n}
\newacronym{WS}{WS}{Wigner--Seitz}
\newacronym{KC}{KC}{Koopmans-compliant}
\newacronym{PBC}{PBC}{periodic boundary condition}
\newacronym{BZ}{BZ}{Brillouin zone}
\newacronym{BTE}{BTE}{Boltzmann transport equation}
\newacronym{TDF}{TDF}{transport distribution function}
\newacronym{BCC}{BCC}{body-centred cubic}
\newacronym{QP}{QP}{quasiparticle}
\newacronym{QSGtildeW}{$QSG\tilde{W}$}{quasi-particle self-consistent GW with vertex correction}
\newacronym{QSGW}{QSGW}{self-consistent GW} 
\newacronym{API}{API}{application programming interface}
\newacronym{FFT}{FFT}{fast-Fourier-transform}
\newacronym{QRCP}{QRCP}{QR factorization with column pivoting}
\newacronym{HPC}{HPC}{high-performance computing}
\newacronym{LAPW}{LAPW}{linearized augmented-plane wave}
\newacronym{LO}{LO}{local orbital}
\newacronym{OPF}{OPF}{optimized projection function}
\newacronym{SLWF}{SLWF}{selectively localized Wannier function}
\newacronym{SCDM}{SCDM}{selected columns of the density matrix}
\newacronym{IBZ}{IBZ}{irreducible Brillouin zone}
\newacronym{FBZ}{FBZ}{full Brillouin zone}
\newacronym{MLWF}{MLWF}{maximally localized Wannier function}
\newacronym{POWF}{POWF}{partly occupied Wannier function}
\newacronym{SAWF}{SAWF}{symmetry-adapted Wannier function}
\newacronym{e-ph}{e-ph}{electron-phonon}
\newacronym{TB}{TB}{tight-binding} % putting dash as it's needed in the first occurrence
\newacronym{PDWF}{PDWF}{projectability-disentangled Wannier function}
\newacronym{MRWF}{MRWF}{manifold-remixed Wannier function}
\newacronym{PAO}{PAO}{pseudo-atomic orbital}
\newacronym{SCF}{SCF}{self-consistent field}
\newacronym{NSCF}{NSCF}{non-self-consistent field}
\newglossaryentry{WF}
{
  name={WF},
  description={Wannier function},
  first={\glsentrydesc{WF} (\glsentrytext{WF})},
  plural={WFs},
  descriptionplural={Wannier functions},
  firstplural=Wannier functions (WFs)   %{\glsentrydescplural{WF} (\glsentryplural{WF})}
} 
\newacronym{MLXWF}{MLXWF}{maximally localized exciton Wannier function}
\newglossaryentry{AHC}
{
  name={AHC},
  description={anomalous Hall conductivity},
  first={\glsentrydesc{AHC} (\glsentrytext{AHC})},
  plural={AHCs},
  descriptionplural={anomalous Hall conductivities},
  firstplural={\glsentrydescplural{AHC} (\glsentryplural{AHC})}
}
\newacronym{SHC}{SHC}{spin Hall conductivity}

\newglossaryentry{HWF}
{
  name={HWF},
  description={hybrid Wannier function},
  first={\glsentrydesc{HWF} (\glsentrytext{HWF})},
  plural={HWFs},
  descriptionplural={hybrid Wannier functions},
  firstplural={hybrid Wannier functions (HWFs)}
}

\makeglossaries % needed to generate glossaries

\newcommand\code[1]{\texttt{#1}}

\def\a{{\bf a}}

\def\a{\alpha}
\def\k{\kappa}
\def\w{\omega}

\def\D{\partial}
\def\boe{{\bf 0}_e}
\def\bop{{\bf 0}_p}
\def\bRe{{\bf R}_e}
\def\bRp{{\bf R}_p}
\def\bt{{\boldsymbol \tau}}

\def\bk{{\bf k}}
\def\bq{{\bf q}}
\def\br{{\bf r}}
\def\bG{{\bf G}}

\def\pnkc{{\psi_{n \bk}}}
\def\pmkqc{{\psi_{m \bk+\bq}}}
\def\pmkGc{{\psi_{m \bk''+\bG}}}

\def\pVc{{\Delta_{\bq \nu}V^{\rm KS}}}

\def\gmnvkqc{g_{mn\nu}(\bk,\bq)}
\def\gmnvkqf{g_{mn\nu}(\bk',\bq')}
\def\gmnvkqcp{g_{mn\nu}(\bk,\bq)}

\def\<{\langle}
\def\>{\rangle}

\def\la{\langle\kern-2.5pt\langle}
\def\ra{\rangle\kern-2.5pt\rangle}
\def\vt{\vert\kern-1.5pt\vert}
\def\lket#1{\vt#1\ra}
\def\lip#1#2{\la#1\vt#2\ra}
\def\lme#1#2#3{\la#1\vt#2\vt#3\ra}

\newcommand{\RR}{{\bf R}}

\newcommand{\bnk}{_{n\kk}}

\newcommand{\ibz}{\frac{V_\text{cell}}{(2\pi)^3}\int_{\rm BZ}}

\newcommand{\kk}{{\bf k}}

\newcommand{\nbw}{J}
\newcommand{\nwann}{J}
\newcommand{\nbands}{{\cal J}}

\newcommand{\ke}{\mathbf{k}}

\def\ip#1#2{\langle#1\vert#2\rangle}
\def\me#1#2#3{\langle#1\vert#2\vert#3\rangle}
\def\rr{{\bf r}}
\def\kk{{\bf k}}
\def\bb{{\bf b}}

\def\RR{{\bf R}}
\def\ttau{\boldsymbol{\tau}}

\makeindex
\begin{document}

\title{The Wannier Function Software Ecosystem for Materials Simulations}

\newcommand{\theos}{Theory and Simulation of Materials (THEOS), and National Centre for Computational Design and Discovery of Novel Materials (MARVEL), \'Ecole Polytechnique F\'ed\'erale de Lausanne, CH-1015 Lausanne, Switzerland}
\newcommand{\trieste}{Dipartimento di Fisica, Universit\`a di Trieste,  I-34151 Trieste, Italy}
\newcommand{\unimore}{Dipartimento di Scienze Fisiche, Informatiche e Matematiche, University of Modena and Reggio Emilia, I-41125 Modena, Italy}
\newcommand{\cnrnano}{Centro S3, CNR-Istituto Nanoscienze, I-41125 Modena, Italy}
\newcommand{\paulscherrer}{Laboratory for Materials Simulations (LMS), and National Centre for Computational Design and Discovery of Novel Materials (MARVEL),
  Paul Scherrer Institut (PSI), CH-5232 Villigen PSI, Switzerland}
\newcommand{\flatiron}{Center for Computational Quantum Physics, Flatiron Institute, 162 5th Avenue, New York, New York 10010, USA}
\newcommand{\suny}{Department of Physics, Applied Physics, and Astronomy, Binghamton University-SUNY, Binghamton, New York 13902, USA}
\newcommand{\imperial}{Departments of Materials and Physics, and the Thomas Young Centre for Theory and Simulation of Materials, Imperial College London, London SW7 2AZ, UK}
\newcommand{\cfm}{Centro de F\'{\i}sica de Materiales, Universidad del Pa\'\i s Vasco, 20018 San Sebasti\'an, Spain}
\newcommand{\ikerb}{Ikerbasque Foundation, 48013 Bilbao, Spain}
\newcommand{\oxford}{Department of Materials, University of Oxford, Parks Road, Oxford OX1 3PH, UK}
\newcommand{\sissa}{Scuola Internazionale Superiore di Studi Avanzati (SISSA), I-34136 Trieste, Italy}

\author{Antimo Marrazzo}
\affiliation{\sissa}
\email{amarrazz@sissa.it}
\affiliation{\trieste}

\author{Sophie Beck}
\affiliation{\flatiron}
\author{Elena R. Margine}
\affiliation{\suny}

\author{Nicola Marzari}
\affiliation{\theos}
\affiliation{\paulscherrer}

\author{Arash A. Mostofi}
\affiliation{\imperial}

\author{Junfeng Qiao}
\affiliation{\theos}

\author{Ivo Souza}
\author{Stepan S. Tsirkin}
\affiliation{\cfm}
\affiliation{\ikerb}
\author{Jonathan R. Yates}
\affiliation{\oxford}
\author{Giovanni Pizzi}
\affiliation{\paulscherrer}
\email{giovanni.pizzi@psi.ch}

\begin{abstract}
Over the last two decades, following the early developments on maximally localized Wannier functions, an ecosystem of electronic-structure simulation techniques and software packages leveraging the Wannier representation has flourished. This environment includes codes to obtain Wannier functions and interfaces with first-principles simulation software, as well as an increasing number of related post-processing packages. Wannier functions can be obtained for isolated or extended systems (both crystalline and disordered), and can be used to understand chemical bonding, to characterize electric polarization, magnetization, and topology, or as an optimal basis set, providing very accurate interpolations in reciprocal space or large-scale Hamiltonians in real space.
In this review, we summarize the current landscape of techniques, materials properties and simulation codes based on Wannier functions that have been made accessible to the research community, and that are now well integrated into what we term a \emph{Wannier function software ecosystem}.
First, we introduce the theory and practicalities of Wannier functions, starting from their broad domains of applicability to advanced minimization methods using alternative approaches beyond maximal localization. Then we define the concept of a Wannier ecosystem and its interactions and interoperability with many quantum simulations engines and post-processing packages. We focus on some of the key properties and capabilities that are empowered by such ecosystem\textemdash from band interpolations and large-scale simulations to electronic transport, Berryology, topology, electron-phonon couplings, dynamical mean-field theory, embedding, and Koopmans functionals\textemdash concluding with the current status of interoperability and automation. The review aims at highlighting basic theory and concepts behind codes, providing relevant pointers to more in-depth references. It also elucidates the relationships and connections between codes and, where relevant, the different motivations and objectives behind their development strategies. Finally, we provide an outlook on future developments, and comment on the goals of biodiversity and sustainability for the whole software ecosystem.
\end{abstract}
\maketitle
\tableofcontents

\section{Introduction}
\glspl{WF} \cite{wannier37}, and in particular \glspl{MLWF} \cite{Marzari1997}, provide an accurate, compact, and localized representation of the electronic-structure problem, and have become widely used in computational condensed-matter physics and materials science~\cite{Marzari2012}. 

Thanks to developments in theory, algorithms and implementations over the past few decades, summarized in Sec.~\ref{sec:howwegothere}, it has now become possible to apply widely the concept of \glspl{MLWF} to single-particle theories and in particular to \gls{KS} \gls{DFT} simulations, to obtain localized orbitals from Bloch states; the latter can be themselves represented with localized or extended basis sets, such as plane waves. 
On one hand, these developments have benefited from profound connections between \glspl{WF} and physical quantities such as electric polarization, orbital magnetization and topological invariants~\cite{VanderbiltWC1993,Niu2005,Thonhauser2005,Soluyanov2011a,Vanderbilt2018}.
On the other hand, the ability to obtain \glspl{MLWF} from \gls{DFT}  simulations can enable the calculation of physical quantities with high accuracy, but at a fraction of the computational cost, thanks to their role as very accurate interpolators \cite{Souza2001,Lee2005,Yates_et_al:2007}. Finally, although not discussed here, localized representations have long been pioneered by the quantum chemistry community to interpret coordination and bonding~\cite{edmiston-ruedenberg-1963}, and \glspl{MLWF} extend to periodic systems the concept of Foster-Boys localized orbitals~\cite{boys66}, thanks to algorithmic breakthroughs in calculating the position operator in solids \cite{blount-ssp62,Zak1989,Nenciu1991,Resta1992,KingSmith1993}.

Wannier functions are typically localized or even exponentially
localized~\cite{Panati2007,Brouder2007,Panati2013}, and due to the
nearsightedness of interacting
electrons~\cite{desCloizeaux_PR1964_1,desCloizeaux_PR1964_2,Kohn1996},
local electronic properties only depend on the nearby
environment~\cite{Bianco2011,Bianco2013,Marrazzo2016,Marrazzo2019}.
As a consequence, the resulting Hamiltonian matrix expressed in a
localized basis set (such as
\glspl{MLWF}~\cite{Calzolari2004,Lee2005}) becomes sparse, i.e., it
displays negligible matrix elements\textemdash or hoppings, in the language of
a \gls{TB} formalism\textemdash if the distance between the corresponding
localized basis functions exceeds a given threshold. In this sense,
\glspl{MLWF} constitute an optimal choice as they decay exponentially in real
space~\cite{Panati2013} and they minimize a localization functional by
design~\cite{Marzari1997,Marzari2012}. The resulting \glspl{MLWF} can be
used as a basis set to build,
LEGO\textsuperscript{\texttrademark}-like, the electronic structure of
large-scale nanostructures~\cite{Lee2005} (that, thanks to the sparsity of the resulting Hamiltonian matrix, could be
solved with linear-scaling
methods~\cite{Mauri1993,Ordejon1993,Li1994}), or as a remarkably
accurate interpolators of electronic properties, operators and
quantities defined as integrals over the \gls{BZ} of periodic
systems~\cite{Souza2001,Yates_et_al:2007}. Interpolation on dense
grids becomes essential when very fine features need to be resolved,
as happens when integrals are restricted to lower-dimensional
manifolds (such as the Fermi surface, in the case of transport
properties of metals).

Nowadays, \glspl{MLWF} are routinely used in many research areas of
condensed-matter physics and materials science.  In
Sec.~\ref{sec:howwegothere} we summarize the past and current
challenges, discussing how we reached the current state.  The flourishing of this field is not only due to theoretical advances,
but also strongly driven by the concerted development of
accessible and efficient software.  Indeed, thanks to the availability
of robust software packages (often open-source, encouraging further
contributions), and to the user support provided by developers, researchers can now not only easily
compute \glspl{MLWF}, but also use them as core ingredients for
advanced simulations.  As more codes appear, they adopt the
\emph{de facto} standardization of input and output formats, resulting
in a set of interacting and interoperating codes that we will call
here the ``Wannier function software ecosystem.''

This review does not aim to provide an
extensive discussion of the theory of \glspl{MLWF}, for which we refer
to~\textcite{Marzari2012}, although we do provide a general introduction to the 
field in Sec.~\ref{subsec:primer}. Instead, the goal is to discuss the nature of the ecosystem and the capabilities of existing codes, focusing in
Sec.~\ref{sec:ecosystem} on a selection of physical phenomena or
quantities that can be efficiently predicted thanks to \glspl{WF}, and
on how \glspl{WF} are used as an ingredient to extend the accuracy of beyond-\gls{DFT} simulations.
Nevertheless, we will still mention
a few notable developments of the past decade, whenever useful to
contextualize the theoretical and software developments.  Our aim is
to provide a reference that can help newcomers and existing practitioners alike 
navigate the ecosystem: which properties can be computed by
which codes, when and why the use of \glspl{WF} is beneficial, which quantities
are exchanged between codes, and how interoperability is being
addressed.

To facilitate the discussion of the codes, in this review we group them into three major categories:
\emph{Wannier engines}, i.e., codes to obtain \glspl{WF};
\emph{interface codes} between the first-principles engines (e.g.,
\gls{DFT} or GW codes) and the Wannier engines; and
\emph{Wannier-enabled codes}, which range from relatively simple
post-processing tools to more advanced codes that use \glspl{WF} as
one of the ingredients to accelerate accurate simulations.  A fourth
category of codes that we discuss in Sec.~\ref{subsec:automation} are
\emph{automation workflows}. 
 Indeed, until very recently the generation of \glspl{WF} typically required human intuition by
experienced researchers to provide initial trial orbitals. This barrier has been largely removed by recent algorithmic
and automation efforts (see Sec.~\ref{subsec:advancedminimisation}),
enabling the use of \glspl{WF} both by new users and for high-throughput materials discovery
and characterization. In the latter case, managing thousands (or more)
simulations poses new challenges, which require not only the use of
robust workflow engines, but also the implementation of WF-specific
workflows to effectively interconnect multiple codes within the ecosystem.

We will conclude in Sec.~\ref{sec:conclusions} with some perspectives on the field related to the 
sustainability of the whole effort, current challenges that still need to be addressed, and possible future developments.

\subsection{\label{subsec:primer}\textit{Ab initio} electronic structure and Wannier functions}
Electronic-structure simulations aim to determine the behavior of electrons in materials and molecules, as governed by the Schr\"odinger or Dirac equations. Electrons feel Coulomb interactions among themselves and with the nuclei, in addition to  couplings with external fields (e.g., electrical, magnetic, or electromagnetic/photons) or perturbations (e.g., strain, phonons). The core electrons of heavy chemical elements can reach relativistic speeds, requiring the Schr\"odinger equation to be corrected with terms obtained from an expansion of the Dirac equation in powers of $1/c^2$, where $c$ is the speed of light. The most relevant relativistic correction is spin-orbit coupling, which is responsible for several important phenomena related to magnetism and to geometrical and topological properties of the electronic manifold (see Sec.~\ref{subsec:Berry} and Sec.~\ref{subsec:topoinv}). 

An exact solution of the Schr\"odinger equation (either the bare one, or with relativistic corrections) would give access to essentially all property of materials. This problem was already clear in 1929, when Paul Dirac declared: ``The underlying physical laws necessary for the mathematical theory of a large part of physics and the whole of chemistry are thus completely known, and the difficulty is only that the exact application of these laws leads to equations much too complicated to be soluble''~\cite{Dirac_1929}.

For more than fifty years, \textit{ab initio} or first-principles methods have been developed to approximately solve the Schr\"odinger equation in realistic settings~\cite{MartinBook2020,GiustinoBook2014}, with more accurate strategies and theories being developed. In parallel, the exponential growth of computational power (Moore's law) has allowed to deploy these new theoretical instruments through numerical simulations, constantly seeking not only to improve accuracy but also targeting more complex and realistic systems. Quickly, numerical solutions became sufficiently accurate to be predictive for a number of properties in relevant systems: the era of first-principles materials modeling had begun~\cite{YinCohenPRL1980,Marzari2021}.

An iconic example is given by \gls{DFT}, which made it possible to determine the electronic structure of complex materials with reasonable accuracy at low cost. In DFT, the total energy of the electrons is expressed as a functional of the electronic charge density~\cite{MartinBook2020,GiustinoBook2014}. The theory is supported by two pillars, the Hohenberg--Kohn theorems, which state not only a one-to-one correspondence between the ground-state many-body wavefunction and the ground-state charge density, but also formulate a variational total-energy functional: the solution of the Schr\"odinger equation can be recast as a minimization problem for the charge density, a remarkably simpler object (a real function of $\mathbf{r}$) than the many-body wavefunction we started from (a complex function of $3N_e$ variables, where $N_e$ is the number of electrons in the system). 

In \gls{KS} DFT, the interacting many-body problem is mapped onto a non-interacting problem sharing the same ground-state charge density but in presence of a suitable local external potential; the latter is in general unknown. This \emph{ansatz} enables the kinetic energy contribution to be calculated accurately through the second derivatives of the KS orbitals; this quantity is hard to calculate directly from the charge density alone. The success of DFT has been possible also thanks to the discovery and development of simple functionals that approximate the exact, but unknown, total-energy functional.
Hence, the KS-DFT approach to solving the many-body Schr\"odinger equation translates into solving a set of non-interacting one-particle Schr\"odinger equations in presence of an external potential that depends, self-consistently, upon the charge density only.
Although the outstanding importance of DFT has been recognized by the 1998 Nobel Prize in chemistry awarded to Walter Kohn (for DFT) and John Pople (for computational methods in quantum chemistry), its impact on the physics community has been, if possible, even greater: the top ten most highly cited articles published by the American Physical Society deal with DFT and related applications~\cite{Talirz_Ghiringhelli_Smit_2021}.

The electronic structure of materials and molecular systems is, at the same time, very similar and yet very different. To some extent, extended bulk materials can be seen as very large molecules and one could focus on the local electronic structure in real space, which is periodically repeated in the case of a crystalline materials (e.g., metals, semiconductors, oxides). This viewpoint is supported by the mathematical structure of the Schr\"odinger equation and its solutions, as observed by Kohn ~\cite{Kohn1996}: the electronic structure is fundamentally a local property, ``nearsighted'' to what happens at further distance in real space. The effect of chemical bonding and the presence of the lattice can be seen as perturbations to the case of isolated atoms, with their well-defined $s$, $p$, $d$ and $f$ orbitals. This perspective is powerful and foundational for linear-scaling methods, which target the simulation of large systems and leverage a description based on localized orbitals. Yet materials are not just very large molecules, and ``more is different''~\cite{Anderson1972}. Extended systems are practically infinite and can thus be described using \glspl{PBC}. As will be discussed in more formal terms in Sec.~\ref{sec:howwegothere}, the electronic structure of a material 
under \glspl{PBC} is more naturally described in terms of Bloch orbitals, which are not localized in real space. Indeed, a different perspective often drives the discussion of the electronic structure of periodic crystals: the behavior in reciprocal (i.e., Fourier-transformed) space. A textbook example is semiconductor physics, that is in general much better understood (and taught) by studying solutions of the Schr\"odinger equation in such reciprocal space.
This approach, somewhat orthogonal to the large-molecule perspective, is indeed also very powerful both at a conceptual and practical level. As a side note, many electronic-structure codes for materials actually adopt a completely delocalized basis (e.g., plane waves) to describe the periodic part of the Bloch orbitals~\cite{pickett_cpc_1989,MartinBook2004,MartinBook2020}.

How can we reconcile these two, almost opposite, perspectives? Remarkably, one can use the fundamental ``gauge freedom'' of quantum mechanics:
First, any quantum state is defined modulo a phase factor. Second, if one considers a set of single-particle states separated in energy from other states, then any trace operation on this manifold is invariant with respect to any unitary transformation among the orbitals; we call this a ``generalized'' gauge freedom.
However, and this is the crucial aspect, the localization properties of that set of states strongly depend on their gauge. 

\glspl{WF} provide a rigorous and insightful way to reconcile the real-space and localized perspective with the reciprocal-space (Fourier) and delocalized one. As it will be clear in the next section, \glspl{MLWF} in particular exploit the generalized gauge freedom to transform delocalized orbitals into localized ones (and vice versa), by constructing the proper unitary matrices.

\section{\label{sec:howwegothere}Wannier functions fundamentals}
The electronic structure of periodic crystals is most commonly
described in terms of Bloch waves
\mbox{$\psi\bnk({\bf r})=\;u_{n{\bf k}}({\bf r})\,e^{i{\bf k\cdot r}}$},
where $\kk$ is the crystal momentum, and $n$ is the band index. This
is the case in textbooks on solid-state theory, but also for the many
software packages that solve the \gls{KS} equations for crystalline
solids. A few years after Felix Bloch developed the theory of electron
waves in periodic crystals~\cite{bloch1929}, Gregory Wannier
introduced an alternative representation in terms of an orthonormal
set of localized functions~\cite{wannier37}, the Wannier functions. 
Given an isolated Bloch band $n$, the \gls{WF}
$w_{n\RR}(\rr)=\langle \rr\vert \RR n\rangle=w_{n{\bf 0}}(\rr-\RR)$
associated with the unit cell labelled by lattice vector $\RR$ is
defined as~\cite{wannier37}
\begin{equation}
\label{eq:wanniertransform}
\ket{\RR n}\;=\; \ibz d{\bf k}\;e^{-i{\bf k\cdot {\bf R}}} \,
\ket{\,\psi\bnk\,},
\end{equation}
where $V_\text{cell}$ is the unit-cell volume.

Since their inception, \glspl{WF} have been employed as a conceptual tool to
tackle problems in solid-state physics (see, e.g.,~\textcite{kivelson82}). However, in the first sixty years following
Wannier's paper, there were few actual calculations of \glspl{WF} for real
materials (see, e.g.,~\textcite{Callaway67,Satpathy88,Sporkmann94}). 
 The main obstacle
was the fact that \glspl{WF} are strongly non-unique, being sensitive to the generalized gauge freedom discussed earlier, e.g., to
$k$-dependent phase changes
$\ket{\psi\bnk}\rightarrow e^{-i\beta\bnk}\ket{\psi\bnk}$ with $\beta\bnk \in \mathbb{R}$
(gauge transformations) of the Bloch eigenstates. In addition, the
energy bands of real materials typically become degenerate at points,
lines, or even entire planes in the \gls{BZ}. The presence of degeneracies
leads to poor localization properties of the \glspl{WF} obtained from
Eq.~\eqref{eq:wanniertransform}, because no matter how the phase factors
$e^{-i\beta\bnk}$ are chosen, the Bloch eigenfunctions are
non-differentiable functions of $\kk$ at the degeneracy points.

\subsection{\label{subsec:mlwf}Maximally localized Wannier functions}
In the following, we discuss the widely used Wannierization methods introduced in~\textcite{Marzari1997,Souza2001}, while we refer the reader to Sec.~\ref{subsec:advancedminimisation} for an overview of more recent and advanced minimization methods.
\subsubsection{Isolated composite groups of bands} \label{sec:isolated_bands}

Consider a group of $J$ Bloch bands of orthonormal $\ket{\psi_{n{\bf k}}}$ Bloch states that may be connected among
themselves by degeneracies, but are isolated from all lower or higher
bands, e.g., the six valence bands in Fig.~\ref{fig:HfSe2}.  Given
such a composite group, the most general expression for the associated
\glspl{WF} is~\cite{Marzari1997} 
\begin{subequations}
\begin{align}
\ket{{\bf R}j}&= \;\ibz d{\bf k}
\,e^{-i{\bf k\cdot {\bf R}}}\ket{\psi^\text{W}_{j{\bf k}}}\,,
\label{eq:wannier}\\
\ket{\psi^\text{W}_{j{\bf k}}}&=\sum_{n=1}^\nbw\,\ket{\psi_{n{\bf k}}}U_{\kk,nj}\,,
\label{eq:psi-W-k}
\end{align}
\label{eq:wannier-psi-W}%
\end{subequations}
where the $U_\kk$ are $\nbw\times\nbw$ unitary matrices that describe
the generalized (multiband) gauge freedom within the Bloch manifold at
each $\kk$. The superscript W denotes a Wannier gauge, as opposed to a Hamiltonian gauge (later denoted by H) where the Hamiltonian matrix is diagonal. Note that at variance with
Eq.~\eqref{eq:wanniertransform}, in Eq.~\eqref{eq:wannier-psi-W} there
is not a one-to-one correspondence between the band index~$n$ and the
intra-cell Wannier index~$j$.

Marzari and Vanderbilt (MV) introduced the concept of
\glspl{MLWF}, in which the $U_\kk$ matrices are chosen so as to
minimize the total quadratic spread of the
\glspl{WF}~\cite{Marzari1997}:
\begin{eqnarray}
\Omega&=&\sum_{j=1}^\nwann\left[    
\langle\,{{\bf 0}j}\,\vert\,r^2\,\vert\,{{\bf 0}j}\,\rangle\,-\,
\left|\langle\,{{\bf 0}j}\,\vert\,{\bf r}\,\vert\,{{\bf 0}j}\,\rangle
\right|^2\,\right].
\label{eq:omega}
\end{eqnarray}
As discussed later (see Sec.~\ref{subsubsec:loc_func}) the spread (a.k.a. localization) functional $\Omega$ and its gradient with respect to an infinitesimal gauge transformation can be expressed in reciprocal space; furthermore, the BZ
integration in Eq.~\eqref{eq:wannier} is replaced by a discrete sum $(1/N)\sum_\kk$ where $N$ is the number of $k$ points in the finite grid used in the numerical simulations, and the optimal $U_\kk$ matrices are found by iteratively minimizing the functional $\Omega$ (see also~\textcite{Marzari1997} for the mathematical details).

From general Fourier-transform considerations~\cite{Duffin1953}, the good real-space localization properties of the \glspl{MLWF} on the left-hand side of
Eq.~\eqref{eq:wannier} mean that the Bloch-like states
$\ket{\psi^\text{W}_{j\kk}}$ appearing on the right-hand side are
smooth functions of $\kk$ for the optimal choice of $U_\kk$ matrices
in Eq.~\eqref{eq:psi-W-k} (or for any other choice leading to
well-localized \glspl{WF}).

The details of the MV methodology can be found in~\textcite{Marzari1997,Marzari2012}; in
the case of single $k$-point sampling (large unit cells), it is
equivalent to the Foster-Boys scheme used in quantum chemistry to
construct localized molecular orbitals~\cite{boys66}. It should be
noted that other localization criteria can be used for the purpose of
obtaining localized orbitals, e.g., the
Edmiston--Ruedenberg~\cite{edmiston-ruedenberg-1963} and
Pipek--Mizey~\cite{pipek-mezey-1989} approaches, based on maximizing
the Coulomb self-repulsion of the orbitals and the sum of the squares
of the Mulliken charges~\cite{mulliken-1955} associated with the
orbitals, respectively. Whilst these are more challenging to adapt to
a periodic, multi-$k$-point formulation, there has been recent work to
obtain \glspl{WF} for periodic systems using the Pipek--Mezey
localization criterion~\cite{jonsson2017,clement2021}.  Nevertheless,
the MV approach of minimizing the quadratic spread is still the most
widely used approach for periodic systems.

\begin{figure*}[tbp]
  \centering
  % {left bottom right top}
  %\fbox{
    \includegraphics[trim={0 0 0 320px},clip,width=0.8\textwidth]{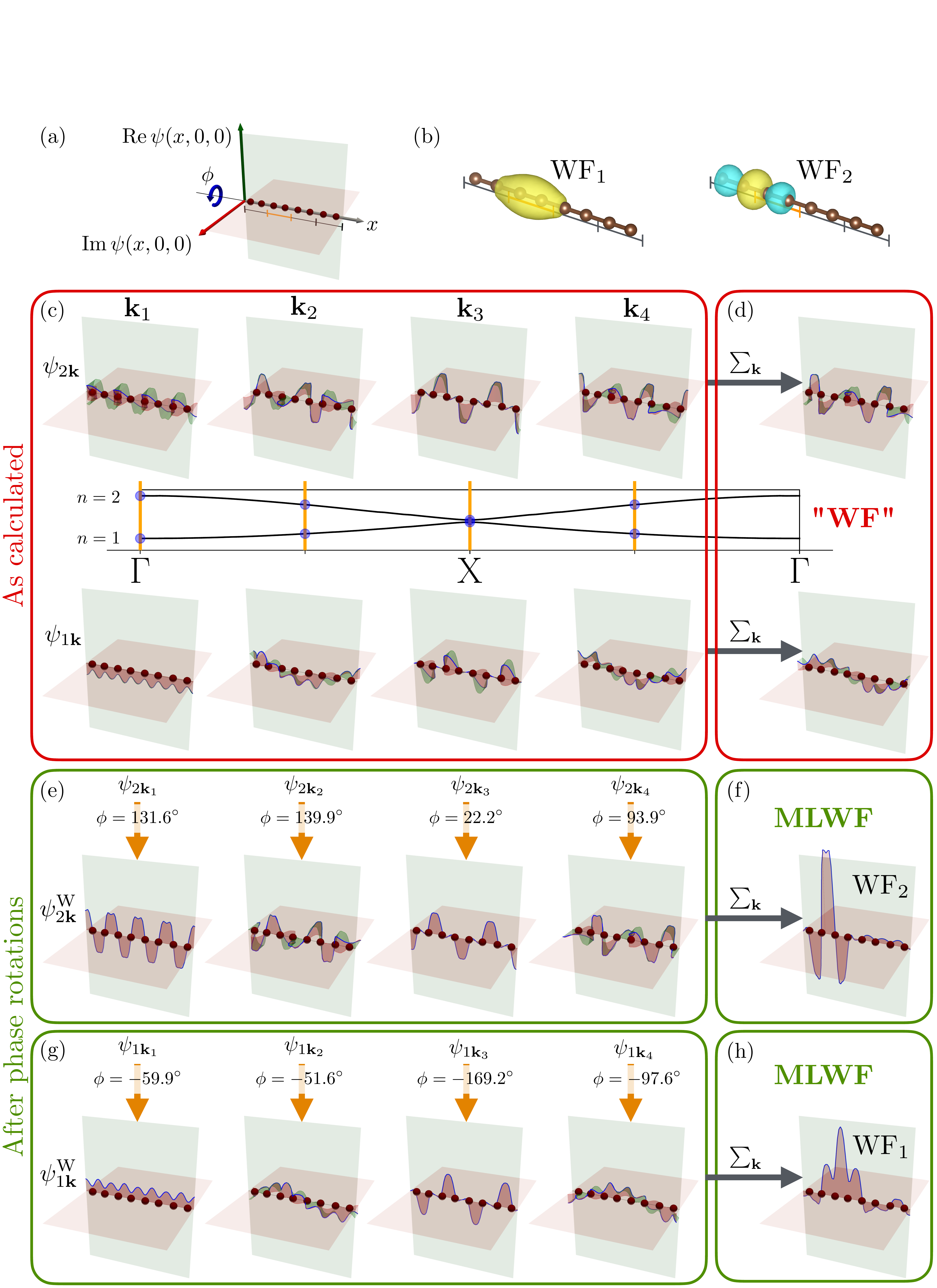}%
  %}
  \caption{\label{fig:wannierisation-singlebands}
  Band-by-band Wannierization of the two lowest bands of a dimerized
carbon chain.
(a) Graphical representation of the 4-times longer supercell where all
quantities will be plotted in panels (c)--(h); given our sampling of the \gls{BZ} with 4
$k$ points, both the Bloch states and the \glspl{WF} have the
supercell periodicity.  The eight carbon atoms are represented as dark
red spheres, and the primitive cell of length $a$ is marked by the
orange segment. One axis is the real-space $x$ axis of the carbon
chain, while the other two are used to plot the real (green) and
imaginary (red) components of complex wavefunctions $\psi(x, 0, 0)$
along $x$ (the profiles are highlighted in blue in subsequent
panels). Multiplying $\psi$ by a phase factor $e^{i \phi}$ corresponds
to a rotation by an angle $\phi$, as indicated by the blue arrow.
(b) 3D isosurfaces of the resulting \glspl{MLWF} for each of the two bands (band $n=1$ leads to WF$_1$, $n=2$ leads to WF$_2$).
(c) The Bloch orbitals $\psi_{n\mathbf k}$ as calculated by the
\textit{ab initio} (\gls{DFT}) code on the $k$ points $\mathbf k_1 =
\Gamma = (0, 0, 0)$, $\mathbf k_2 = (\pi/2a, 0, 0)$, $\mathbf k_3 =
\text{X} = (\pi/a, 0, 0)$, $\mathbf k_4 = (3\pi/2a, 0, 0)$ indicated
by the blue dots and vertical orange lines in the center of the panel, together with the full \gls{DFT} band structure.
(d) The sum of these Bloch orbitals, as in the original Wannier
definition of Eq.~\eqref{eq:wanniertransform}, delivers Wannier functions ``WF'' that are not
well localized.
(e) The construction of a \gls{MLWF} associated with the second lowest
band ($n=2$). For an isolated band in one dimension, the procedure amounts to a
complex phase rotation at each $k$ point; the optimal values of
these complex phases for this specific example are indicated.
(f) The sum of these rotated Bloch states results in the \gls{MLWF}
WF$_2$, also shown in panel (b).
(g) and (h) are the same as (e) and (f), but for the lowest band ($n=1$).
}
  \end{figure*}

To provide an illustrative example of the Wannierization procedure, we briefly discuss here the simple example of a slightly dimerized polyyne-like carbon chain, i.e., a chain of carbon atoms with two atoms per unit cell and alternating distances $d_1$ and $d_2$. Fig.~\ref{fig:wannierisation-singlebands}
displays the results of actual \gls{DFT} calculations, where carbon--carbon distances are $d_1=1.245\text{\AA}$ and $d_2=1.345\text{\AA}$, and the lattice parameter is thus $a=2.6\text{\AA}$.
We compute the electronic band structure of this linear chain with the \code{Quantum ESPRESSO} code~\cite{Giannozzi2009,Giannozzi2017}, which uses pseudopotentials with a plane-wave basis set, where the bands originating from the $1s$ orbitals of carbon are not explicitly computed. If these $1s$ core orbitals were computed explicitly, they would be almost-flat bands (since the orbitals are very localized) at lower energy.
At the center of panel (c) of Fig.~\ref{fig:wannierisation-singlebands} we show the two lowest bands considered in the \gls{DFT} calculation. A projected density of states calculation (not shown here) shows that these two lowest bands originate from a combination of $2s$ and $2p_x$ orbitals centered on the two atoms in the unit cell. In addition to being separated from all other bands, these two bands are also separated from each other by a small gap at X (the two bands would instead be degenerate at X for a non-dimerized chain, i.e., when $d_1=d_2$). Therefore, they are isolated, and each of them can be Wannierized separately.

In Figure~\ref{fig:wannierisation-singlebands}(c), we also plot the wavefunctions $\psi_{n\mathbf k}$ as provided by the \textit{ab initio} engine, computed on a regular $4\times 1\times 1$ grid in the \gls{BZ} composed of the four points $\mathbf k_1 = \Gamma = (0, 0, 0)$, $\mathbf k_2 = (\pi/2a, 0, 0)$, $\mathbf k_3 = \text{X} = (\pi/a, 0, 0)$, $\mathbf k_4 = (3\pi/2a, 0, 0)$.
The legend to interpret these plots is provided in panel (a) of Fig.~\ref{fig:wannierisation-singlebands}: in particular, we represent the real (green) and imaginary (red) components of complex wavefunctions (e.g., $\psi_{n\mathbf k}$, $\psi^{\text W}_{n\mathbf k}$, $\ldots$) along the $x$ axis passing through the carbon chain, i.e., $\psi_{n\mathbf k}(x, 0, 0)$. 
Given the $4\times 1\times 1$ sampling of the \gls{BZ}, both the wavefunctions and the resulting \glspl{WF} are periodic in a real-space $4\times 1\times 1$ supercell, i.e., the same one shown in the panels of Fig.~\ref{fig:wannierisation-singlebands}.
Note that, while in most use cases the $\psi_{n\mathbf k}$ are eigenstates of the Hamiltonian, this might not be true for all codes. 
In general, the phases of each state $\psi_{n\mathbf k}$ are random (e.g., they often come from independent diagonalizations at each $k$ point). In addition, if we consider multiple bands with degeneracies, then the degenerate states would also be randomly mixed among each other. (This would happen, for instance, at the $k$ point $\text{X}$ for a non-dimerized chain, as discussed earlier.) As a consequence, the simple sum of the $\psi_{n\mathbf k}$ according to Eq.~\eqref{eq:wanniertransform} 
does not provide a well-localized function, as shown in panel (d) of Fig.~\ref{fig:wannierisation-singlebands}.

However, each $\psi_{n\mathbf k}$ at every $k$ point can be rotated by the optimized $U_{\mathbf k}$ matrices obtained from a Wannierization procedure.
In this case, since $J=1$, the $U_{\mathbf k}$ at each $k$ point is a $1\times 1$ matrix, i.e., simply a complex phase $U_{\mathbf k} = e^{i \phi_{\mathbf k}}$ that can be visualized as a rotation in the plot; this is marked by the blue circular arrow in Fig.~\ref{fig:wannierisation-singlebands}(a).
In this simple one-dimensional case (i.e., for a single band) the Wannierization is essentially trivial, as discussed e.g. in Sec.~IV.C.1 of Ref.~\cite{Marzari1997}: one has to just impose that the same-band overlaps between adjacent $k$ points are equal to $e^{i\phi_n/4}$, where $\phi_n$ is the the Berry phase of the entire band $n$ and the factor $\frac 1 4$ accounts for the fact that there are four $k$ points (see later Eq.~\eqref{eq:berry-phase} in Sec.~\ref{subsec:topoinv} for a discussion of Berry phases). This condition is realized by making the appropriate phase rotations $e^{i \phi_{\mathbf k}}$ at every $k$ point that counterbalance the random phases coming from the \textit{ab initio} code, thus delivering maximal smoothness in reciprocal space, corresponding to the so-called twisted parallel-transport gauge~\cite{Vanderbilt2018}.
The rotated states $\psi^{\text W}_{n\mathbf k}$ are shown in panel (e) and (g) for the second lowest ($n=2$) and lowest ($n=1$) bands, respectively.
When these $\psi^{\text W}_{n\mathbf k}$ are summed as prescribed by Eq.~\eqref{eq:wannier}, localization emerges and we obtain the final \glspl{MLWF} WF$_1$ and WF$_2$, shown both in panels (f) and (h), respectively, and in panel (b) as 3D isosurfaces.
Note that a final global phase rotation might be required to ensure real-valued \glspl{MLWF}.
Intuitively, WF$_1$ can be interpreted as (predominantly) originating from a linear combination of $s$ orbitals on the C atoms, while WF$_2$ as (predominantly) originating from a linear combination of $p_x$ orbitals on the C atoms (with opposite signs, so that the positive part of the $p_x$ orbitals centered on two neighboring C atoms sums constructively in the middle of the C--C bond).

\begin{figure*}[tbp]
\centering
% {left bottom right top}
%\fbox{
  \includegraphics[trim={0 0 0 500px},clip,width=0.8\textwidth]{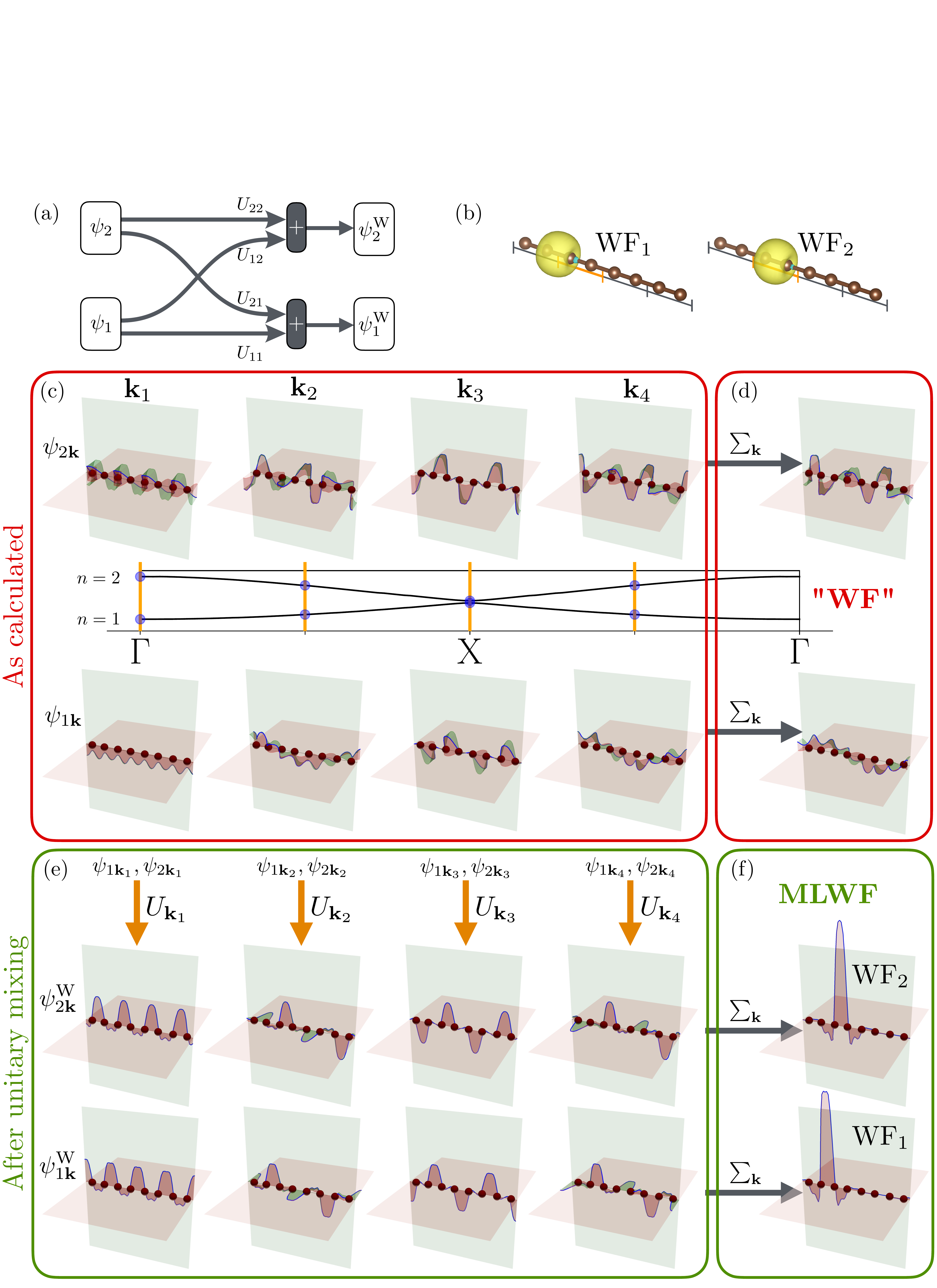}%
%}
\caption{\label{fig:wannierisation-intro}
Same as Fig.~\ref{fig:wannierisation-singlebands},
but now treating the two lowest bands as a composite
group. (a) Graphical representation of the generalized gauge
transformation $U_{\mathbf k}$ in Eq.~\eqref{eq:2x2-unitary-action}
that mixes the Bloch states from the two bands.
As the result of this mixing, it is no longer possible to
associate each MLWF obtained from Eq.~\eqref{eq:wanniertransform}
with a specific band.
By taking advantage of the generalized gauge freedom, the two composite
\glspl{MLWF} in the $\mathbf R = 0$ cell, shown in panels (b) and (f), are better
localized than their single-band counterparts in
Fig.~\ref{fig:wannierisation-singlebands}. Note that WF$_1$ and WF$_2$ are not identical, i.e., they are not related by a rigid
translation, even though they are both bond centered and look similar.
This is because the chain is dimerized, and thus the
respective bond lengths are different.
}
\end{figure*}

We can also treat the two bottom energy bands as a composite group. 
In this case the $U_\mathbf k$ are $2\times 2$ unitary matrices, whose
action on the Bloch wavefunctions,
  \begin{equation}
     (\psi^\text W_1\quad \psi^\text W_2) = (\psi_1\quad \psi_2)\; 
     \left(
    \begin{array}{cc}
    U_{11} &    U_{12} \\
    U_{21} &    U_{22}
    \end{array}
    \right) ,
    \label{eq:2x2-unitary-action}
\end{equation}
is represented schematically in Fig.~\ref{fig:wannierisation-intro}(a).
We note that, in such a simple case with only two bands, the action of a unitary matrix could be generally interpreted as a sequence of complex phase rotations of the two initial states, a rotation that mixes the two states, and a final additional complex phase rotation between the two final states.
In this case, each pair of $\psi_{n\mathbf k}$ at every $k$ point can be mixed by the optimized $U_{\mathbf k}$ matrices obtained from a Wannierization procedure. The rotated states $\psi^{\text W}_{n\mathbf k}$ are shown in panel (e) (for convenience of the reader, we repeat in panels (c) and (d) of Fig.~\ref{fig:wannierisation-intro} the same band structure and $\psi_{n\mathbf k}$ wavefunctions as in panels (c) and (d) of Fig.~\ref{fig:wannierisation-singlebands}).
When these $\psi^{\text W}_{n\mathbf k}$ are summed as prescribed by Eq.~\eqref{eq:wannier}, we obtain the final \glspl{MLWF} $\ket{\mathbf Rj}$ ($j=1, 2$) shown in panel (f), and localization is now apparent (the figure shows the \glspl{MLWF} for $\mathbf R = \mathbf 0$). Also in this case, a final global phase rotation might be required to ensure real-valued \glspl{MLWF}.
We finally note that, since we now give more freedom to the states to mix, the resulting \glspl{MLWF} are more localized than those of Fig.~\ref{fig:wannierisation-singlebands}(b). Also, the two \glspl{WF} of Fig.~\ref{fig:wannierisation-singlebands} (separate Wannierization) can be obtained as a linear combination of the eight \glspl{WF} of Fig.~\ref{fig:wannierisation-intro} (combined Wannierization), i.e., WF$_1$ and WF$_2$ translated by $\mathbf R_{-1}=(-a, 0, 0)$, $\mathbf R_0=(0, 0, 0)$, $\mathbf R_1=(a, 0, 0)$ and $\mathbf R_2=(2a, 0, 0)$. The opposite is also true: WF$_1$ and WF$_2$ of Fig.~\ref{fig:wannierisation-intro} are a linear combination of the eight \glspl{WF} of Fig.~\ref{fig:wannierisation-singlebands} translated by $\mathbf R_{-1}$, $\mathbf R_{0}$, $\mathbf R_{1}$, $\mathbf R_{2}$.

\subsubsection{Entangled bands} \label{sec:entangled_bands}

\begin{figure*}[tb]
  \includegraphics[width=15cm]{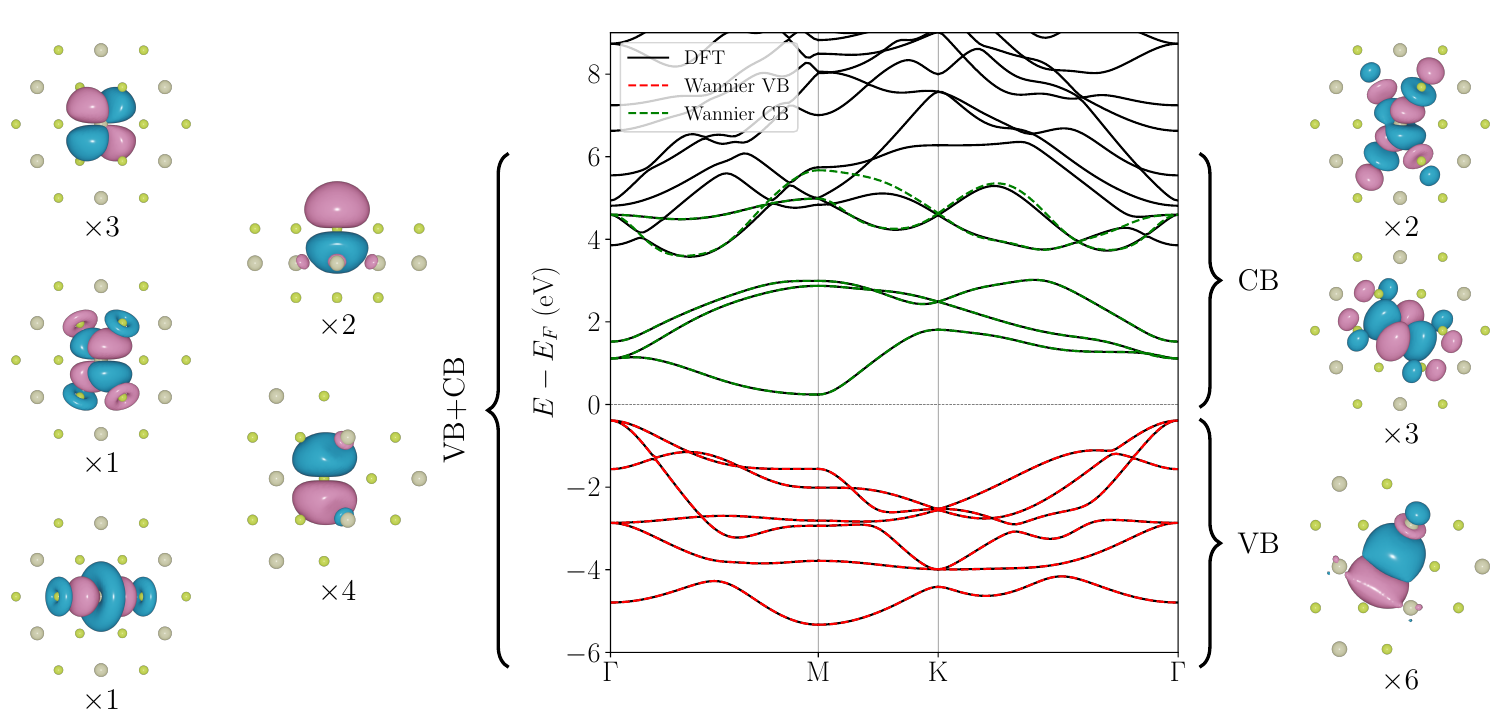}
  \caption{\glspl{MLWF} and band structure of the 2D material
    HfSe\textsubscript{2}. In the center, the comparison between the
    \gls{DFT} band structure (black lines) and the
    Wannier-interpolated band structure from valence \glspl{MLWF} only
    (VB, red dashed lines) and from low-lying
    conduction \glspl{MLWF} only (CB, green dashed lines) are shown.
    Note that the Wannier-interpolated bands from both valence +
    conduction \glspl{MLWF} together (VB+CB) are not shown since they
    are visually indistinguishable from the combination of the VB and
    CB \glspl{MLWF}.  On the left, we show the real-space shapes of
    (in total 11) valence + conduction \glspl{MLWF}, obtained starting
    from Hf $d$ and Se $p$ initial guess orbitals, followed by
    disentanglement (see Sec.~\ref{sec:entangled_bands}) from
    high-energy conduction states: specifically, three of them
    resemble $d_{xy,xz,yz}$ orbitals, one resembles a $d_{z^2}$
    orbital, one resembles a $d_{x^2-y^2}$ orbital, and the remaining
    six resemble $p$ orbitals. Some small hybridization with orbitals
    from nearby atoms is visible.  On the right, we show the
    real-space shapes of (in total 6) valence \glspl{MLWF} (lower
    panel) and (in total 5) conduction \glspl{MLWF} (upper panel).
    The valence \glspl{MLWF} span an isolated group of bands (see
    Sec.~\ref{sec:isolated_bands}) and are composed by six hybridized
    bonding orbitals, where the Hf $d$ and Se $p$ orbitals overlap
    constructively.  The conduction \glspl{MLWF} are instead five
    hybridized anti-bonding orbitals, where Hf $d$ and Se $p$ orbitals
    overlap destructively, forming nodal planes close to bond
    centers. (The notation $\times n$ below each shape denotes the
    multiplicity of the corresponding \gls{MLWF}, i.e., $n$
    \glspl{MLWF} having similar shapes but different spatial
    orientation.)  } \label{fig:HfSe2}
\end{figure*}

The MV approach described above provides a means to construct
well-localized \glspl{WF} from isolated groups of bands, such as the valence
bands of insulators.  However, it is often useful to obtain \glspl{WF} from
non-isolated (or ``entangled'') groups of bands. Typical examples
include the low-lying conduction bands or
the valence plus conduction bands of insulators (see Fig.~\ref{fig:HfSe2}),
and the bands crossing the Fermi level in metals.

A possible strategy to deal with such cases is to first identify an
appropriate $\nbw$-dimensional Bloch manifold at each $k$ point from a
larger set of $\nbands_\kk$ Bloch eigenstates $\ket{\psi_{m\kk}}$,
e.g., the ones within some energy window. Formally, this
band-disentanglement step can be expressed as
\begin{equation}
\label{eq:Wannierization-matrix}
\ket{\tilde{\psi}_{n\kk}}=\sum_{m=1}^{\nbands_\kk}\,\ket{\psi_{m\kk}}
\tilde V_{\kk,mn}\,,
\end{equation}
where the $\tilde V_\kk$ are $\nbands_\kk\times J$ matrices satisfying
\mbox{$\tilde V_\kk^\dagger\tilde V_\kk =\mathds{1}_{\nwann\times\nwann}$}.
In 2001, Souza, Marzari and Vanderbilt~\cite{Souza2001} (SMV)
introduced a practical scheme to extract an optimally smooth
Bloch-like subspace
$\hat P=\sum_{n=1}^\nwann\,\ket{\tilde u\bnk}\bra{\tilde u\bnk}$ across
the \gls{BZ}, from which a set of \glspl{MLWF} could then be obtained using the MV
prescription. The resulting ``disentangled \glspl{WF}'' are given by
Eq.~\eqref{eq:wannier-psi-W} with the \textit{ab initio} Bloch
eigenstates $\ket{\psi_{n\kk}}$ therein replaced by the disentangled Bloch-like states
$\ket{\tilde{\psi}_{n\kk}}$, that is,
\begin{subequations}
\begin{align}
\ket{{\bf R}j}&= \;\frac{1}{N}\sum_\kk
\,e^{-i{\bf k\cdot {\bf R}}}\ket{\psi^\text{W}_{j{\bf k}}}\,,
\label{eq:wannier-dis}\\
\ket{\psi^\text{W}_{j{\bf k}}}&=
\sum_{n=1}^{\nbands_\kk}\,\ket{\psi_{n{\bf k}}}V_{\kk,nj}\,,
\label{eq:psi-smooth}
\end{align}
\label{eq:wannier-psi-dis}%
\end{subequations}
where the $\nbands_\kk\times J$ matrices $V_\kk=\tilde V_\kk U_\kk$
encode the net result of the disentanglement (subspace-selection) and
maximal localization (gauge-selection) steps.  As in the case of
Eq.~\eqref{eq:wannier-psi-W}, the states
$\ket{\psi^\text{W}_{j{\bf k}}}$ in Eq.~\eqref{eq:wannier-psi-dis} are
smooth functions of $\kk$ whenever the associated \glspl{WF} are well
localized.

\begin{figure*}[tbp]
  \centering
  %\fbox{
    \includegraphics[trim={0 50px 0 250px},clip,width=0.9\textwidth]{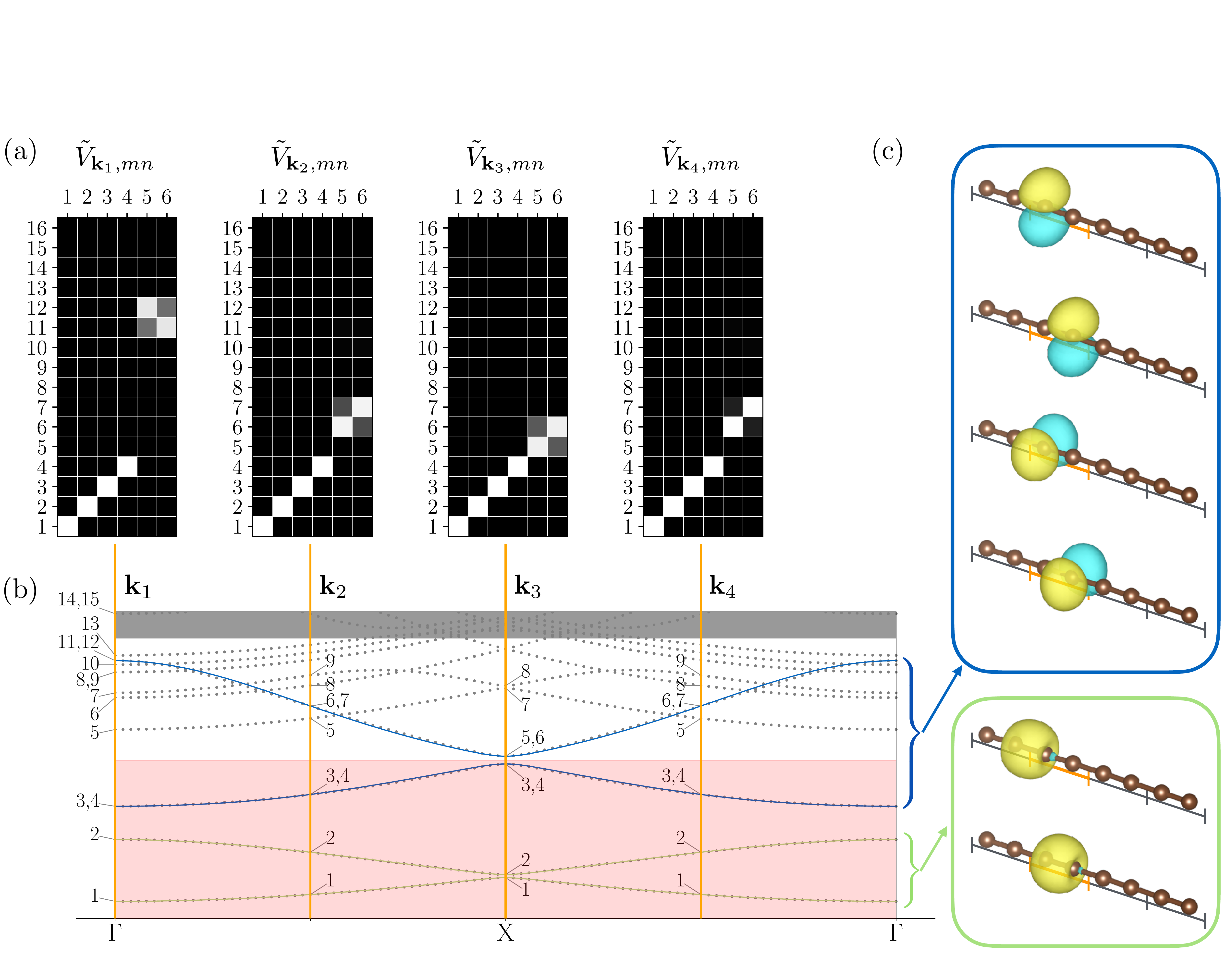}%
  %}
  \caption{\label{fig:disentanglement-intro}
  Illustration of the disentanglement procedure for the same linear
  carbon chain as in Figs.~\ref{fig:wannierisation-singlebands}
  and~\ref{fig:wannierisation-intro}.
  (a) Graphical representation of the $\tilde V_{\mathbf k, mn}$
  matrices in Eq.~\eqref{eq:Wannierization-matrix}
  at $k$ points $\mathbf k = \mathbf k_1$, $\mathbf k_2$, $\mathbf
  k_3$, $\mathbf k_4$. The color scale represents the absolute values
  of the matrix elements from black (zero) to white (maximum absolute
  value). The index $m$ (from 1 to 16) labels the bands, and the index
  $n$ (from 1 to 6) labels the disentangled Bloch states $\ket{\tilde
  \psi_{n\mathbf k}}$.
  (b) Band structure of the carbon chain. Dotted gray points denote the
  \gls{DFT} bands, solid lines the disentangled
  (Wannier-interpolated) bands.  The frozen
  energy window is shown with a red background, while the energy range with
  gray background is outside the disentanglement window. The relevant
  band indices at $\mathbf k_1$, $\mathbf k_2$, $\mathbf k_3$ and
  $\mathbf k_4$ are indicated. The six disentangled bands are shown in
  green (bottom two, the same bands as in
  Figs.~\ref{fig:wannierisation-singlebands}
  and~\ref{fig:wannierisation-intro}) and in blue (top four,
  doubly degenerate).  To construct the topmost blue bands, the
  disentanglement procedure correctly picks up the optimally connected
  \gls{DFT} bands as a function of $\mathbf k$: bands 11 and 12 at
  $\mathbf k_1$, bands 6 and 7 at $\mathbf k_2$ and $\mathbf k_4$, and
  bands 5 and 6 at $\mathbf k_3$. Notably, in this example the topmost
  blue bands can be exactly disentangled from all other \gls{DFT} bands,
  such that the disentangled bands coincide at every $\mathbf k$ with the
  corresponding \gls{DFT} bands. We note, however, 
  that this is not possible in general for an arbitrary band structure (compare,
  for instance, with the conduction bands in Fig.~\ref{fig:HfSe2}).
  (c) Resulting \glspl{MLWF} on a $4\times 1 \times 1$ supercell (the
  primitive unit cell is indicated by the orange segment).  Since in
  this specific example the disentanglement matrices $\tilde V$ are
  block diagonal (as well as the $U$ matrices from the Wannierization
  procedure, not shown here), the two bonding-like \glspl{MLWF}
  originate only from the bottom two (green) bands and do not mix with
  the four $p_y$ and $p_z$ \glspl{MLWF} from the other four (blue)
  bands.   
  }
\end{figure*}

The disentanglement step can be carried out in such a way that the
\textit{ab initio} eigenstates are described exactly within a
``frozen'' or ``inner'' energy window that is contained by the
``outer'' energy window mentioned earlier~\cite{Souza2001}.  This is
useful, for example, when studying transport properties, for which one
would like to obtain a faithful description of the states within some
small energy range around the Fermi level. Note that
  because of these energy windows, the required input from the
  first-principles calculation includes the energy eigenvalues
  $\varepsilon\bnk$ in addition to the overlap matrices
  Eq.~\eqref{eq:Mkb}.
For illustrative purposes, in Fig.~\ref{fig:disentanglement-intro} we display the $\tilde V_{\mathbf{k}}$ matrices as calculated for the carbon chain discussed earlier. We also stress that, in this simple illustrative example, we can exactly disentangle all six bands from the rest, but this is not true in general (see, for example, the conduction bands in Fig.~\ref{fig:HfSe2}).

Over the years, many
alternative approaches and algorithms have been developed\textemdash from
partially occupied Wannier functions~\cite{Thygesen2005,Fontana2021}
to quasi-atomic orbitals~\cite{Qian2008}, to the
\gls{SCDM}~\cite{Damle2015,Damle2017}, and to projectability
disentanglement and manifold remixing~\cite{Qiao2023,Qiao2023a}; these
and others will be discussed in
Sec.~\ref{subsec:advancedminimisation}.

\subsubsection{The projection method} \label{sec:projections}
The MV and SMV approaches leverage iterative minimization, hence a good starting guess for the unitary matrices $U_{\mathbf{k}}$ and $\tilde{V}_{\mathbf{k}}$ is crucial to avoid being trapped in local minima of the spread functional. A popular approach is the \emph{projection method}, where a set of $J$ localized ``trial functions'' $g_n({\mathbf{r}})$ are chosen by guessing the orbital character and location of the target \glspl{WF}. These functions are typically Gaussians, atomic-like orbitals with angular character such as $s,p,d$, or hybrid orbitals such as $sp^3$. The first step is to project the Bloch manifold onto these trial orbitals:
\begin{equation}
\ket{\phi_{n\kk}}=\sum_{m=1}^{\nbands_\kk}\ket{\psi_{m\kk}}\braket{\psi_{m\kk}|g_n}.
\end{equation}
Then, L\"owdin orthonormalization is performed by inverting the overlap matrix $S_{\mathbf{k},mn}=\braket{\phi_{m\kk}|\phi_{n\kk}}=(A^{\dagger}A)_{\mathbf{k},mn}$
\begin{equation}
\ket{\tilde{\psi}_{n\kk}}=\sum_{m=1}^{\nbands_\kk}\,\ket{\psi_{m\kk}}
(A_{\kk}S^{-\frac{1}{2}}_{\kk})_{mn}
\end{equation}
where the $A_{\kk,mn}=\braket{\psi_{m\kk}|g_n}$ is called the projection matrix. 
The matrix $A_{\kk}S^{-\frac{1}{2}}_{\kk}$ is unitary and can be computed through the singular value decomposition of $A=ZDW$:
\begin{equation}
  A_{\kk}S^{-\frac{1}{2}} = Z\mathds{1}W,
\end{equation}
where the diagonal matrix $D$ is replaced with the identity~$\mathds{1}$.

The choice of the trial orbitals for composite bands is less critical with respect to entangled bands and, for simple compounds, even a set of Gaussians randomly centered in the cell might work. We emphasize that if the manifold of composite bands coincide with the valence band of an insulator or semiconductor, then the \glspl{MLWF} will reflect the local chemistry: for instance, in covalent materials \glspl{MLWF} are typically bond-centered as in Si or GaAs~\cite{Marzari1997}, with some notable exceptions such as MoS$_2$~\cite{gibertini_natcomm_2014} where one \gls{WF} is centered in the middle of the hexagonal cell due to the hybridization of several orbitals. On the contrary, the SMV disentanglement hinges on a careful choice of trial functions, that define the orbital character of the bands to be extracted. Whilst disentangling at once the valence and conduction manifold often yields atom-centered \glspl{WF}, this is not true in general: \glspl{MLWF} for the low-lying bands of copper result in five Cu d-like \glspl{WF} and two additional \glspl{WF} centered at the tetrahedral-interstitial locations~\cite{Souza2001}.

Different aspects of the projection method are discussed in Sec.~\ref{subsec:practitioner} and~\ref{subsec:advancedminimisation}.
\subsection{Major applications of Wannier functions}
\paragraph{Interpolation} The efficient interpolation in reciprocal space of $k$-dependent quantities is arguably the most common application of \glspl{WF}, enabling the calculation of simple (e.g., the band structure) or complex (e.g., electron-phonon coupling) electronic-structure properties. A large part of this review is devoted to the fundamentals of \gls{WF} interpolation (Sec.~\ref{subsec:tb}) and their applications, including ballistic transport (Sec.~\ref{sec:ballistic}), Berry-phase related properties (Sec.~\ref{subsec:Berry}) and electron-phonon interactions (Sec.~\ref{subsec:elph}). As discussed in more detail in Sec.~\ref{subsec:tb}, the reason for such widespread set of applications (not all of them covered in this review) is that \glspl{WF} can be easily applied to any generic operator that is local in reciprocal space, i.e., any lattice-periodic operator. More generally, we note that even some non-local operators in reciprocal space (e.g. containing the position operator, which is not lattice periodic and transforms into $k$ derivatives) can also be interpolated, see e.g. Sec.~\ref{subsec:Berry} on Berryology. Equally important is that \glspl{WF} allow reproducing the correct band connectivity: in particular, avoided crossings are not mistaken for actual crossings. This distinguishes Wannier interpolation from other methods based on direct Fourier interpolation of the energy eigenvalues. In other words, \glspl{WF} allow to exploit the fundamental locality (``nearsightedness'' according to Kohn~\cite{desCloizeaux_PR1964_1,desCloizeaux_PR1964_2,Kohn1996}) of the electronic structure and the related exponential localization of \glspl{WF} to construct a potentially exact representation of an operator in real space, such that any interpolation back to reciprocal space is exact as well. The procedure is also systematic as \glspl{WF} are guaranteed to exist and the convergence is exponential with the linear sampling density~\cite{Panati2007,Brouder2007,Panati2013}; prefactors and coefficients might depend on electronic-structure properties such as the band gap, and on the specific operator under consideration.

\paragraph{Geometry and Topology} \glspl{WF} have several profound connections with quantum-geometrical and topological aspects of the electronic structure~\cite{Vanderbilt2018}; some of them are discussed in Sec.~\ref{subsec:Berry} and~\ref{subsec:topoinv}. In the following, we refer to topological properties as a subset of geometrical properties that are quantized and hence represented by integer topological invariants robust to certain classes of perturbations. A prime example of geometrical\textemdash and in some circumstances also topological\textemdash quantity is the electric polarization of periodic solids, which can be calculated in reciprocal space as a Berry phase~\cite{Vanderbilt2018} (see Sec.~\ref{subsec:Berry}). 
Electric polarization can also be equivalently computed by summing over \glspl{WF} centers in real space~\cite{Marzari1997,Marzari2012,Vanderbilt2018}, which provides a more intuitive formulation of the modern theory of polarization~\cite{Resta1992,KingSmith1993,Resta_RMP_1994,Vanderbilt2018} and restores some justification to the classical Clausius--Mossotti~\cite{clausius,mossotti} viewpoint. While electronic-structure geometry in reciprocal space speaks the language of differential geometry (e.g., curvatures, parallel transport, smoothness of manifolds), \glspl{WF} allow to express the same quantities in terms of matrix elements of the Hamiltonian and position operator $\hat{\mathbf{r}}$. The reciprocal-space smoothness, which is measured by the quantum geometric tensor~\cite{provost_cmp_1980}, can be equivalently analyzed in real space by measuring the degree of \glspl{WF} localization.

To some extent, the connection between \glspl{WF} and topological invariants is even stronger, where the former provide not only powerful approaches to calculate invariants for real materials (see Sec.~\ref{subsec:topoinv} for a discussion) but also fundamental understanding of topological phases. In fact, topological insulators are essentially systems that cannot be connected adiabatically to atomic insulators, hence it is impossible to truly represent their ground state with \glspl{WF}~\cite{timo_prb_2006,Vanderbilt2018}. In reciprocal space, non-trivial topological invariants translate into unavoidable obstructions to choosing a smooth gauge over the \gls{BZ}~\cite{Vanderbilt2018,bernevig_book_2013}. The fundamental connection between non-trivial topology of electronic bands and the corresponding absence of a Wannier representation has been recently generalized and made systematic in the context of elementary band representation~\cite{zak_prb_1982,michel_zak}, leading to the so-called topological quantum chemistry \cite{Bradlyn2017,graph_theory,cano_TQC} (see also related efforts on symmetry-based indicators by~\textcite{Po_2017,SI_anomalous_surface,Mapping_symmetry-topology,Combinatorics_Slager}), allowing to screen materials databases and identify non-trivial materials of various classes~\cite{Nature_TQC1,Nature_both,Nature_SI1,Wieder2022}.

\paragraph{Advanced electronic-structure methods} \gls{DFT} simulations of periodic solids can be conveniently (but definitely not necessarily) performed by adopting a plane-wave basis set, in conjunction with smooth pseudopotentials that reproduce the interaction between valence electrons and nuclei plus core electrons~\cite{MartinBook2020}. The resulting \gls{KS} eigenstates are also not particularly localized functions, and \gls{DFT} is invariant under unitary rotations of the occupied electronic states. However, several electronic-structure methods, aiming at improving or complementing the capabilities of \gls{DFT}, fundamentally require to be formulated in terms of localized orbitals (see Sec.~\ref{subsec:dmft}). In addition, several of these beyond-\gls{DFT} methods are not deployed directly on the crystal structure, but operate more as corrections to starting~\gls{DFT} calculations. Also, beyond-DFT methods can be computational rather intensive, and it is common practice to apply them only on a subset of bands extracted from the entire manifold. In this context, \glspl{WF} provide a robust way to bridge \gls{DFT} with advanced electronic-structure methods by allowing to systematically construct orthogonal localized states that represent the manifold of interest:~\glspl{WF} are first constructed on the \gls{KS} \gls{DFT} solution and then fed into beyond-DFT methods; a technical overview of how this is carried out in practice is the subject of Sec.~\ref{subsec:dmft}.
\begin{figure*}[tb]
  \centering\includegraphics[width=16cm]{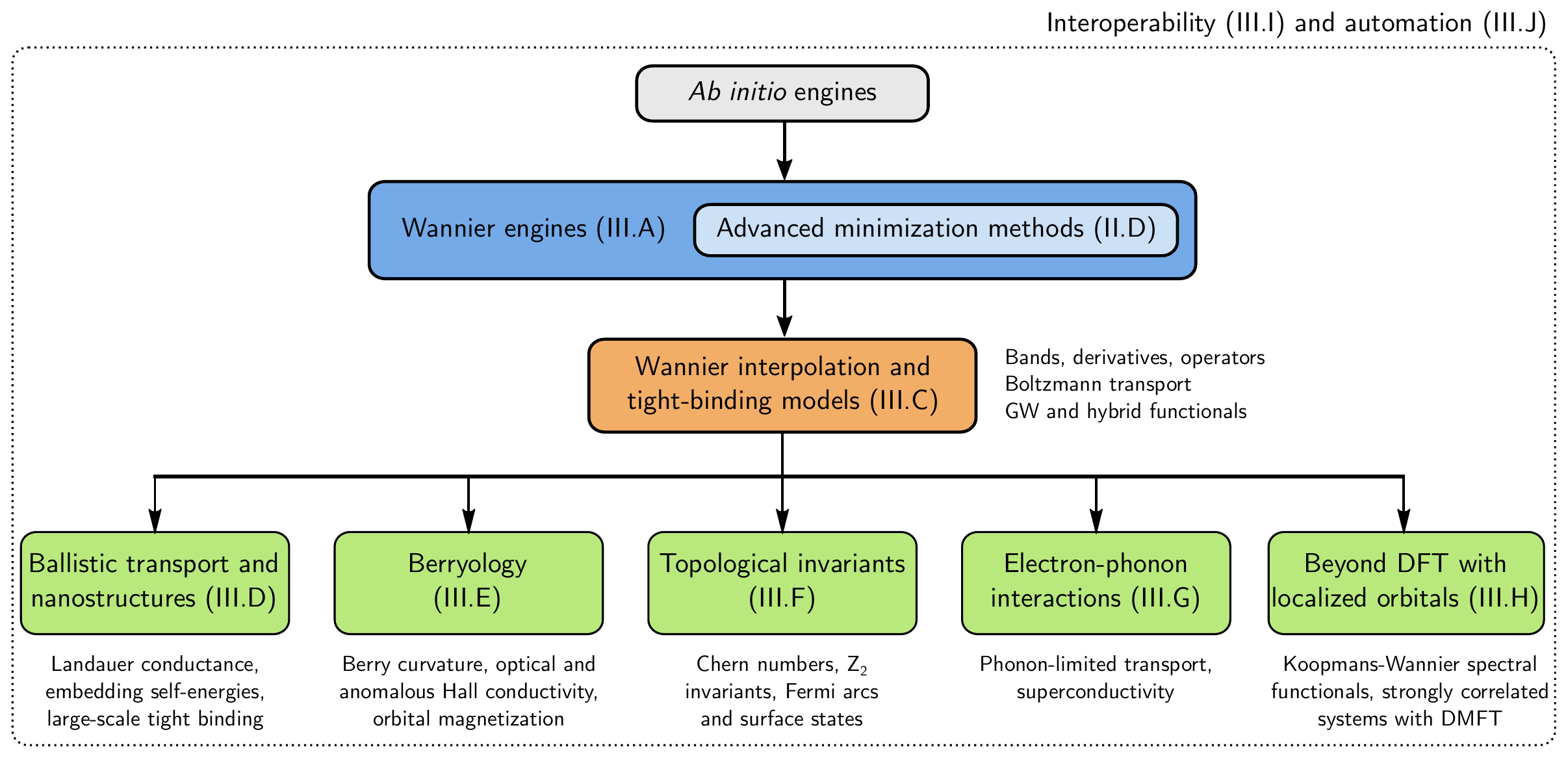}
  \caption{\label{fig:schematics}Overall schematics of how different
    codes in the ecosystem interact. This figure serves also as a
    ``table of contents'' to this review, where sections describing the
    relevant blocks are indicated in the figure.  At the top, the
    \emph{ab initio} engines (grey block) generate the data that is
    transferred to a Wannier engine (blue block), that might implement
    also advanced methods discussed in
    Sec.~\ref{subsec:advancedminimisation}. Once \glspl{WF} are
    obtained, they are typically used to generate a \gls{TB} model and
    perform interpolation of the Hamiltonian and other operators (orange
    block). Such interpolated quantities are used for a number of
    different applications; the ones that are discussed in more detail
    in this review are indicated in the green blocks.  As schematically
    indicated by the outer dotted rectangle, all these codes forming the
    Wannier ecosystem may be automated with workflow tools,
    which also coordinate data transfer
    between them.}
  \end{figure*}

\subsection{\label{subsec:practitioner}Wannier functions for the practitioner}
\subsubsection{\label{subsubsec:loc_func}The spread functional in reciprocal space}

The Blount identities~\cite{blount-ssp62} provide the matrix elements of the position operator between WFs, and, remarkably, prefigure the link between macroscopic properties and integrals (Berry phases) of Berry connections~\cite{KingSmith1993}:
\begin{equation}
\langle{\bf R}i\vert{\bf r}\vert{\bf 0}j\rangle = i\,{\frac{V_\text{cell}}{(2\pi)^3}}
\int d{\bf k} \, e^{i{\bf k}\cdot{\bf R}}
\langle u_{i{\bf k}}\vert\nabla_{\bf k}\vert u_{j{\bf k}}\rangle
\label{eq:rmatel}
\end{equation}
and
\begin{equation}
\langle{\bf R}i\vert r^2 \vert{\bf 0}j\rangle = -
{\frac{V_\text{cell}}{(2\pi)^3}} \int d{\bf k}\,
e^{i{\bf k}\cdot{\bf R}}
\langle u_{i{\bf k}}\vert\nabla_{\bf k}^2\vert u_{j{\bf k}}\rangle
\;\;.
\label{eq:rrmatel}
\end{equation}
It is through these identities that one can recast the spread functional 
$\Omega$ using reciprocal-space expressions, where the gradients and higher derivatives are obtained from finite differences. The building blocks for these finite-difference expressions are the overlap matrices
\begin{equation}
M_{ij}^{(\bf k,b)} = \langle u_{i\bf k}\vert u_{j\bf k+b}\rangle
\label{eq:Mkb}
\end{equation}
between cell-periodic Bloch eigenstates $\vert u_{n\bf k}\rangle$ at neighboring points on a
regular grid in the \gls{BZ} (the {\bf b} vectors connect one $k$ point to its neighbors on a regular discrete grid); in the limit of very dense meshes the {\bf b} vectors tend to zero and the gradient in ${\bf k}$ is recovered. This finite-difference construction remains valid even in the case of $\Gamma$-sampling (e.g. for molecular systems treated with \glspl{PBC} or when considering large supercells), where the neighboring $k$ points are given by the primitive reciprocal lattice vectors ${\bf G}$, 
with the Bloch orbitals differing just by phase factors
$\exp(i{\bf G\cdot r})$. Now, we note that the gradient in $\bf k$ of a function $f(\bf k)$ can be written as
\begin{equation}
\nabla f({\bf k})=\sum_{\bf b}\, w_b \, {\bf b}
 \,[f({\bf k+b})-f({\bf k})]
+{\cal O}(b^2)
 \;\;
\label{eq:grad}
\end{equation}
using stars (``shells") of neighboring $k$ points, where each shell has a weight $w_b$ (see Appendix B in ~\onlinecite{Marzari1997}
and ~\onlinecite{Mostofi2008} for a detailed
description); for a linear function $f({\bf k})=f_0 + {\bf g}\cdot{\bf k}$ it can be easily verified that the exact result $\nabla_{\alpha} f({\bf k}) = g_{\alpha}$ is recovered. In the simple case of cubic Bravais lattices, the first shell of reciprocal-space nearest neighbors (6, 8 or 12 for Bravais lattices in direct space that are simple cubic, FCC, or BCC) is sufficient; the general case where several shells need to be chosen automatically is detailed in~\textcite{Mostofi2008}. While the procedure is automated, for unusual cases (e.g., very elongated cells) it might be convenient to find manually the most symmetric choice of shells~\cite{posternak2002}.
With these definitions, the diagonal matrix elements of the position operator can be evaluated by finite differences as
\begin{equation}
\langle{\bf 0}i\vert{\bf r}\vert{\bf 0}i\rangle  = - \frac{1}{N} \sum_{\bf k,b}\, w_b \, {\bf b} \,
{\rm Im}\,\ln M_{ii}^{(\bf k,b)},
\label{eq:r_fd}
\end{equation}
where $N$ is the number of $k$ points of the reciprocal-space grid.
More complex expressions for the second derivatives and for the entire spread functional can be obtained~\cite{Berghold:2000}, all equal to the leading order in $b$; the choices made in~\textcite{Marzari1997} were driven by the need to provide the same value for the localization functional under a
transformation that shifts $\ket{\RR i}$ by a lattice vector.
We also note that when using such finite-difference formalism, the spread
functional converges slowly (polynomially) with reciprocal-space sampling, and hence some care must be paid in comparing its actual values in case of calculations performed with different discrete samplings.

Importantly, the finite-differences construction is particularly convenient for constructing \glspl{MLWF} in a code-agnostic form, as the only input needed from the original first-principles calculation is encoded in the overlap matrices $M_{ij}^{(\bf k,b)}$. Thus, once the $M_{ij}^{(\bf k,b)}$ have been calculated,
no further interaction is necessary with the electronic-structure code that calculated
the ground state wavefunctions, making the entire
Wannierization procedure a code-independent post-processing step (see, e.g., ~\onlinecite{Ferretti_2007} for the extension to ultrasoft pseudopotentials and the projector-augmented wave method,
and~\onlinecite{posternak2002,Freimuth_PRB_2008,Kunes_CPC_2010} for the full-potential linearized augmented
planewave method).
As regards the disentanglement procedure, note that because of energy windows the needed input from the
  first-principles calculation includes the energy eigenvalues
  $\varepsilon\bnk$ in addition to the overlap matrices
  Eq.~\eqref{eq:Mkb}.

\subsubsection{\label{subsec:acc_conv}Accuracy and convergence}
While the main focus of this review is on the powerful applications of \glspl{WF}, their successful use relies on the Wannierization process being done correctly: in the following, we briefly comment on fundamental tests and established procedures to assess and improve the quality of \glspl{WF}.

Two major convergence parameters control the quality (and the cost) of
the Wannierization procedure: the spread minimization and the
$k$-point grid used to obtain the initial Hamiltonian eigenstates
(e.g., the cell-periodic part of Bloch states, if working with
\glspl{PBC}).

The spread minimization is generally performed with an iterative steepest-descent 
or conjugate gradients algorithm until results do not change within a certain tolerance.
While the iterative algorithm is in general robust,
the minimization can become trapped in local
minima. As introduced in Sec.~\ref{sec:projections}, the strategy to avoid that is to select a very good starting
point: if the initial spread is sufficiently close to the absolute
minimum, it is more likely to reach it by following the local
gradient. Hence, particular care needs to be paid to select good
projection functions to obtain the initial unitary matrix of
Eq.~\eqref{eq:psi-smooth}, to be then iteratively optimized. In absence of
chemical intuition, a common strategy is to calculate a projected
  density of states on the pseudoatomic orbitals and identify the
  orbital character in the energy region of interest: the atomic
  orbitals which project more on the relevant bands can be used as
  initial projection. Note that
~\glspl{MLWF} are often not atom-centered, and atomic orbitals are not
always good starting projections, as it the case of the valence bands
of monolayer MoS$_2$~\cite{gibertini_natcomm_2014}.  In
Sections~\ref{subsec:advancedminimisation} and~\ref{subsec:automation}
we cover advanced methods to automate the selection of the starting
point for the minimization procedure.

The spread functional measures the degree of localization in real space and, to some extent, the efficiency of the interpolation: more localized \glspl{WF} decay faster in real space, hence they require a smaller \gls{BvK} supercell to include all non-vanishing matrix elements of the Hamiltonian and the other operators, which in turn allows adopting coarser $k$-point grids in the starting electronic-structure simulation that is performed in reciprocal space. Indeed, the accuracy of band interpolation can be considered a proxy for the quality of the underlying \glspl{WF} not only as regards $k$-point convergence: especially in the case of entangled bands (see Sec.~\ref{sec:entangled_bands}), poor interpolation might signal problems in the disentanglement procedure. In addition, it is worth to emphasize that in general the Wannierization procedure is not forced to preserve symmetries (unless dedicated methods designed to do so are employed, see Sec.~\ref{subsec:advancedminimisation}). Hence, the spurious splitting of symmetry-protected degeneracies in the interpolated band structure might signal convergence problems related to the minimization, to the $k$-point convergence or to the choice of projection functions. This holds true not only for crystalline symmetries, but also for time-reversal symmetry, which is particularly relevant in noncollinear simulations of non-magnetic materials in presence of spin-orbit coupling (e.g., topological insulators).

Among other indicators of the quality of~\glspl{WF} we mention the
ratio between their imaginary (Im) and real (Re) part: for isolated
bands (and not considering spin-orbit coupling), \glspl{MLWF} at the global minimum should be real functions
~\cite{Marzari1997}. Note that the
calculation of the Im/Re part, and anything related to \glspl{WF}
themselves and their visualization, requires to have access to the
full Bloch orbitals, not just the
  overlap matrices. 
It is also important to emphasize that different quantities derived from ~\glspl{WF}\textemdash such as the \gls{WF} spread and centers, as well as unitary matrices $\tilde V_\kk$, $U_\kk$\textemdash converge in general with different speeds, also depending on the specific formulation adopted~\cite{stengel_prb_2006}.

Finally, we remark that many of the complications related to producing \glspl{WF} for periodic solids are related to construction of a smooth gauge across the BZ. Hence, supercell Wannierizations with $\Gamma$-only sampling are typically more straightforward and less prone to be trapped in local minima. The challenge there is more on algorithmic efficiency due the large size of the systems involved; a number of $\Gamma$-only dedicated methods have been developed~\cite{silvestrelli_PRB_1999,gygi_cpc_2003,stubbs_prb_2021}.

\subsection{\label{subsec:advancedminimisation}Advanced minimization methods, and beyond maximally localized Wannier functions}
As discussed in section~\ref{sec:howwegothere}, there is in principle large freedom in choosing the recipe to obtain well-localized \glspl{WF}. Not only one can replace the MV spread functional with other cost functions, but also different minimization procedures and their starting points can be chosen, hence affecting the resulting \glspl{WF} and their localization properties. Over the years, a number of methods have been developed to address all these different aspects of the Wannierization procedure.
It is worth emphasizing that for many of these methods, the initial guess already provides well-localized \glspl{WF}, so that an iterative minimization can in principle be avoided. The unitary matrices $U_\kk$ of these ``projection-only'' \glspl{WF} are set directly by the
 initial projection functions (see Sec.~\ref{sec:projections}). While this choice cannot
guarantee optimal localization properties, it has the advantage of
enforcing some degree of symmetry induced by the choice of atomic
orbitals used as projection functions. However, in all these cases it is possible---and in some cases even recommended---to minimize the MV spread or some other functional as a final step. 

The prime decision deals with the functional that is to be minimized in order to determine the unitary matrices of Eq.~\eqref{eq:wannier-psi-dis}. The most popular choice is the MV \gls{MLWF} procedure~\cite{Marzari1997} for composite bands (see Sec.~\ref{sec:isolated_bands}) and the SMV disentanglement scheme for entangled bands~\cite{Souza2001} (see Sec.~\ref{sec:entangled_bands}). The minimization of the spread functional leads to very well localized \glspl{WF}, hence reducing the size of the \gls{BvK} supercell needed to represent operators (e.g., the Hamiltonian) in a \gls{WF} basis. While \glspl{MLWF} and disentanglement represent the most convenient choice in the vast majority of applications, substantial work has been done to augment the \gls{MLWF} scheme or develop alternatives which satisfy the needs of specific applications.

The \gls{MLWF} iterative algorithm leads to localized \glspl{WF} in real space, but
is not guaranteed to yield orbitals that preserve desirable
crystal symmetries. This is only partially relieved using symmetric
initial projections, as typically obtained by a proper selection of
atomic-like orbitals.  Symmetry-preserving \glspl{WF} are appealing in
providing the correct orbital or site symmetries for many-body
approaches like \gls{DMFT}.  Hence, it is not surprising that several
non-\gls{MLWF} procedures directly or indirectly include crystal
symmetries in the functional to minimize. In the \gls{SAWF}
method~\cite{Sakuma2013}, symmetric \glspl{WF} are obtained through
additional constraints on the unitary matrices $U_\kk$ which are based
on symmetry operations of the site-symmetry group. The \gls{SAWF}
method is fully compatible with the maximal-localization procedure and
the SMV disentanglement (and has also been recently extended to the case where a frozen window is used~\cite{Koretsune2023}), although the additional constraints imply a
possibly larger total spread, even if some individual \glspl{WF} can
actually be more localized than in the \gls{MLWF}
procedure. Currently, the implementation of the \gls{SAWF} method in the code \code{Wannier90} 
(more about software in Sec.~\ref{subsec:wannier-engines}) is interfaced with the \code{Quantum
  ESPRESSO} distribution and, once the site positions and the orbital symmetries
  of the \glspl{SAWF} are chosen (through the initial projection functions),
  the site-symmetry group can be automatically
  computed by the interface code. If needed, the site-symmetry group can also be
  manually specified by the user, to construct \glspl{SAWF} with target symmetries.

While the \gls{SAWF} method provides a rigorous way to include
symmetries in the maximal-localization procedure, it requires quite
some prior knowledge of the electronic structure of the material under
study. An alternative and simpler approach is to construct
\glspl{SLWF}, where the \gls{MLWF} procedure is applied only
to a subset of the entire \glspl{WF}
considered~\cite{Wang2014}. In addition, some \gls{WF} centers
can be constrained (SLWF+C) to specific positions by adding a
quadratic penalty term to the spread functional. While the SLWF+C
approach does not enforce symmetries, it has been observed that the
resulting \glspl{WF} typically exhibit the site symmetries
corresponding to the constrained centers~\cite{Wang2014}. The SLWF+C
can be used in the case of entangled bands, where the SMV disentanglement step is
performed as usual, while the selective localization and constrained
centers are applied only to the final Wannierization step. A more
in-depth review of the \gls{SAWF} and SLWF+C methods, including their
implementation and usage in \code{Wannier90}, can be found in~\textcite{Pizzi2020}.
Dedicated tools for symmetry analysis and symmetrization of the real-space \gls{WF} Hamiltonian are available~\cite{wannsymm_csc_2022}.

The localization and possibly the symmetry can crucially depend on the
number of \glspl{WF} considered in a given energy range. The so-called
\glspl{POWF}~\cite{Thygesen2005} formalize this observation by
including the relevant unoccupied states which lead to the minimal
spread functional, essentially implementing a bonding--antibonding
closing procedure. \glspl{POWF} can have a high degree of symmetry,
while the bonding-antibonding criterion has been shown to correspond
to the condition of maximal average
localization~\cite{Thygesen2005}. Notably, in the \gls{POWF} scheme
the total spread functional $\Omega=\Omega_I+\tilde\Omega$ is
minimized at once, at variance with the SMV scheme where first the
gauge-invariant part $\Omega_I$ is minimized through the
disentanglement step, and only after the
gauge-dependent part $\tilde{\Omega}$ is minimized through the usual
MV scheme.  The iterative minimization of the total spread $\Omega$
has been further developed by Damle, Levitt and Lin in~\textcite{Damle2019}. They reformulated the Wannierization, which is
a constrained non-linear optimization problem, as unconstrained
optimization on matrix manifolds, where the SMV disentanglement
procedure can be interpreted as a splitting method which represents an
approximate solution.

We stress that even if minimization of the full
   spread functional $\Omega$ guarantees the highest degree of overall
  localization, several of the methods discussed
here can actually produce \glspl{WF} such that a subset of them might
individually be more localized than their maximally localized
counterparts. Along those lines, Fontana et
al. developed the
spread-balanced \glspl{WF}~\cite{Fontana2021}, where they added a
penalty term to the spread functional, proportional to the variance of
the spread distribution among all \glspl{WF} of the system. This
scheme could be less prone to produce solutions with one or several
poorly localized \glspl{WF}, at the price of an increased total spread
for the whole set.  The addition of terms to the spread functional can
also be used to preserve some degree of locality in energy, such as in
the case of mixed Wannier-Bloch functions~\cite{giustino_prl_2006} and
dually localized Wannier functions~\cite{yang_prb_2022}. These
approaches are based on a generalized spread
functional~\cite{gygi_cpc_2003} designed to carry both spatial
localization (Wannier character) and limited spectral broadening
(Bloch character), by minimizing a functional that contains not only a
spatial variance (as for \gls{MLWF}) but also an energy variance.

Once a choice for the functional to be minimized is made (the total
spread as in the \gls{MLWF} scheme, or any other choice), there is
still a lot of flexibility on the choice of algorithm to perform the minimization.  First, one needs to define a
starting guess for the unitary matrices $U_\kk$, which is customarily
obtained by specifying a set of localized projection functions through the projection method introduced in Sec.~\ref{sec:projections}. While
for composite bands a set of randomly centered spherically symmetric
Gaussian orbitals might work, in general more sophisticated choices
are required. Typically, atomic orbitals are used as projection
functions, such as $s,p,d$-orbitals as well as hybrid orbitals (e.g.,
$sp^3$), which are often centered either on atoms or along bond
directions. As discussed in Sec.~\ref{subsec:acc_conv}, the choice of the right atomic orbitals is typically
  based on chemical intuition and can be partly informed by inspecting
  the projected density of states in the energy region of interest.
   Still, the choice of the right
projection orbitals---i.e., those providing a good starting point for
a successful minimization of the target functional---can often be a
non-trivial task, especially in the context of automated
high-throughput materials screening and, more generally, when one
wants to study a novel material never investigated before (especially
in case of unfamiliar orbital composition). Hence, in the last decade
substantial efforts have been targeted at developing automated
algorithms removing the need for users to define appropriate initial
projections.  A first approach in this direction is the \glspl{OPF}
method~\cite{Mustafa2015} for composite bands. In \gls{OPF}, a larger
set of functions that overspan the space of \glspl{MLWF} is built and
used as a starting point. While in plane-wave codes the \gls{OPF}
approach~\cite{Mustafa2015} still needs the user to provide a list of
initial projections, for instance atomic-like \glspl{LO}, \textit{ab
  initio} codes operating with localized (or mixed
plane-wave/localized) basis sets can leverage the built-in localized
orbitals.  For instance, the full-potential \gls{LAPW} method can be
extended by adding the so-called \glspl{LO}, which are atomic-like,
very localized, and can be employed in the construction of
\glspl{WF}. \textcite{Tillack2020} combined the SMV
disentanglement with the \gls{OPF} method~\cite{Mustafa2015} to
construct initial guesses for \glspl{MLWF} from a set of \glspl{LO} in
an automated way.  Finally, another set of parameters that require
tuning in the standard SMV disentanglement scheme are the inner and
outer energy windows.  \textcite{Gresch2018} targeted the
removal of the need for manual input by focusing on the automated
optimization of both windows.

On the algorithmic side, in the quest for fully automating the
generation of localized \glspl{WF}, various general and practical
approaches have been recently proposed, targeting the construction of
well-localized \glspl{WF} using algorithms that are often non-iterative;
this not only makes them more automatable, but also provides an excellent starting point for a
final Wannierization, if required: the \gls{SCDM}
approach~\cite{Damle2015,Damle2017,Damle2018}, the continuous Bloch
gauges~\cite{Cances2017,Gontier2019}, and the projectability
disentanglement and manifold remixing
approaches~\cite{Qiao2023,Qiao2023a}.

\gls{SCDM} is based on \gls{QRCP} of the reduced single-particle
density matrix. The approach can be used either to produce
well-localized \glspl{WF} without performing an iterative
minimization, or it can be considered a linear-algebra method to
identify a good starting point for a \gls{MLWF} procedure. The
\gls{SCDM} method is implemented~\cite{Vitale2020} in the interface
code to \code{Quantum ESPRESSO}, with an algorithm for the \gls{QRCP}
factorization that works on a
smaller matrix instead of the full density
matrix~\cite{Damle2015,Damle2017,Damle2018}. For a set of composite
bands, \gls{SCDM} is parameter-free. A comprehensive study on 81
insulators~\cite{Vitale2020} has shown how the \gls{MLWF} procedure applied to \gls{SCDM} initial
  projections (SCDM+MLWF) improves the interpolation
  accuracy and localization of the resulting \glspl{WF}, although
  \gls{SCDM}-only \glspl{WF} perform already remarkably well, both in
  terms of accuracy of band structure interpolation and in terms of
  localization. In the case of entangled bands, \gls{SCDM} requires to
  specify an energy-window function, its center and width, as well as
  the number of \glspl{WF} to consider. \textcite{Vitale2020}
  introduced a recipe to automatically select those parameters (more
  details in Sec.~\ref{subsec:automation}), which was tested on 200
  bulk materials.
Unlike in the case of isolated groups
  of bands, for entangled bands the \gls{SCDM}+\gls{MLWF}
method greatly improved the localization of the
\glspl{WF} with respect to \gls{SCDM}-only. Notably, however, the
reduced spread induced by the MV procedure might result in lower
band-interpolation accuracy. Although the \gls{SCDM} projections could
be used together with the SMV disentanglement scheme, this more
complex procedure does not provide systematic gain in accuracy, such
that a \gls{SCDM}-only or \gls{SCDM}+\gls{MLWF} (with MV minimization
for $\tilde{\Omega}$) is recommended~\cite{Vitale2020}.  In addition,
we note that \gls{SCDM} requires real-space
wavefunctions as input, and therefore has a higher computational and memory
cost with respect to other methods discussed later in this section,
 which are implemented in reciprocal space.

 Another non-iterative approach for composite bands is based on
 ensuring a continuous Bloch gauge over the entire
 \gls{BZ}~\cite{Cances2017,Gontier2019}, resulting in good
 localization properties of the \glspl{WF} and not requiring any
 chemical intuition for their construction. 
 The main idea is that one can construct a sequence of
 gauge matrices that are not only continuous across the \gls{BZ} but also satisfy
 its periodicity at the BZ edges.
 This is achieved by first adopting parallel transport for the gauge matrix, starting from a chosen $k$ point
 (usually the $\Gamma$ point) and propagating along a line (e.g. the $k_x$ line). This enables the
 periodicity to be fixed at the two end points of the line, while preserving the continuity.
 Then, for each $k$ point on the line, parallel transport is applied to each of the gauge matrices
 along an orthogonal direction (e.g. in the $k_y$ direction), and all gauge matrices are fixed again at the endpoints, to ensure periodicity of the 2D plane. Finally, for each $k$ point on
 the 2D plane, one can apply a similar procedure to construct gauge matrices
 along the third direction, therefore obtaining a global continuous gauge across the full
 \gls{BZ}.
 Often, the resulting gauge is continuous but not smooth
 enough: a subsequent conventional MV iterative minimization
 can improve the localization and reach the \gls{MLWF} gauge.  Such
 algorithm is able to construct \glspl{MLWF} for difficult cases such
 as $\mathbb{Z}_2$ topological insulators~\cite{Gontier2019}.

Lastly, a very robust approach has emerged in the form of
projectability disentanglement~\cite{Qiao2023}, where the inner and
outer energy windows are replaced by projectability thresholds. For
each state $\vert u_{n\bf k}\rangle$, a
projectability~\cite{Agapito2013,Vitale2020} onto localized atomic
orbitals (typically, those coming from the pseudopotentials) is
calculated. Then, states that have very high projectability are
retained identically (exactly as done for states inside the inner
frozen window in the SMV method); states that have low projectability
are discarded altogether (since they do not provide useful
contributions to \glspl{MLWF}); and states that span the intermediate
projectability values are treated with the standard SMV
disentanglement.  This approach leads naturally and
robustly to atomic-like \glspl{PDWF}, spanning both occupied and
unoccupied states corresponding to Bloch sums of bonding/anti-bonding
combinations of atomic orbitals.
These \glspl{PDWF} can in turn be remixed into linear combinations
that aim to describe target submanifolds; e.g., the valence states
only, the conduction states only, or certain groups of bands that are
separated in energy from the rest. This may be beneficial for finding optimal target states for beyond-DFT methods (see Sec.~\ref{subsec:dmft}).

This remixing is particularly valuable for Koopmans
functionals (see Sec.~\ref{subsubsec:KF}), which require separate sets of \glspl{MLWF} for the
valence and conduction manifolds, or for transport calculations.
For this purpose, the
\glspl{MRWF}~\cite{Qiao2023a} are obtained by starting from the
\glspl{PDWF} spanning the whole manifold (valence plus conduction), which is then split by rotating the gauge
matrices into a block-diagonal structure across all the $k$ points,
while simultaneously maintaining the gauge smoothness for each
block. This is achieved by a combination of automated Wannierization
of the whole manifold, diagonalization of the Wannier Hamiltonian,
parallel transport, and maximal localization.  The automated
Wannierization of the whole manifold can be obtained using the
\gls{PDWF} method; the Hamiltonian diagonalization splits the manifold
into desired submanifolds (e.g., two for valence and conduction,
respectively); the parallel transport fixes the gauge randomness to
construct two sets of localized \glspl{WF}; the final maximal
localization smoothens the gauge, leading to subsets of \glspl{MLWF}
for the respective submanifolds.
\textcite{Qiao2023a} demonstrates that,
when combined with \glspl{PDWF}, the \gls{MRWF}
method can be fully automated, and can also
be extended to other types of band manifolds gapped in energy, such as
the single top valence band
of MoS\textsubscript{2}, or the $3d$ and $t_{2g}/e_g$ submanifolds of
SrVO\textsubscript{3}. For high-throughput results of \gls{PDWF} and
\gls{MRWF}, see Sec.~\ref{subsec:automation}.

\section{\label{sec:ecosystem}The Wannier ecosystem: theory and software packages}
\subsection{\label{subsec:wannier-engines}Development of widely available Wannier engines}

The MV and SMV methods described in~\textcite{Marzari1997,Souza2001}
were originally implemented in Fortran 77. The code would compute the
overlaps in Eq.~\eqref{eq:Mkb} and the projection of the periodic part
of the Bloch orbitals onto trial localized states, by reading the
former evaluated on a regular $k$-point grid by a \gls{DFT} code\textemdash
originally by an early version of \code{CASTEP}~\cite{Marzari1997PRL,
  castep}. To provide a more general model, driven by the need to
interface with a \gls{DFT} code based on the \gls{LAPW}
method~\cite{posternak2002}, the choice was made to keep the
calculation of all the scalar products involving Bloch orbitals needed
by the Wannier code within the electronic-structure code of choice,
typically as a postprocessing step. Well-defined
protocols were established to exchange this information writing/reading files to/from
disk and the format of those files fully documented.
The resulting \code{Wannier77} code was released under a GPL
v2 license in March 2004.

In 2005, two of the current authors (AAM and JRY), then working in the
groups of NM and IS respectively, rewrote the routines using modern
modular Fortran, relying on their experience of software development
gained from working on the \code{ONETEP}~\cite{onetep} and
\code{CASTEP}~\cite{castep} \gls{DFT} programs. The resulting
program, \code{Wannier90}~\cite{Mostofi2008}, was released under a
GPL license in April 2006. Following the early layout in the
\code{Wannier77} code, \code{Wannier90} was designed to be easily
interfaced to any electronic-structure code, irrespective of its
underlying basis set.  The first release of \code{Wannier90} came with
extensive documentation, tutorials, and two validation
tests. Development used CVS as a version control system. GP joined the
development effort in 2012, and a new parallel post-processing code
(\code{postw90}) was developed and released in \code{Wannier90}
version 2~\cite{mostofi-cpc14} in October 2013.

\begin{figure}[tb]
  \includegraphics[width=\columnwidth]{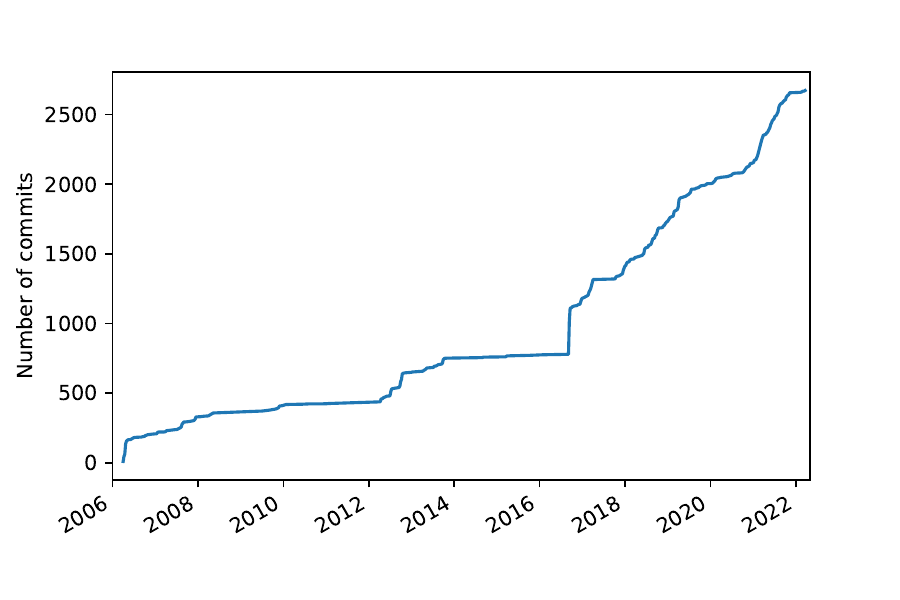}
  \caption{Total number of commits over time for the \code{Wannier90}
    repository (CVS until August 2016, then transferred to Git and hosted on GitHub). Note the
    significant increase during and immediately after the community developer
    workshop in September 2016.} \label{fig:commits}
\end{figure}

While the development of \code{Wannier90} as an open-source
interoperable code was innovative in 2006, by 2016 it was clear that
the development tools being used did not make use of what was then
considered best practice. For example, having only a few developers
with access to the main repository presented a barrier to adding new
functionality to the program.  A decision was made to move to a
community development model, and the \code{Wannier90} repository was
migrated to GitHub with the adoption of a ``fork and pull request''
approach. This new model was launched with a community developer
workshop held in San Sebasti\'an in September of 2016. This event is
clearly recognizable in Fig.~\ref{fig:commits}, which shows the number of commits to the
code repository over time, with a large number of commits contributed
during (or immediately after) the 2016 event. Moreover, once the code made
its transition to a community development model, the rate of commits
significantly increased, as well as the number of individual
contributors (over 35 different people had contributed code, tests or
documentation via commits by the end of 2022).  Essential to this
change was the development of an extensive suite of tests, which run
automatically to validate each pull request. In 2019,
\code{Wannier90} version 3 was released, including all community
contributions to the code~\cite{Pizzi2020}.

In addition to \code{Wannier90}, there are a few electronic-structure software
packages that nowadays implement internal functionality for computing \glspl{WF} in
periodic systems. These include \code{GPAW}~\cite{Enkovaara-GPAW-2010},
\code{QBOX}~\cite{gygi-qbox-2008}, \code{CP2K}~\cite{hutter-cp2k-2020},
\code{exciting}~\cite{Tillack2020}, \code{OpenMX}~\cite{terakura-mlwfs-openmx-2009},
\code{RESPACK}~\cite{Nakamura_et_al:2021} and
\code{CRYSTAL}~\cite{zicovich-wilson-crystal-2001}. 
However, in this review we want to use the term ``Wannier engines'' to describe software packages
that generate \glspl{WF} \emph{and} are designed to be used interoperably
with other software packages: e.g., with the electronic-structure codes for
solving the electronic ground state, used as input to a Wannier
engine, and with post-processing codes that use the \glspl{WF} from a
Wannier engine to calculate advanced electronic properties.
\code{Wannier90} is clearly an example of a Wannier engine, but it is not the only readily available one\textemdash another notable example is the \code{Atomic Simulation Environment (ASE)}~\cite{Larsen2017}, which also implements routines based on the minimization of the quadratic spread of the \glspl{WF}, but uses a different approach~\cite{Thygesen2005, Fontana2021} to that described above (see also Sec.~\ref{subsec:advancedminimisation}).
In addition, the recent \code{Wannier.jl} package~\cite{wannierjl} implements several Wannierization algorithms
using manifold optimization techniques, and brings the methodology of \glspl{WF} to the Julia~\cite{julia} community.
In this review, we focus on those codes that interface to the \code{Wannier90} code; nevertheless, 
we note that often the other Wannier engines have adopted the same file formats first defined by \code{Wannier77}/\code{Wannier90} (see discussion in Sec.~\ref{sec:ecosystem-concept}), thus being fully compatible with the ecosystem.

\subsection{The concept of a Wannier function software
  ecosystem} \label{sec:ecosystem-concept}

To discuss the modular approach that has catalyzed the formation of a Wannier function software ecosystem, we start with a brief and general overview of modularization strategies in software programs. 
We also mention some related efforts on code modularization, discussing the aspects that apply to the Wannier ecosystem.
Complex software can adopt a variety of architecture design
approaches, often differing substantially in the level of modularity
(or lack thereof) of their components.  Historically, most computer
programs started as monolithic applications: self-contained and
independent codes made of tightly coupled functions.  This is a
natural choice when writing new software from scratch, and it reduces
the installation burden for users, who do not need to deal with the
management of many dependencies.  Over time, however, features and
post-processing tools tend to get added, making the codebase very
large and complex.  This results in serious challenges for development
and maintenance, which become critical when the code needs to be
adapted and optimized for newer hardware architectures.  Furthermore,
this leads to reimplementation of common routines in each code, 
which could instead be written and optimized
only once, and then used as a
library.  The library approach is already common in the
electronic-structure community for linear-algebra and diagonalization
routines, where the code calls functions via standard interfaces
defined by the BLAS~\cite{blackford2002updated} and
LAPACK~\cite{lapack99} libraries, and the executables are linked to
performance-optimized versions on \gls{HPC} clusters.  While a similar
approach is often used for other low-level routines, such as
\gls{FFT} computation~\cite{FFTW05} or to support file formats such as
netCDF~\cite{Rew1990} or HDF5~\cite{hdf5}, it was until recently far
less common for higher-level materials-science-oriented routines.

To address the challenges of monolithic codes, many electronic-structure codes are being redesigned or rewritten using a more modular approach, where core modules are\textemdash when possible\textemdash generalized and separated into a library of reusable routines, then called by higher-level functions to execute complex tasks.
Some of these codes have evolved into distributions, i.e., a set of relatively independent but interoperable executables reusing common core routines.
However, even with this approach, the different modules can often operate only within the distribution, and the development of all modules needs to be constantly in sync.

Ultimate interoperability is obtained when code (such as core routines
or full functionality) is reused by different independent software
distributions, maintained by non-overlapping developer groups.  A
crucial challenge to enable such a level of interoperability is to
design a clear \gls{API} defining which data needs to be transferred
between codes, and in which format. This requires discussions and
coordination, which can be catalyzed via
targeted coordination efforts. 

We stress that most of these challenges related to code modularization and interoperability are not specific to materials simulations and have been discussed since the early days of scientific computing~\cite{roberts_cpc_1969}.  In the field of electronic structure, an example worth mentioning is the
CECAM Electronic Structure Library project~\cite{cecamESL}.  At an
even higher level, one can address code interoperability by defining
common interfaces (e.g., input/output schemes) for workflows computing
a quantity of interest, independent of the underlying simulation code,
such as the common workflow interface of~\textcite{Huber:2021} to
perform crystal-structure relaxation and to compute equations of
state. The workflow only requires as input, in a
common format, the crystal structure and
a few basic input parameters, and is then interfaced
with 11 different \gls{DFT} codes to run the actual simulations.  Such
universal interfaces make workflows accessible to a broader audience
and codes fully interoperable, allowing researchers to switch between
them without the need to learn from scratch the details of each
one. In addition, they can be seamlessly applied to perform cross-code
verification studies~\cite{Bosoni2023}.

When codes have to exchange data, the interfaces between them can be
actual code \glspl{API} (e.g., in C or Fortran), where the library is
directly compiled and linked with the main code, but also simply files
in a well-documented format, written by the first application and read
by the second one (see also later discussion of this approach in
Sec.~\ref{sec:interoperability-in-ecosystem}).  The actual choice
depends on the interdependency between the algorithmic steps and on
performance considerations.  The use of files is typically favored
when the corresponding simulation workflows imply a sequential
execution of codes rather than interconnected loops between them, when
the exchanged data is small (up to few GB), and when the individual
steps are computationally demanding, so that I/O overhead is only a
small fraction of the total execution time. (In a few
advanced cases, other interfaces such as network sockets have been
used to keep the applications decoupled while still reducing the I/O
overhead for simulations that are not very computationally
demanding~\cite{Kapil2019}.)  In addition, by writing intermediate
results to files, the steps do not need to be combined in the same
run, but can be executed at differed points in time (e.g., days later)
or by different researchers.

In this context, \glspl{WF} happen to represent a remarkable and elegant method
to decouple the \emph{ab initio} simulation of the electronic
structure from the calculation of the physical properties.  This is
possible thanks to two core aspects of \glspl{WF}.  First, \glspl{WF}
are independent of the basis set used in the first-principles
electronic-structure code: the \gls{MLWF} algorithm requires the sole
knowledge of a handful of vectors and matrices, such as the overlap
matrices on a coarse grid of $k$ points. Wavefunctions, typically
stored in very large files, are not required during Wannierization,
but only used optionally in few post-processing steps, e.g., when
representing the \glspl{WF} on a real-space grid.  Second, many
physical quantities can be obtained efficiently once a \gls{WF} basis
is constructed, only with the knowledge of relevant operators
represented as small matrices directly in the Wannier basis, such as
the Hamiltonian or the position operator.  Indeed, while extended
basis sets such as plane waves are particularly convenient to obtain
charge densities and wavefunctions of periodic systems, reciprocal
space integrals can be more efficiently calculated using a
Fourier-interpolated basis set, originating from a compact
maximally localized representation in real space.  From a
computer-science perspective, we can say that these two aspects of the
Wannierization process make it an effective data-compression encoding,
avoiding the need to transfer large wavefunctions between the \emph{ab
  initio} codes and the property calculators, while retaining an
equivalent level of accuracy.

Thanks to the first aspect, i.e., basis-set independence, the Wannier
code~\cite{Marzari1997} evolved from being a standalone code focused
on the minimization procedure to one with a well-defined format for
the input data (overlap and projection matrices), which also defined and
documented the corresponding files (e.g., for the overlap matrices in
\texttt{.mmn} format and the projection matrices in \texttt{.amn}
format). The calculation of the latter was delegated to specific interfaces
implemented within the corresponding first-principles
packages~\cite{posternak2002}.  This design persisted in the
\code{Wannier90} code~\cite{Mostofi2008,Pizzi2020} and as a result the
\code{Wannier90} engine can now be interfaced with virtually any electronic-structure code as already discussed
in Sec.~\ref{subsec:wannier-engines}, with interfaces currently
available for many widespread codes, including
\code{ABINIT}~\cite{Gonze2020}, \code{BigDFT}~\cite{Ratcliff2020},
\code{Elk}~\cite{elk}, \code{FLEUR}~\cite{FLEURZenodo},
\code{GPAW}~\cite{Enkovaara-GPAW-2010}, \code{Octopus}~\cite{Octopus},
\code{OpenMX}~\cite{openmx}, \code{pySCF}~\cite{Sun2018,Sun2020},
\code{Quantum ESPRESSO}~\cite{Giannozzi2009,Giannozzi2017},
\code{SIESTA}~\cite{Soler2002}, \code{VASP}~\cite{Kresse:1999} and
\code{WIEN2k}~\cite{Blaha2020}.

Because of the second aspect, i.e., the possibility of efficiently
obtaining many physical quantities in the Wannier basis,
\code{Wannier90} started to include a large number of efficient
post-processing utilities for materials properties, ranging from
simple band-structure interpolation to more complex properties such as
the ordinary and anomalous Hall conductivities, Seebeck coefficients,
orbital magnetization, and many more~\cite{Pizzi2020}.  However, in
the last decade the community has spontaneously moved towards a
decentralized software ecosystem (as opposed to a centralized, albeit
modular, Wannier distribution), where different packages interact
through \glspl{API} and a common data format.  The
decentralized model was again facilitated by a clear and documented
interface to generate data as input to the next steps (e.g., the
\texttt{\_tb.dat} file containing the full \gls{TB} model: \gls{WF}
centers, on-site energies, and hopping energies).  The community has
been rapidly growing, and several independent packages exploiting
\glspl{MLWF} exist nowadays, targeting diverse properties such as
\gls{TB} models (Sec.~\ref{subsec:tb}), ballistic transport
(Sec.~\ref{sec:ballistic}), Berry-phase related properties
(Sec.~\ref{subsec:Berry}), topological invariants
(Sec.~\ref{subsec:topoinv}), electron-phonon coupling
(Sec.~\ref{subsec:elph}), beyond-\gls{DFT} methods
(Sec.~\ref{subsec:dmft}), high-throughput calculations
(Sec.~\ref{subsec:automation}), and more.

This review article describes such a community of
symbiotic packages, forming a research and software ecosystem built
upon the concept of \glspl{MLWF}. We illustrate this schematically in
Fig.~\ref{fig:schematics}. To make the codes of the ecosystem as easy to find as possible, we also started in 2024 an online Wannier Software Ecosystem Registry~\cite{Wannier_registry}. Such a registry lists software packages that form the ecosystem, and as of May 2024 it already includes 53 entries. The repository provides key information including a short description, a domain tag (e.g., ``Ab initio engines'', ``Tight-binding'', ``Berryology and topology'') and links to the code homepage, documentation and source code (if available). The registry is dynamic: developers and users can add new entries or modify existing ones by submitting a pull request through the corresponding GitHub repository~\cite{GitHub_Wannier_registry}, which also includes detailed instructions for contribution.

\subsection{\label{subsec:tb}Wannier interpolation and tight-binding models}
A very common application of \glspl{WF} is to evaluate various $k$-space
quantities and \gls{BZ} integrals by ``Wannier interpolation''. This name
has come to refer to a type of Slater--Koster interpolation where the
required on-site and hopping integrals are calculated explicitly in the
\gls{WF} basis~\cite{Souza2001,Calzolari2004,Lee2005,Yates_et_al:2007}, as
opposed to being treated as fitting parameters as done in empirical
\gls{TB} theory. Here we review the
basic procedure as it applies to energy
bands and other simple quantities, leaving more sophisticated
applications to later sections.  Before proceeding, let us mention
that the Wannier interpolation scheme has been adapted to work with
non-orthogonal localized orbitals instead of (orthogonal)
\glspl{WF}~\cite{naderlli-cms18,lee-prb18,wang-njp19,jin-jpcm21}.

\subsubsection{Band interpolation}\label{sec:band-inter}

To interpolate the band structure, one needs the matrix elements of
the \gls{KS} Hamiltonian in the \gls{WF} basis,
\begin{equation}
H^\text{W}_{ij}(\RR) = \me{{\bf 0}i}{\hat{H}}{\RR j};
\label{eq:H-R-def}
\end{equation}
here $H_{ii}({\bf 0})$ are on-site energies, and the remaining matrix
elements are hoppings. One way to evaluate these matrix elements is to
start from Eq.~\eqref{eq:wannier-dis} for the \glspl{WF} in terms of the \gls{KS}
Bloch eigenstates on the \textit{ab initio} $k$-grid. Inserting
that expression in Eq.~\eqref{eq:H-R-def} gives
\begin{equation}
H^\text{W}_{ij}(\RR)=\frac{1}{N}\sum_\kk\,e^{-i\kk\cdot\RR}
\sum_{n=1}^{\nbands_\kk}\,V^*_{\kk,ni}\varepsilon_{n\kk}V_{\kk,nj}.
\label{eq:H-R}
\end{equation}
This procedure is particularly convenient in the framework of the MV
and SMV Wannierization schemes, which are formulated as post-processing
steps after a conventional \textit{ab initio} calculation is carried out on a uniform $\{\kk\}$
grid; Eq.~\eqref{eq:H-R} only involves the $V_\kk$ matrices generated by
the Wannier engine starting from the \textit{ab initio} overlap
matrices and the energy eigenvalues themselves (see Sec.~\ref{subsec:mlwf}).
An alternative to Eq.~\eqref{eq:H-R}
would be to express the \glspl{WF} in a real-space basis, e.g., localized
orbitals or a grid, and then evaluate Eq.~\eqref{eq:H-R-def} directly on
that basis.

In view of the localized character of the \glspl{WF},
$|H^\text{W}_{ij}(\RR)|$ is expected to become negligibly small when
the distance $|\RR+\ttau_j-\ttau_i|$ between the centers of the two
\glspl{WF} becomes sufficiently large (here,
$\ttau_j=\me{{\bf 0}j}{\hat\rr}{{\bf 0}j}$). However, due to the
finite size $N$ of the \textit{ab initio} grid, the \glspl{WF}
obtained from Eq.~\eqref{eq:wannier-dis} are actually periodic over a
real-space supercell of volume $NV_\text{cell}$; accordingly, the
matrix elements given by Eq.~\eqref{eq:H-R} are also
supercell-periodic:
$H^\text{W}_{ij}(\RR+{\bf T})=H^\text{W}_{ij}(\RR)$, for any supercell
lattice vector ${\bf T}$. To minimize spurious effects associated with
this artificial periodicity, the hopping matrix should be truncated
by setting $H^\text{W}_{ij}(\RR)=0$ whenever the vector
$\RR+\mathbf{T}+\ttau_j$ lies outside the \gls{WS} supercell
centered at the origin.  Provided that this supercell is sufficiently
large to ensure negligible overlap between a \gls{WF} and its periodic
images, the truncation error will be insignificant. This means that in
practice one can achieve well-converged numerical results with a
relatively coarse \textit{ab initio} grid. Note, however, that the
matrix elements do not decay exactly to zero for finite-size \gls{WS}
supercells. Therefore, when multiple $\RR$ vectors lie on the boundary
of the \gls{WS} supercell and are connected by a supercell vector
$\mathbf{T}$, it is better to consider all these equivalent vectors
with appropriate weights, rather than picking only one of them, 
which would introduce spurious symmetry
breaking in the Hamiltonian.  The details of this approach and its
implementation in \code{Wannier90} are discussed in Sec. 4.2 of~\textcite{Pizzi2020}.

Once the on-site energies and hoppings have been tabulated, the
Hamiltonian matrix is interpolated onto an arbitrary \gls{BZ} point $\kk'$ by
performing an inverse Fourier transform,
\begin{equation}
H^\text{W}_{\kk',ij}=\sum_\RR^{\text{WS}}\,
{\frac{1}{\mathcal{N}_{\RR,ij}}}\sum_{l=1}^{\mathcal{N}_{\RR,ij}}
e^{i\kk'\cdot{(\RR+{\bf T}_{\RR,ij}^{l})}} H^\text{W}_{ij}(\RR)\,.
\label{eq:H-k}
\end{equation}
The summation runs over the lattice vectors $\RR$ (which lie in the \gls{WS}
supercell centered at the origin. as discussed above) with
$\mathcal{N}_\RR>1$ whenever $\RR+{\bf T}_{\RR,ij}^{l}+\ttau_j$  falls on the boundary of the WS supercell centered at $\ttau_i$. To improve the quality of the interpolation, for each
combination of $i$, $j$, and $\RR$ the supercell lattice vector
${\bf T}$ appearing in Eq.~\eqref{eq:H-k} is chosen as the one that
minimizes $|\RR+{\bf T}+\ttau_j-\ttau_i|$~\cite{Pizzi2020}.  Finally,
the interpolated energy eigenvalues are obtained by diagonalizing the
above matrix,
\begin{equation}
\left[ {\mathcal U}^\dagger_{\kk'} H^\text{W}_{\kk'}{\mathcal U}_{\kk'}\right]_{mn}=
\delta_{mn}\varepsilon^\text{H}_{n\kk'}\,,
\label{eq:eig-int}
\end{equation}
so that the column vectors of the unitary matrix ${\mathcal U_{\kk'}}$
are eigenvectors of $H^\text{W}_{\kk'}$.

Since the interpolation steps~\eqref{eq:H-k} and~\eqref{eq:eig-int}
only involve Fourier transforming and diagonalizing $J\times J$
matrices that are typically small, the overall procedure tends to be
much less expensive than a direct \gls{DFT} calculation at every
interpolation point, especially when a dense interpolation grid
$\{\kk'\}$ is needed.  The efficient evaluation of the Hamiltonian
matrix and band derivatives (see
  below) enables
\gls{BZ} integration methods beyond the standard equispaced scheme to be explored. These are of particular use when fine
features in $k$ space need to be resolved using adaptive integration
methods~\cite{Assmann_et_al:2016,Kaye2023,Munoz_et_al:2024}.

\begin{figure}[tb]
  \centering\includegraphics[width=0.45\textwidth]{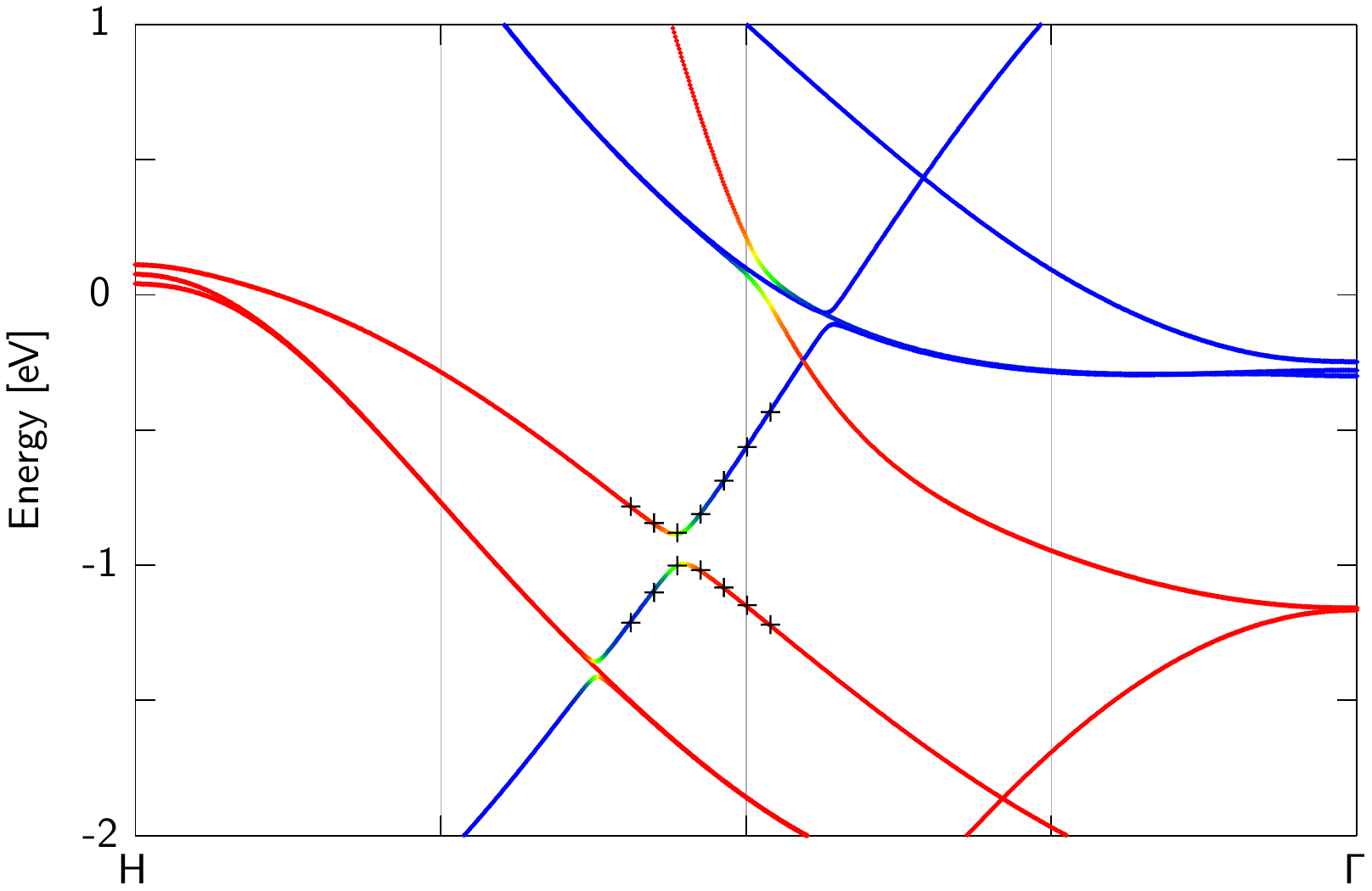}
  \caption{\label{fig:interpolated-bands} Wannier-interpolated bands of
    \gls{BCC} Fe along the H--$\Gamma$ line. The bands are color-coded according to the
    value of the spin expectation value
    $\me{\psi\bnk}{\hat S_z}{\psi\bnk}$: red for spin up, and blue for
    spin down. The energies are given in eV, and the Fermi level is at
    0~eV. The vertical gray lines indicate $k$ points on the \emph{ab initio} mesh
    used for constructing the \glspl{WF}.
  }
  \end{figure}

The above interpolation scheme has been shown to accurately reproduce\textemdash within the frozen energy window\textemdash the energy eigenvalues obtained
by a direct \gls{DFT}  calculation.  As an example, we show in
Fig.~\ref{fig:interpolated-bands} a detail of the interpolated band
structure of ferromagnetic \gls{BCC}
Fe along the H--$\Gamma$
line~\cite{Yates_et_al:2007}.  
The vertical gray lines indicate
points on the $\{\kk\}$ mesh used for constructing the \glspl{WF}. For
comparison, we plot as plus symbols the \textit{ab initio} dispersion
around a weak spin-orbit-induced avoided crossing between two bands of
opposite spin. It is apparent that the Wannier interpolation procedure
succeeds in resolving details on a scale much smaller than the spacing
between those points. In particular, the correct band connectivity is
obtained, so that avoided crossings, no matter how weak, are not
mistaken for actual crossings.  This characteristic, which
distinguishes Wannier interpolation from methods based on direct
Fourier interpolation of the energy
eigenvalues~\cite{madsen-cpc06,madsen-cpc18}, makes it a powerful tool
for studying topological properties (Sec.~\ref{subsec:topoinv}), and
for evaluating \gls{BZ} integrals involving quantities that change rapidly
over small regions of $k$ space, such as the Berry curvature
(Sec.~\ref{subsec:Berry}) and electron-phonon matrix elements
(Sec.~\ref{subsec:elph}).

Wannier interpolation works for the same reason that empirical
\gls{TB} does: the short range of the
real-space Hamiltonian matrix~\eqref{eq:H-R} ensures that its Fourier
transform~\eqref{eq:H-k} is a smooth function in reciprocal space.
This can also be seen by writing the left-hand side of
Eq.~\eqref{eq:H-k} as
\begin{equation}
H^\text{W}_{\kk',ij}=\me{\psi^\text{W}_{i\kk'}}{\hat H}{\psi^\text{W}_{j\kk'}}\,,
\end{equation}
where 
\begin{equation}
\ket{\psi^\text{W}_{j\kk'}}=\sum_{\RR}\,e^{i\kk'\cdot\RR}\ket{\RR j}
\label{eq:psi-W}
\end{equation}
interpolates the smooth Bloch functions defined on the \textit{ab
  initio} grid by Eq.~\eqref{eq:psi-smooth}.
We may also write the left-hand side of Eq.~\eqref{eq:eig-int} as
\begin{equation}
H^\text{H}_{\kk',mn}=\me{\psi^\text{H}_{m\kk'}}{\hat H}{\psi^\text{H}_{n\kk'}}\,,
\end{equation}
where 
\begin{equation}
\ket{\psi^\text{H}_{n\kk'}}=\sum_{j=1}^J\,\ket{\psi^\text{W}_{j\kk'}}
    {\mathcal U}_{\kk',jn}
\label{eq:psi-H}
\end{equation}
describes a unitary transformation from the Wannier gauge W to the
Hamiltonian gauge H. Inside the frozen energy window, the states
$\ket{\psi^\text{H}_{n\kk'}}$ interpolate\textemdash up to arbitrary phase
factors\textemdash the \textit{ab initio} eigenstates $\ket{\psi\bnk}$.

In summary, performing Fourier interpolation in the W gauge followed
by a unitary transformation to the H gauge allows interpolating
quantities\textemdash band energies, Bloch eigenstates, and matrix elements
thereof (see below)\textemdash that can vary rapidly in $k$ space, and even
become non-analytic at degeneracies. This strategy retains the accuracy
of a full-blown \textit{ab initio} calculation, while benefiting from
the efficiency of Slater--Koster interpolation.

\subsubsection{Band derivatives and Boltzmann transport}
\label{sec:band-der}
The interpolation procedure outlined above can be adapted to evaluate
band velocities, inverse effective-mass tensors, and higher $k$-derivatives of the energy eigenvalues~\cite{Yates_et_al:2007}; as in
the empirical \gls{TB}
method~\cite{graf-prb95}, this is achieved without relying on
finite-difference methods, which become problematic in the vicinity of
band crossings and weak avoided crossings, where the band ordering can
change from one grid point to the next.

Band derivatives are needed, for instance, to evaluate transport
coefficients such as the electrical conductivity $\sigma$, the Seebeck
coefficient $S$, or the electronic contribution to the thermal
conductivity $K$.  Within the semiclassical \gls{BTE} framework, one 
defines a scattering time $\tau_{n\bf{k}}$ for an
electron on band $n$ at wavevector $\bf{k}$
(incidentally, the contributions from
electron--phonon scattering to $\tau_{n\bf{k}}$ can be efficiently computed exploiting Wannier
functions, see Sec.~\ref{subsec:elph}). Then, the expressions for the
transport tensors are given by~\cite{Ziman:1972}:
\begin{widetext}
\begin{align}\label{eq:eq-sigma} {\sigma}_{ab}(\mu,T)&=
  e^{2}\int_{-\infty}^{+\infty}dE\left(-\frac{\partial
      f(E,\mu,T)}{\partial E}\right)\Sigma_{ab}(E),\\
\label{eq:eq-seebeck}  [{\sigma}{S}]{}_{ab}(\mu,T)&=
  \frac{e}{T}\int_{-\infty}^{+\infty}dE\left(-\frac{\partial
      f(E,\mu,T)}{\partial E}\right)(E-\mu)\Sigma_{ab}(E),\\
\label{eq:eq-thermal}  {K}{}_{ab}(\mu,T)&=
  \frac{1}{T}\int_{-\infty}^{+\infty}dE\left(-\frac{\partial
      f(E,\mu,T)}{\partial E}\right)(E-\mu)^{2}\Sigma_{ab}(E),
\end{align}
\end{widetext}
where $\mu$ is the chemical potential, $T$ is the temperature, $a$ and
$b$ are Cartesian indices, $\sigma S$ denotes the matrix product of the two
tensors, $\partial f/\partial E$ is the derivative of the Fermi--Dirac
distribution function with respect to the energy, and $\Sigma_{ab}(E)$
is the transport distribution function. The latter
is defined by
\begin{equation}
\Sigma_{ab}(E)=\frac{1}{V_\text{cell}}\sum_{n{\bf{k}}}v^a_{n\bf k} v^b_{n\bf k}\tau_{n{\bf k}}\delta(E-E_{n\bf{k}}),
 \label{eq:sigma-ij}
\end{equation}
where the summation is over all bands $n$ and over all the \gls{BZ}, $\varepsilon_{n\bf k}$ is the energy for band $n$
at $\bf k$ and $v^a_{n\bf k}$ is the $a$ component of the
band velocity at $(n,\bf k)$.

Obtaining converged quantities for Eqs.\eqref{eq:eq-sigma}--\eqref{eq:eq-thermal}, therefore, requires to compute the band derivatives ${\bf v}_{n\bf k}$ while sampling the \gls{BZ} over dense $k$-point
grids~\cite{schulz-prb92,uehara-prb00,madsen-cpc06}, since the term $\partial f/\partial E$ is non-zero only in a narrow energy region (of typical size $k_B T$, where $k_B$ is the Boltzmann constant) around the chemical potential $\mu$.
Wannier
interpolation allows carrying out this task efficiently and
accurately even when (avoided) crossings occur close to the Fermi
level: band derivatives at a given $k$ point are obtained with an analytical expression, without resorting to finite-difference methods~\cite{Yates_et_al:2007}. Moreover, computation on dense $k$-point grids is very efficient, as already discussed earlier for band interpolation.
This WF-based Boltzmann-transport
methodology is implemented in \code{Wannier90} and used to compute transport tensors in its \code{BoltzWann} module~\cite{Pizzi2014}, as well as in other codes (see, e.g., the electron-phonon section, Sec.~\ref{subsec:elph}), and is also used for post-processing calculations in many-body theory (see Sec.~\ref{subsubsec:DMFT}). 

Furthermore, many transport coefficients (e.g., linear and non-linear \glspl{AHC}~\cite{yao-prl04,sodemann-prl15}, anomalous Nernst
thermoelectric conductivity~\cite{xiao-prl06},
magnetoresistance~\cite{gao-prb17}, and magnetochiral
anisotropy~\cite{yokuchi-prl23}) depend on the Berry curvature and
other quantum-geometric
quantities~\cite{xiao-rmp10,Vanderbilt2018,gao-fp19}.  As they involve
$k$-derivatives of the Bloch states themselves, such quantities cannot be obtained
from the energy dispersions. Moreover, those quantities tend to become
strongly enhanced when weak avoided crossings occur near the Fermi
level; when that happens, very dense $k$-point grids must be employed
to converge the calculation~\cite{yao-prl04}.  Compared to band
interpolation, the interpolation of Berry-type quantities is more
involved because it requires setting up matrix elements of the
position operator $\hat\rr$, which is non-local in reciprocal
space, i.e., not lattice periodic~\cite{blount-ssp62}.  We defer a discussion of that case to
Sec.~\ref{subsec:Berry}, and below we describe how to interpolate the
matrix elements of a generic lattice-periodic operator $\hat X$.

\subsubsection{Interpolation of a generic lattice-periodic operator}

Replacing $\hat H\rightarrow \hat X$ in Eq.~\eqref{eq:H-R-def} and using
Eq.~\eqref{eq:wannier-psi-dis} yields
\begin{equation}
X^\text{W}_{ij}(\RR)=\frac{1}{N}\sum_\kk\,e^{-i\kk\cdot\RR}
\sum_{m,n=1}^{\nbands_\kk}\,
V^*_{\kk,mi}\me{\psi_{m\kk}}{\hat X}{\psi_{n\kk}}V_{\kk,nj}\,,
\label{eq:X-R}
\end{equation}
which reduces to Eq.~\eqref{eq:H-R} for $\hat X=\hat H$.
The considerations made earlier regarding the spatial decay and
truncation of the $H^\text{W}(\RR)$ matrix apply equally well to
$X^\text{W}(\RR)$. The Fourier-transform step is also analogous to
Eq.~\eqref{eq:H-k},
\begin{equation}
X^\text{W}_{\kk',ij}=\sum_\RR\,e^{i\kk'\cdot\RR} X^\text{W}_{ij}(\RR)=
\me{\psi^\text{W}_{i\kk'}}{\hat X}{\psi^\text{W}_{j\kk'}}\,,
\label{eq:X-k}
\end{equation}
and the final step is to apply to $X^\text{W}_{\kk'}$ the same unitary
transformation that was used in Eq.~\eqref{eq:eig-int} to diagonalize
$H^\text{W}_{\kk'}$. Using Eq.~\eqref{eq:psi-H}, we find
\begin{equation}
X^\text{H}_{\kk',mn}=
\left({\mathcal U}^\dagger_{\kk'} X^\text{W}_{\kk'}{\mathcal U}_{\kk'}\right)_{mn}=
\me{\psi^\text{H}_{m\kk'}}{\hat X}{\psi^\text{H}_{n\kk'}}\,.
\label{eq:X-H}
\end{equation}
In particular, diagonal elements of $X^\text{H}_{\kk'}$ give the
expectation values of $\hat X$ in the interpolated states.  With
$\hat X=\hat {\bf S}$, for example, one obtains their spin
polarization, which is how the color-coding in
Fig.~\ref{fig:interpolated-bands} was generated.

Note that in addition to the overlap matrices and energy eigenvalues,
interpolating a generic operator $\hat X\not=\hat H$ requires setting
up its matrix elements $\me{\psi_{m\kk}}{\hat X}{\psi_{n\kk}}$ on the
\textit{ab initio} grid; this should be done by the same interface
code that computes the overlap matrices.

\subsubsection{Wannier function perturbation theory}

Several important materials properties can be calculated as the linear
response of the system to an external perturbation $\hat{V}$. A common
example is the calculation of phonon dispersions and electron-phonon coupling through
\gls{DFPT}~\cite{Baroni2001}, the latter case is
discussed in Sec.~\ref{subsec:elph}.

We follow~\textcite{Park_PRX_2021}, and write the Hamiltonian
eigenstates of the perturbed system in terms of the ones of the
unperturbed system plus the wavefunction perturbation,
\begin{equation}
\ket{\psi_{n\mathbf{k}}}=
\ket{\psi_{n\mathbf{k}}}^{(0)}+\lambda \ket{\psi_{n\mathbf{k}}}^{(1)}+
\mathcal{O}(\lambda^2)\,,
\end{equation}
where the wavefunction perturbation can be calculated with a sum over
empty states,
\begin{equation}
\ket{\psi_{n\mathbf{k}}}^{(1)} =
\sideset{}{'}\sum_{n'\mathbf{k}'}
\ket{\psi_{n'\mathbf{k}'}}^{(0)}
\frac{\braket{\psi_{n'\mathbf{k}'}|\hat{V}|\psi_{n\mathbf{k}}}^{(0)}}{\epsilon_{n\mathbf{k}}-\epsilon_{n'\mathbf{k}'}}.
\end{equation}
The primed sum means that terms for which the denominator vanishes are excluded.

Alternatively, the wavefunction perturbation can be calculated
without summing over high-energy states by solving the Sternheimer
equation~\cite{Baroni2001}.  Lihm and Park~\cite{Park_PRX_2021} have
shown that such perturbation theory can be formulated in the Wannier
representation, where the \glspl{WF} of the perturbed system can be
written as
\begin{equation}
\ket{\mathbf{R}j}=\ket{\mathbf{R}j}^{(0)}+\lambda \ket{\mathbf{R}j}^{(1)}+\mathcal{O}(\lambda^2).
\end{equation}
The expression for $\ket{\mathbf{R}j}^{(1)}$, reported in Eqs.~(8)
and~(9) of \textcite{Park_PRX_2021}, consists of two terms: the first
can be calculated with the Sternheimer equation to obtain the states
$\ket{\psi_{n\mathbf{k}}}^{(1)}$, while the second contains matrix
elements of $\hat{V}$ and the projector over the \glspl{WF} of the
unperturbed system. Crucially, both terms only require energies and
matrix elements within the Wannier outer window introduced in
Sec.~\ref{sec:entangled_bands}.  Thus, the Wannier function
perturbation can be calculated without making explicit use of the
states outside that energy window.

For a monochromatic static perturbation with wavevector ${\bf q}$, the
first-order wavefunction perturbation can be interpolated
as~\cite{Park_PRX_2021}
\begin{eqnarray}
\ket{\psi_{n\mathbf{k}',\mathbf{q}}^{\text{H}}}^{(1)}&=&
\frac{1}{\sqrt{N}}\sum_{j,\mathbf{R}}e^{i\mathbf{k}'\cdot\mathbf{R}}
\ket{\mathbf{R}_{\mathbf{q}}j}^{(1)}{\mathcal U}_{\mathbf{k}',jn} \nonumber \\
&+&\sideset{}{'}\sum_{m=1}^{N_{W}}
\ket{\psi_{m\mathbf{k}'+\mathbf{q}}^\text{H}}^{(0)}
\frac{\tilde{g}_{mn\mathbf{k}',\mathbf{q}}^\text{H}}
{\epsilon_{n\mathbf{k}'}^{(0)}-\epsilon_{m\mathbf{k}'+\mathbf{q}}^{(0)}},
\end{eqnarray}
where the WF perturbations are expanded as a sum of monochromatic perturbations
\begin{equation}
  \ket{\mathbf{R}j}^{(1)} = \sum_{\mathbf{q}} \ket{\mathbf{R}_{\mathbf{q}}j}^{(1)}
\end{equation}
while the superscript H marks the Hamiltonian gauge
of the unperturbed ($0$) and perturbed ($1$) system,
${\mathcal U}_{\kk'}$ is
  the unitary matrix that diagonalizes the unperturbed Hamiltonian
  according to Eq.~\eqref{eq:eig-int}, and
$\tilde{g}_{mn\mathbf{k},\mathbf{q}}^{\text{H}}$ is obtained by
performing the Fourier transform of $\tilde{g}_{ij\mathbf{R},\mathbf{q}}$
and then rotating to the
Hamiltonian gauge using ${\mathcal U}_{\kk'}$. The quantity $\tilde{g}_{ij\mathbf{R},\mathbf{q}}$ is related to the perturbation and is made of two terms ($\tilde{g}_{ij\mathbf{R},\mathbf{q}} = g_{ij\mathbf{R},\mathbf{q}} +  \delta{g}_{ij\mathbf{R},\mathbf{q}}$), the first accounts for the matrix elements of $\hat{V}$ between \glspl{WF} of the unperturbed states
\begin{equation}
  g_{ij\mathbf{R},\mathbf{q}} = \braket{\mathbf{0}i^{(0)}|\hat{V}_{\mathbf{q}}|\mathbf{R}j^{(0)}},
\end{equation}
while the second is a correction stemming from the change of the~\glspl{WF}
\begin{equation}
  \delta{g}_{ij\mathbf{R},\mathbf{q}} = \braket{\mathbf{0}i^{(0)}|\hat{H}^{(0)}|\mathbf{R}_{\mathbf{q}}j^{(1)}}+\braket{\mathbf{0}_{-\mathbf{q}}i^{(1)}|\hat{H}^{(0)}|\mathbf{R}j^{(0)}}.
\end{equation}
The correction term $\delta{g}_{ij\mathbf{R},\mathbf{q}}$ is not required for the scattering matrix elements, but it is relevant for the perturbed wave functions.
A key aspect is that Wannier function perturbations are localized in
real space, so the perturbed Hamiltonian and its eigenstates can be
efficiently interpolated by considering coarse $k$-point grids. This permits the efficient interpolation of various matrix elements involving
the wavefunction perturbation, such as in the case of electron-phonon
self-energies and Kubo formulas. \gls{WFPT} has been applied to
describe temperature-dependent electronic band structures and indirect
optical absorption, shift spin currents and spin Hall
conductivities~\cite{Park_PRX_2021}. \gls{WFPT} has recently been made 
available in the EPW
code version 5.8~\cite{EPW2023}.

\subsubsection{Porting Wannier Hamiltonians to TB codes}

In \gls{TB} theory, two phase
conventions are commonly used to perform the Fourier transforms from
real to reciprocal space~\cite{Vanderbilt2018}: the one adopted in
Eqs.~\eqref{eq:H-k}, \eqref{eq:psi-W} and~\eqref{eq:X-k} (``original convention''), and the
alternative one (``modified convention'') where the phase factors in those equations are
modified as
\begin{equation}
e^{i\kk'\cdot\RR}\rightarrow e^{i\kk'\cdot(\RR +\ttau_j-\ttau_i )}\,.
\label{eq:phase-convention}
\end{equation}

Although the interpolated eigenvalues $\varepsilon^\text{H}_{n\kk'}$
and matrix elements $X^\text{H}_{\kk',mn}$ come out the same with both
conventions (as they should), the modified convention is more natural
for the purpose of evaluating quantities, such as Berry connections
and curvatures, that are sensitive to the real-space embedding of the
\gls{TB} model via the position matrix
elements $\me{{\bf 0}i}{\hat\rr}{\RR j}$ (see
Sec.~\ref{subsec:Berry}). This has to do with the fact that with the
original convention, \gls{TB}
eigenvectors (the column vectors of ${\mathcal U}_{\kk'}$) behave like
full Bloch eigenstates $\ket{\psi_{n\kk'}}$, whereas with the modified
convention they behave like their cell-periodic parts
$\ket{u_{n\kk'}}$~\cite{Vanderbilt2018}; and it is in terms of the
latter that Berry-type quantities are most naturally expressed.

The modified phase convention is the one adopted in the \gls{TB} codes
\code{PythTB}~\cite{pythtb} and \code{TBmodels}~\cite{tbmodels}; both
are able to import the Wannier Hamiltonian $H^\text{W}_{ij}(\RR)$,
along with the Wannier centers $\{\ttau_j\}$, from the {\tt
  seedname}{\_}{\tt tb.dat} file written by
\code{Wannier90}. \code{PythTB} was originally written for pedagogical
purposes, as part of a course on Berry phases in electronic-structure
theory that was later turned into a textbook~\cite{Vanderbilt2018}. It
is feature-rich but is not optimized for speed, as it was designed
with \gls{TB} toy models in mind (however, a high-performance
implementation is available~\cite{yeet-pythtb}).  Instead,
\code{TBmodels} has fewer post-processing functionalities, but it
delivers critical speed-up and improved scaling. Among other
Wannier-\gls{TB} codes, it is worth mentioning \code{Wannier Tools}~\cite{Wu2018}, which
implements sparse Hamiltonians for large systems, band unfolding, and
several other features related to Berry-type quantities (see
Sec.~\ref{subsec:Berry}).
\subsubsection{\label{subsubsec:inter_beyond_DFT}Wannier interpolation beyond density-functional theory}
As discussed above, one of the most powerful and effective
applications of \glspl{WF} is the interpolation of band structures and
other electronic-structure properties. While this is already very useful in the context of DFT
calculations, it becomes even more compelling
for beyond-DFT methods, such as hybrid
functionals~\cite{MartinBook2020,Kummel2008,Becke2014,Lee1988,Becke1988,Perdew1996,Heyd2003},
\gls{MBPT} such as $G_0W_0$~\cite{Golze2019}, and non-diagrammatic
approaches such as the Koopmans-compliant
functionals~\cite{Dabo2010,Borghi2014,Colonna2018,Colonna2019,Linh2018,Elliott2019,DeGennaro2022,Colonna2022,linscott_koopmans_2023,Marrazzo_arxiv_2024}. In
fact, in DFT the potential can always be recalculated from the sole
knowledge of the ground-state electronic charge
density; therefore, the corresponding
\gls{KS} Hamiltonian can be directly calculated at any arbitrary
$\mathbf{k}$-point. Instead, for most beyond-DFT
methods this is no longer the case, and band structure calculations
on a high-symmetry path cannot be performed as a series of independent
diagonalizations. For hybrid functionals and GW, the eigenvalues at a
given $\mathbf{k}$-point requires the knowledge of the wavefunctions
and eigenenergies on all $(\mathbf{k}+\mathbf{q})$-points, where the
$\mathbf{q}$-points are defined on a uniform grid which has to be
converged. In other words, reasonably dense sampling on high-symmetry
paths can only be obtained with some form of interpolation.
While electronic-structure codes typically offer general-purpose
interpolation methods, often based on Fourier
series~\cite{Koelling1986,Pickett88}, \glspl{WF} provide two concrete
advantages. First, they are a physically motivated basis
set, which exhibits exponential convergence and is guaranteed to
deliver the exact result for a sufficiently dense $\mathbf{k}$-point
grid, so the accuracy can be systematically increased simply by
considering denser grids. If \glspl{MLWF} are chosen, the efficiency
is maximal and rather coarse grids are
often sufficient to faithfully reproduce the entire band
structure. Second, once a \gls{WF} basis is constructed, it not only yields
interpolated eigenvalues (e.g., the band structure) but it
also enables the Hamiltonian and
many other operators to be described in a compact real-space representation. Once the relevant operators in a \gls{WF} basis are available, one gets access to the full spectrum of theories and software packages
that are part of the Wannier ecosystem, capable of much more complex tasks than just band interpolation. Notably, once a
Wannierization is performed with hybrid functionals or at the $G_0W_0$
level, all other Wannier-interpolated quantities can be obtained at
the same level of theory with no extra effort or cost.  Finally,
thanks to recent work in advanced minimization techniques
(Sec.~\ref{subsec:advancedminimisation}) and automation
(Sec.~\ref{subsec:automation}), the Wannier interpolation does not
require much more human intervention than other standard methods such
as smooth Fourier interpolation.  In the following, we briefly outline
the motivation and the corresponding procedure to deploy Wannier
interpolation for two of the most popular excited-state approaches:
hybrid functionals, and many-body perturbation theory at the $G_0W_0$
level.

\paragraph{Hybrid functionals.} A very popular approach to improve the
accuracy of \textit{ab initio} band structures is to combine explicit
density-dependent functionals with Hartree--Fock terms, which leads to
orbital-dependent functionals called
``hybrids''~\cite{MartinBook2020,Kummel2008,Becke2014,Lee1988,Becke1988,Perdew1996,Heyd2003}. The
procedure to obtain \glspl{WF} is similar to that in vanilla
\gls{DFT}, except that \gls{NSCF} calculations cannot be performed, as the potential is not a functional of the total density only but requires the knowledge of  single-particle orbitals.  Hence, only \gls{SCF} calculations are performed with hybrid functionals, including some higher-energy empty states (if any) that might be needed to obtain Wannier functions through disentanglement~\cite{Souza2001}. This is different with respect to DFT, where typically a \gls{SCF} calculation is performed on the occupied states only (plus some lower-lying conduction bands in the case of metals) and a \gls{NSCF} calculation is performed including higher-energy empty states, possibly on a different $k$-point grid. 
In addition, ground-state calculations are typically performed on the
\gls{IBZ} by exploiting crystalline symmetries, while \code{Wannier90}
requires a uniform grid on the \gls{FBZ}. As performing the
self-consistent calculation on the \gls{FBZ} is certainly possible but
rather inefficient, the typical procedure involves unfolding the
ground-state orbitals and band structure from the \gls{IBZ} to the
\gls{FBZ}. This is done as a post-processing step  performed after the
self-consistent calculation and before producing the overlap matrices
and the other input required to obtain \glspl{WF}. For example, in the
\code{Quantum ESPRESSO}
distribution~\cite{Giannozzi2009,Giannozzi2017} this is done through
the \verb|open_grid.x| code.  Notably, \glspl{WF} can be used to speed
up the core hybrid-functionals calculations, as they allow reducing
the number of exchange integrals to be
computed~\cite{E_cms_2007,Car_PRB_2009,E_PRB_2009,Car_jcp_2014,Mountjoy_jcp_2017,Carnimeo_ES_2019}.

\paragraph{$G_0W_0$.} Most of what has been discussed for hybrid
functionals also holds for \gls{MBPT} calculations in the $G_0W_0$
approximation, with two important remarks. First, $G_0W_0$ is a
one-shot approach in the \gls{QP} approximation which is typically
performed on top of a \gls{DFT} calculation: so the orbitals remain at
the \gls{KS} level and only the energy eigenvalues are corrected,
hence neglecting off-diagonal elements of the self-energy in the
\gls{KS} basis. Second, as only the energies are changed at
the $G_0W_0$ level, the \gls{KS} states might swap
their band indices and not
be ordered in energy anymore. A typical case
where this might manifest clearly are topological insulator
candidates (and systems with band inversions in general) such as
monolayer TiNI~\cite{MarrazzoNano2019}, which is topological in
\gls{DFT} and trivial at the $G_0W_0$ level. The practicalities of
obtaining $G_0W_0$ \glspl{WF} and related quantities depend on whether
the \gls{DFT} and \gls{MBPT} calculations are performed with the same
distribution (e.g., \code{VASP}~\cite{Merzuk:2015}) or with two
separate codes (as for example with \code{Quantum
  ESPRESSO}~\cite{Giannozzi2009,Giannozzi2017} and
\code{Yambo}~\cite{Sangalli2019}).  In general, $G_0W_0$ \gls{QP}
corrections
$\Delta\epsilon_i^{\text{QP}} = \epsilon_i^{G_0W_0}-\epsilon_i^{\text{DFT}}$ need to
be computed on a uniform grid in the \gls{FBZ}, which can be made efficient by unfolding
from the \gls{IBZ} to the \gls{FBZ} (in \code{Yambo}, this operation
is carried out by the post-processing utility \verb|ypp|). While
orbitals remain at the \gls{DFT}-\gls{KS} level, \code{Wannier90}
requires the states to be ordered with ascending energy and, in
addition, some input matrices (e.g., the \texttt{uHu}) need to be updated with
\gls{QP} corrections. If this is not performed by the \textit{ab
  initio} engine, it can be taken care by the \code{Wannier90} utility
\verb|gw2wannier90.py| if the \gls{DFT} eigenvalues and \gls{QP}
corrections are provided in the standard format
\verb|seedname.eig|. After this step, the Wannierization can proceed
as usual, and it is available in both
\code{Wannier90}~\cite{Pizzi2020} and
\code{WanT}~\cite{Ferretti2012}.  Note that in $G_0W_0$, \gls{QP}
corrections need to be computed on a subset of the $k$-point
grid required to calculate the self-energy: this can be exploited to
speedup the calculation, especially for 2D materials as discussed
in~\textcite{Sangalli2019}, because Wannierization typically requires
relatively coarse $k$-point grids. Finally, we briefly touch
on beyond-$G_0W_0$ development of interest from a \gls{WF}
perspective. While Aguilera et al. have shown that off-diagonal
components of the self-energy---which are not included in standard perturbative approaches---are very relevant in case of band
inversions~\cite{Aguilera2013} (such as for topological insulators),
Hamann and Vanderbilt have found~\cite{Hamann2009} that, in general,
differences between \glspl{MLWF} obtained with local-density
approximation (LDA) and \gls{QSGW} are minimal.

\paragraph{Bethe--Salpeter equation}
In order to address neutral excitations, as opposed to the charged
excitations of GW, one needs to describe the bound state of an excited
electron with the hole that has been created. This is accomplished
either using time-dependent density-functional theory or Green's
function methods~\cite{Onida2002}; in the latter case, the
Bethe--Salpeter equation (BSE) is solved on top of the GW solutions.
Within the Tamm--Dancoff
approximation~\cite{dancoff_non-adiabatic_1950,
  hirata_time-dependent_1999, Onida2002}, the BSE can be recast into
an effective two-particle eigenvalue problem~\cite{PhysRevB.62.4927,
  Onida2002}, which in the electron-hole
(e-h) basis reads
\begin{equation}
    \sum_{v'c'} \left[ D_{vc,v'c'} + 2K^{\rm x}_{vc,v'c'} - K^{\rm d}_{vc,v'c'}\right] A^{\lambda}_{v'c'} = E_{\lambda} A^{\lambda}_{vc},
    \label{eq:BSE_eh}
\end{equation}
where the $v, v'$ and $c, c'$ indices run over valence and conduction states, $E_{\lambda}$ are the neutral excitation energies (the poles of the density-density response function), and $A^{\lambda}_{vc}$ are the coefficients of the excitonic wavefunctions (that are related to oscillator strengths) in the e-h basis. The effective 2-particle Hamiltonian is made by i) a diagonal term $ D_{vc,v'c'} = (\varepsilon^{\rm QP}_c - \varepsilon^{\rm QP}_v)\delta_{vv'}\delta_{cc'}$ representing the ``bare'' e-h transitions (i.e., without accounting for the electron-hole interaction) from the \gls{QP} theory, ii) an exchange-like term 
\begin{equation}
  K^{\rm x}_{vc,v'c'} = \int d\br d\br' \, \phi_c^*(\br) \phi_v(\br) |\br-\br'|^{-1} \phi^*_{v'}(\br')\phi_{c'}(\br'),
\end{equation} 
and iii) a direct screened Coulomb term 
\begin{equation}
  K^{\rm d}_{vc,v'c'} = \int d\br d\br' \, \phi_c^*(\br) \phi_{c'}(\br) W(\br,\br') \phi^*_{v'}(\br')\phi_{v}(\br')
\end{equation} 
responsible for an effective attractive interaction between the electron and the hole. 
Here $\{\phi_i\}$ and $\{\varepsilon^{\rm QP}_i\}$ are the \gls{QP} wavefunctions and \gls{QP} energies, respectively. The solution of Eq.~\eqref{eq:BSE_eh} in the e-h basis would require the explicit computation of a significant number of empty states, which becomes in turn a critical convergence parameter. This can be conveniently avoided~\cite{umari_accelerating_2011, giustino_gw_2010, rocca_ab_2010, marsili_large-scale_2017} resorting to well established techniques from \gls{DFPT}~\cite{Baroni2001} by introducing i) the projector over the conduction manifold $\hat{Q} = \mathds{1} - \hat{P} = \mathds{1}- \sum_v |\phi_v \rangle \langle \phi_v|$, and ii) a set of auxiliary functions $\xi_v(\br) = \sum_{c} A_{cv} \phi_c(\br)$, usually called a \textit{batch} representation~\cite{walker_efficient_2006, rocca_turbo_2008}. These are $N_v$ (where $N_v$ is the number of occupied states) auxiliary functions that live in the unperturbed empty-states manifold and provide an equivalent but more compact representation of the excitonic wavefunction $\Theta(\br ,\br') = \sum_{vc} A_{cv}\phi^*_v(\br)\phi_c(\br') = \sum_v \phi_v(\br)\xi_v(\br')$. In this representation, the effective 2-particle Hamiltonian is completely specified by its action on the components of the batch~\cite{rocca_ab_2010}:
\begin{align}
     \sum_{v'} D_{vv'} |\xi_{v'}\rangle = &  \sum_{v'} (\hat{H}^{\rm QP} - \varepsilon_{v'}\mathds{1})\delta_{vv'} |\xi_{v'}\rangle \\
     \sum_{v'}K^{\rm x}_{vv'} |\xi_{v'}\rangle = & \sum_{v'} \hat{Q}\left( \int \frac{1}{|\br -\br'|} \phi^*_{v'}(\br') \xi_{v'}(\br') \, d\br'\right) |\phi_v\rangle \\
     \sum_{v'}K^{\rm d}_{vv'} |\xi_{v'}\rangle = & \sum_{v'} \hat{Q}\left( \int W(\br,\br') \phi^*_{v'}(\br') \phi_{v'}(\br') \, d\br'\right) |\xi_{v'}\rangle.
     \label{eq:BSE_batch}
\end{align}
The advantage of this formulation is that there is no explicit reference to the empty states (hidden inside the projector $\hat{Q}$ and the batch representation), and only the $N_v$ auxiliary functions $\{|\xi_v\rangle\}$ need to be determined solving the BSE in the batch representation. 

A further computational speed-up (and an improvement on the overall scaling~\cite{marsili_large-scale_2017}) can be achieved by moving from the \gls{KS} orbitals to \glspl{MLWF} and by exploiting their localization to greatly reduce the number of operations needed to evaluate the action of the BSE Hamiltonian on a trial state $|\xi_v\rangle$. This is particularly relevant for the direct term (Eq.~(\ref{eq:BSE_batch})), which represents the real bottleneck of the calculations. In the Wannier representation, this becomes
\begin{equation}
    \sum_{v'} K^{\rm d}_{vv'} |\tilde{\xi}_{v'}\rangle = \hat{Q}\left( \sum_{v'} \int W(\br,\br') \omega^*_{v'}(\br') \omega_{v'}(\br') \, d\br'\right) |\tilde{\xi}_{v'}\rangle,
    \label{eq:BSE_Kd_MLWF}
\end{equation}
where $\{\omega_v(\br)\}$ are the \glspl{MLWF} and $\tilde{\xi}_v(\br)$ is the batch component in the Wannier representation (simply obtained by rotating the original $\xi_v(\br)$ with the unitary matrix rotation that transforms the manifold from the canonical to the \gls{MLWF} representation). Exploiting locality, one can define a threshold for which a given pair of \glspl{MLWF} overlap. By excluding non-overlapping pairs of \glspl{MLWF} from the summation in Eq.~(\ref{eq:BSE_Kd_MLWF}), it becomes possible to lower the scaling of the evaluation of the action of the direct term on trial states from $\mathcal{O}(N^4)$ to $\mathcal{O}(N^3)$~\cite{marsili_large-scale_2017},
an approach that has been established and applied in the community~\cite{umari_accelerating_2011, marsili_large-scale_2017,Elliott2019}. 
Tight-binding and phenomenological models based on localized representation have also recently appeared~\cite{WanTiBEXOS,Alvarez2023}.

The concept of \gls{MLWF} can be actually extended to multiparticle Bloch states and has been recently applied to excitons, which are two-particle correlated e-h excitations and where maximal localization can be defined with respect to an average e-h coordinate in real space~\cite{Haber2023}. 
The benefits of these \glspl{MLXWF} for excitons are essentially the same as those of \glspl{MLWF} for electrons, including providing a compact basis for \emph{ab initio} exciton \gls{TB} models and interpolating key quantities such as the excitonic bands, the exciton-phonon matrix elements and Berry curvatures for the exciton wavefunction, while also providing physical insights on the nature of excitons~\cite{Haber2023}.

\subsection{\label{sec:ballistic}Ballistic transport and nanostructures}
\glspl{MLWF} can be used to build the electronic structure of large
nanostructures~\cite{Lee2005,Lihm2019} and to determine their
ballistic transport when connected to semi-infinite
leads~\cite{Calzolari2004,Lee2005}.  In this latter case, the
formalism of Green's functions is used to embed a conductor into the
surrounding environment. In all cases, the building blocks are
Hamiltonian matrix elements between the localized \glspl{MLWF} that
are used to construct, LEGO\textsuperscript{\texttrademark}-like,
either the desired non-self-consistent Hamiltonian of a much larger
nanostructure, or the self-energies embedding the conductor of
interest into semi-infinite leads. Importantly, any solid (or surface)
can be viewed as an infinite (or semi-infinite, in the case of
surfaces) stack of ``principal layers'' interacting only with neighboring
layers~\cite{Lee1981,Lee1981a}. In this way,
the infinite-dimensional real-space Hamiltonian can be divided into
finite-sized Hamiltonian matrices; for a bulk system (i.e., infinite
and periodic) the only independent components are $H_{00}$ and
$H_{01}$, where the former represents the interaction between
\glspl{MLWF} located in the same principal layer, and the latter the
interaction between orbitals in one principal layer and the next.

As discussed by Nardelli~\cite{Nardelli1999}, one can consider a system composed of a conductor $C$ connected to two
semi-infinite leads, $R$ and $L$ ($C$, $R$, and $L$ are in themselves composed by a finite or infinite number of principal layers).
The conductance through
$C$ is
related to the scattering properties of $C$ itself via the
Landauer formula \cite{Landauer1970}:
\begin{equation}
{\cal G}(E) = {2 e^2 \over h} {\cal T}(E),
\end{equation}
where ${\cal T}$ is the transmission function,
  ${\cal G}$ is the conductance, and ${\cal T}$ is the probability that an electron injected at one end of
the conductor will transmit to the other end. This transmission
function can be expressed in terms of the Green's functions of the
conductors and of its coupling to the leads
\cite{Lee1981,Lee1981a} as
\begin{equation}
{\cal T} = {\rm Tr}(\Gamma_L G_C^r \Gamma_R G_C^a),
\end{equation}
where $G_C^{\{r,a\}}$ are the retarded and advanced Green's functions
of the conductor, respectively, and $\Gamma_{\{L,R\}}$ are functions
that describe the coupling of the conductor to the two leads.  To
compute the Green's function of the conductor, one starts from the
equation for the Green's function of the whole system:
\begin{equation}
(\epsilon - H)G = \mathds{1}
\label{green}
\end{equation}
where $\epsilon = E+{\rm i}\eta$ with $\eta$ arbitrarily small, and
$\mathds{1}$ is the identity matrix. The
  matrix representation of $G$ in a WFs
  basis can be partitioned as
\begin{equation}
\left(
\begin{array}{lll}
G_L  & G_{LC} &  G_{LCR}
\\
G_{CL} & G_C & G_{CR}
\\
G_{LRC} & G_{RC} & G_R
\end{array}
\right)
=
\end{equation}
$$
\left(
\begin{array}{ccc}
(\epsilon -H_L) & H_{LC} & 0
\\
H_{LC}^\dagger & (\epsilon -H_{C}) & H_{CR}
\\
0 & H_{CR}^\dagger & (\epsilon -H_R)
\end{array}
\right)^{-1}.
\label{condmatr}
$$
We can then write the conductor Green's function as
\begin{equation}
G_C(E) = (\epsilon - H_C - \Sigma_L - \Sigma_R)^{-1},
\label{eq:bulkgreen}
\end{equation}
where the effect of the semi-infinite leads on the conductor
is described by the self-energies $\Sigma_{L,R}$, and the coupling
functions $\Gamma_{\{L,R\}}$ are obtained from the self-energies as
\begin{equation}
\Gamma_{\{L,R\}} = {\rm i}[\Sigma_{\{L,R\}}^r -
\Sigma_{\{L,R\}}^a]\,;
\label{eq:gamma}
\end{equation} 
here, the advanced self-energy is the
Hermitian conjugate of the retarded one, and we solve
  for the latter.  Given that the semi-infinite leads are also made of principal layers, one can construct the
self-energies as~\cite{Nardelli1999}
\begin{equation}
\begin{array}{l}
\Sigma_L=H_{LC}^{\dagger} (\epsilon
-H_{00}^{L}-(H_{01}^{L})^{\dagger}\overline T_L)^{-1}H_{LC},\\ \\
\Sigma_R=H_{CR}(\epsilon
-H_{00}^{R}-H_{01}^{R}T_R)^{-1}H_{CR}^{\dagger},
\end{array}
\label{sigma}
\end{equation} where, e.g.,
$H_{00}$ describes the intralayer interactions and
$H_{01}$ the interlayer couplings. 
The transfer matrices $\overline T_{L,R}$ and $T_{L,R}$, defined such that
$G_{10}=TG_{00}$ and $G_{00}={\overline T}G_{10}$, are calculated following
the iterative procedure by Lopez-Sancho et al.~\cite{LopezSancho1984}.
The only inputs are the matrix elements of the Hamiltonian $H_{mn}$ in a localized representation; the accuracy of the results depends on having principal layers 
that do not couple beyond next neighbors, i.e., on having a well-localized 
basis.
The knowledge of the conductor Green's function $G_C$ also gives direct information on the electronic spectrum via the spectral density of the electronic states:
$N_C(E) = -(1/\pi) {\rm Im}({\rm Tr}G_C(E))$.

As an example, we take a (5,5) single-wall carbon nanotube, described with a supercell containing 100 carbon atoms (five times the primitive cell); the disentanglement and Wannierization procedure results in 150 $sp^2$ orbitals and 100 $p_z$
orbitals. Such supercell size is
large enough to allow for $\Gamma$-sampling only in the
\gls{BZ}, and to have negligible overlap with the \glspl{MLWF} belonging to the second next supercell.
\begin{figure}
\centering
\includegraphics[width=0.7\columnwidth,angle=270]{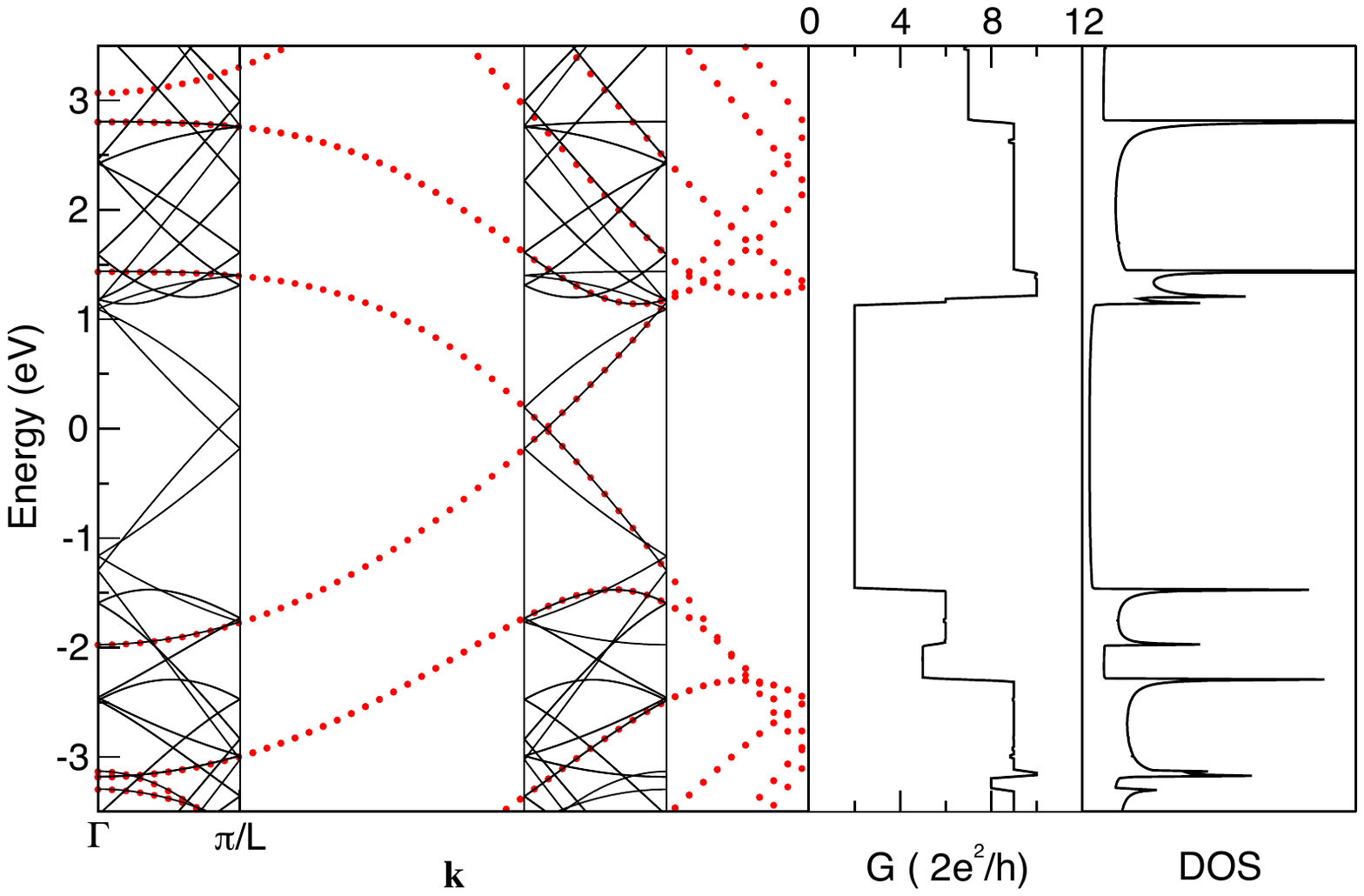}
\caption{Left panel: Comparison of the band structure for a (5,5)
  single-walled carbon nanotube, calculated in the 20-atom primitive cell with a self-consistent Hamiltonian using 5 $k$ points along the 1-dimensional Brillouin zone ($k=n\pi/L, n=0,1,...,4$) and then diagonalized non-self-consistently everywhere (red dots), or calculated in a supercell 5 times longer. In this latter case, the Brillouin zone is 5 times shorter, it displays 5 times more bands, and the Hamiltonian is consistently obtained using $\Gamma$ sampling only. While a pseudogap of more than 2~eV is present at $\Gamma$, non-self-consistent diagonalization everywhere (solid thin lines) captures very faithfully the metallic character of the nanotube. Folding the original bands (red dots) into the smaller Brillouin zone of the supercell also shows perfect agreement between the calculations. These results highlight how disentanglement in the supercell recovers from the empty states at $\Gamma$ what is needed to capture the character of filled and empty states away from $\Gamma$. The center and right panels reinforce this point, showing the ballistic conductance and density of states obtained from the Green's functions calculated using the supercell Wannier functions. Notably, the van Hove singularities and conductance steps are captured with great accuracy even when the bands edges are at arbitrary points, highlighting the role of \glspl{MLWF} as excellent interpolators and building blocks for the non-self-consistent electronic structure of large-scale nanostructures (see also~\textcite{Lee2005}).}
\label{fig:band}
\end{figure}
We show in Fig.~\ref{fig:band}
the results for the band structure, quantum conductance, and density of states, adapted from~\textcite{Lee2005}.
The metallic nanotube, notwithstanding its
$\sim$2 eV pseudo-gap at $\Gamma$, shows a
band structure in perfect agreement with that obtained from a very fine sampling of the BZ, and perfect agreement between the steps in conductance and the peaks in the van Hove singularities, with the maxima and the minima of the band dispersions at arbitrary points in the BZ.
While these methodologies based on \glspl{WF} have been implemented either as an extension of \code{Wannier90}~\cite{Shelley_CPC_2011} or standalone in the \gls{WF}-based \code{WanT} code~\cite{Ferretti2012}, analogous techniques have been developed for other types of localized orbitals and implemented in simulation codes such as \code{TranSiesta/Atomistix Toolkit}~\cite{Brandbyge_prb_2002}, \code{SMEAGOL}~\cite{Rocha_PRB_2006}, \code{OpenMX}~\cite{Ozaki_prb_2010}, \code{NEMO5}~\cite{Fonseca_jce_2013,Wang_JAP_2017,Wang_APL_2021,Sarangapani_PRAPPL_2022}, \code{nextnano}~\cite{Birner_IEEE_2007}, \code{NanoTCAD Vides}~\cite{Fiori_IEEE_2007,Bruzzone2014}, \code{KWANT}~\cite{Groth_IOP_2014}, some of which also provide an interface with the Wannier ecosystem.
In addition, we emphasize that different kinds of \glspl{WF} can be adopted as basis for \textit{ab initio} quantum transport calculations, such as the \glspl{POWF}~\cite{Thygesen2005, Fontana2021} discussed in Sec.~\ref{subsec:advancedminimisation}, which have been used both at the DFT~\cite{thygesen_cp_2005} and GW~\cite{thygesen_prb_2008} level.

\subsection{\label{subsec:Berry}Berryology}
\subsubsection{Motivation}

Berry phases and related quantities are central to the description of
the electronic properties of
crystals~\cite{Vanderbilt2018,xiao-rmp10}.  Here are some
representative examples~\footnote{All the formulas are given for single band occupancy.}.

\begin{enumerate}

\item The electronic contribution to the electric polarization of an
insulator is given by
\begin{equation}
{\bf P}_\text{el}=-e\sum_n^\text{occ}\int_\text{BZ}\frac{d^3k}{(2\pi)^3}\,
{\bf A}_{\kk,nn}\,,
\label{eq:polarization}
\end{equation}
where $-e$ is the electron charge and ${\bf A}_{\kk,nn}$ are
diagonal elements of the Berry connection matrix,
\begin{equation}
{\bf A}_{\kk,mn} = i\ip{u_{m\kk}}{{\boldsymbol\nabla}_{\kk} u_{n\kk}}\,.
\label{eq:connection}
\end{equation}

\item Off-diagonal elements of ${\bf A}_\kk$ describe electric-dipole
transition moments, allowing the interband optical conductivity to be expressed as
\begin{align}
\sigma_{ab}(\omega)&=
\frac{ie^2}{\hbar}\sum_{m,n}\int_\text{BZ}\frac{d^3k}{(2\pi)^3}\,
\left(f_{m\kk}-f_{n\kk}\right)\times\nonumber\\
&\times \frac{\varepsilon_{m\kk}-\varepsilon_{n\kk}} 
{\varepsilon_{m\kk}-\varepsilon_{n\kk}-\hbar(\omega+i0^+)}
A^a_{\kk,nm}A^b_{\kk,mn}\,, 
\label{eq:optical-conductivity}
\end{align}
where $f_{n\kk}$ is the Fermi-Dirac occupation factor.

\item The Berry curvature is defined as the curl of the Berry connection,
\begin{equation}
{\boldsymbol\Omega}\bnk=
{\boldsymbol\nabla}_\kk\times {\bf A}_{\kk,nn}=
-\text{Im}
\bra{{\boldsymbol\nabla}_\kk u\bnk}\times\ket{{\boldsymbol\nabla}_\kk u\bnk}\,,
\label{eq:curv}
\end{equation}
and its integral over the occupied states gives the intrinsic
\gls{AHC},
\begin{equation}
\sigma_{yx}=\frac{e^2}{\hbar}\int_\text{BZ}\frac{d^3k}{(2\pi)^3}\,
\sum_n\,f\bnk\Omega^z\bnk\,.
\label{eq:ahc}
\end{equation}

\item The ground-state orbital magnetization reads
\begin{equation}
{\bf M}_\text{orb}=\int_\text{BZ}\frac{d\kk}{(2\pi)^3}\,\sum_n\,
f\bnk
\left[
{\bf m}^\text{orb}\bnk+
\frac{e}{\hbar}\left(\varepsilon_\text{F}-\varepsilon\bnk\right)
{\boldsymbol\Omega}\bnk
\right]\,,
\label{eq:orbital-magnetization}
\end{equation}
with $\varepsilon_\text{F}$ the Fermi energy and
\begin{equation}
{\bf m}^\text{orb}\bnk=\frac{e}{2\hbar}
\text{Im}\bra{{\boldsymbol\nabla}_\kk u\bnk}\times
\left(\hat H_\kk-\varepsilon\bnk\right)
\ket{{\boldsymbol\nabla}_\kk u\bnk}
\label{eq:orbital-moment}
\end{equation}
the intrinsic orbital moment of a Bloch state.

\item To first order in applied fields ${\bf E}$ and ${\bf B}$, the
semiclassical equations of motion for a wavepacket in a Bloch band
read
\begin{subequations}
\begin{align}
\dot \rr&=\frac{1}{\hbar}{\boldsymbol\nabla}_\kk\tilde\varepsilon\bnk-
\dot\kk\times{\boldsymbol\Omega}\bnk\,,\\
\dot \kk&=-\frac{e}{\hbar}{\bf E}-\frac{e}{\hbar}\dot\rr\times{\bf B}\,,
\end{align}
\end{subequations}
where
\begin{equation}
\tilde\varepsilon\bnk=\varepsilon\bnk-
\left({\bf m}^\text{spin}\bnk+{\bf m}^\text{orb}\bnk\right)\cdot{\bf B}
\end{equation}
is the Zeeman-shifted band energy.

\end{enumerate}

The motivation to apply Wannier interpolation to Berry-type quantities
came from pioneering \textit{ab initio} calculations of the \gls{AHC} in the
ferromagnets SrRuO$_3$~\cite{fang-science03} and \gls{BCC}
Fe~\cite{yao-prl04}, which revealed the integrand of Eq.~\eqref{eq:ahc}
to be strongly peaked in the vicinity of avoided crossings between
occupied and empty bands; resulting in the need to
sample the BZ over millions of $k$ points to reach convergence.  An
efficient Wannier-interpolation scheme for evaluating the \gls{AHC} was
developed in~\textcite{Wang-AHC-PhysRevB.74.195118}, and since then
the methodology has been applied to many other properties.

Wannier interpolation of Berry-type quantities was introduced in
version~2 of \code{Wannier90} as part of its post-processing code
\code{postw90}~\cite{mostofi-cpc14}, with the ability to compute
\gls{AHC}~\cite{Wang-AHC-PhysRevB.74.195118}, interband optical
conductivity~\cite{Yates_et_al:2007}, and orbital
magnetization~\cite{Lopez-orbital-PhysRevB.85.014435}. The
list of available properties has grown considerably since then, and
more recently the \code{WannierBerri} code
package~\cite{Tsirkin2021-WannierBerri} introduced several
methodological improvements including ``pruned FFT'' \cite{Markel-FFT-pruning,Sorensen-FFT} (a combination of fast and slow Fourier transforms),
and the use of symmetries and of the tetrahedron method for BZ
integrals. Other codes also compute Berry-type quantities in different contexts, such as \code{dynamics-w90}~\cite{dynamics-w90} for time-dependent dynamics. We note that some codes, including \code{WannierTools}~\cite{Wu2018} and
\code{linres} \cite{Zelezny-rinres-code}, implement some functionalities but
with additional approximations (see discussion of
Eq.~\eqref{eq:r-matrix-diag} below).  The reader should consult the
documentation of the codes for an up-to-date description of their
capabilities.

In the following, we outline the basic interpolation strategy for
Berry-type quantities, using the off-diagonal Berry connection
as an example (the diagonal Berry connection that enters the Berry
phase requires a separate treatment, see Sec.~\ref{subsec:topoinv}).
For discussion purposes, we adopt the modified phase convention for
Bloch sums indicated in Eq.~\eqref{eq:phase-convention}.

\subsubsection{Wannier interpolation of the interband Berry
  connection}

We wish to evaluate off-diagonal elements of the Berry connection
matrix in the Hamiltonian gauge.  Inserting in
Eq.~\eqref{eq:connection} the relation~\eqref{eq:psi-H} between
interpolated Bloch states in the Hamiltonian and Wannier gauges, we
obtain
\begin{equation}
{\bf A}^\text{H}_{\kk'}=
i{\mathcal U}^\dagger_{\kk'}{\boldsymbol\nabla}_{\kk'}{\mathcal U}_{\kk'}+
{\mathcal U}^\dagger_{\kk'}{\bf A}^\text{W}_{\kk'}{\mathcal U}_{\kk'}\,.
\label{eq:A-H}
\end{equation}
Note the extra (first) term compared to the gauge-transformation
rule~\eqref{eq:X-H} for a lattice-periodic matrix object.
Recalling from Eq.~\eqref{eq:eig-int} that the columns of
${\mathcal U}_{\kk'}$ are eigenvectors of $H^\text{W}_{\kk'}$,
the off-diagonal matrix elements of that term can be evaluated from
non-degenerate perturbation theory as
\begin{equation}
\left(
{\mathcal U}^\dagger_{\kk'}{\boldsymbol\nabla}_{\kk'}{\mathcal U}_{\kk'}
\right)_{mn}=
\frac{
  \left[
  {\mathcal U}^\dagger_{\kk'}
  \left({\boldsymbol\nabla}_{\kk'} H^\text{W}_{\kk'}\right)
  {\mathcal U}_{\kk'}
  \right]_{mn}}
{\varepsilon^\text{H}_{n\kk'}-\varepsilon^\text{H}_{m\kk'}}\,.
\label{eq:A-int}
\end{equation}
All quantities on the right-hand side can be obtained from
Eqs.~\eqref{eq:H-k} and~\eqref{eq:eig-int} starting from
$\me{{\bf 0}i}{\hat{H}}{\RR j}$ and
$\ttau_j=\me{{\bf 0}j}{{\hat\rr}}{{\bf 0}j}$ (the latter appears
in the modified phase factors in Eq.~\eqref{eq:phase-convention}).
For the second term in Eq.~\eqref{eq:A-H}, we also need
\begin{equation}
{\bf A}^\text{W}_{\kk',ij}=
\sum_\RR\,e^{i\kk'\cdot(\RR+\ttau_j-\ttau_i)}{\bf d}_{ij}(\RR)\,,
\label{eq:A-W}
\end{equation}
which follows from inserting in Eq.~\eqref{eq:connection} the Bloch
sum~\eqref{eq:psi-W} with the modified phase factors. Here
${\bf d}_{ij}(\RR)$ are the off-diagonal matrix elements of $\hat\rr$
in the Wannier basis, that is,
\begin{equation}
\me{{\bf 0}i}{\hat\rr}{\RR j}=
\delta_{\RR,{\bf 0}}\delta_{ij}\ttau_j+{\bf d}_{ij}(\RR)\,.
\label{eq:r-matrix}
\end{equation}

The matrix elements $\me{{\bf 0}i}{\hat{H}}{\RR j}$ are evaluated
using Eq.~\eqref{eq:H-R}, and the corresponding procedure for
$\me{{\bf 0}i}{\hat\rr}{\RR j}$ is as follows. First, we can use
Eq.~\eqref{eq:wannier-dis} to write $\me{{\bf 0}i}{\hat\rr}{\RR j}$ as
$(1/N)\sum_\kk\,e^{-i\kk\cdot\RR} A^\text{W}_{\kk,ij}$.
Since the Bloch functions are smooth in the Wannier gauge,
${\bf A}^\text{W}_{\kk,ij}$ can be evaluated on the \textit{ab initio}
grid by discretizing the $\kk$ derivative appearing in
Eq.~\eqref{eq:connection}. Adopting the finite-differences scheme
described in~\textcite{Marzari1997} and~\textcite{Mostofi2008} we obtain
\begin{align}
\me{{\bf 0}i}{\hat\rr}{\RR j}&=
\frac{i}{N}\sum_\kk\,e^{-i\kk\cdot\RR}\sum_{\bf b}\,w_b{\bf b}\,\times\nonumber\\
&\times\sum_{m,n}\,
V^*_{\kk,mi}M_{mn}^{(\kk,{\bf b})}V_{\kk+{\bf b},nj}\,,
\label{eq:r}
\end{align}
where ${\bf b}$ are vectors connecting neighboring grid points, $w_b$
are appropriately chosen weights, $M^{(\kk,{\bf b})}$ are the overlap
matrices defined by Eq.~\eqref{eq:Mkb}, and $V_\kk$ are the
Wannierization matrices in Eq.~\eqref{eq:psi-smooth}; since the
overlap matrices were computed in preparation for the \gls{WF}
construction procedure, and the Wannierization matrices were obtained
at the end of that procedure, both are readily available.

Once $\me{{\bf 0}i}{\hat{H}}{\RR j}$ and
$\me{{\bf 0}i}{\hat\rr}{\RR j}$ have been tabulated, the interband
Berry connection can be evaluated from
Eqs.~\eqref{eq:A-H}--\eqref{eq:A-W}, with the matrices
${\mathcal U}_{\kk'}$ therein (along with the interpolated energy
eigenvalues) given by Eq.~\eqref{eq:eig-int}.  Finally, the Berry
connection and energy eigenvalues are inserted in
Eq.~\eqref{eq:optical-conductivity} to obtain the interband optical
conductivity~\cite{Yates_et_al:2007}.

Equation~\eqref{eq:r} entails a numerical error of order
$(\Delta k)^2$,
where $\Delta k$ is the \textit{ab initio} mesh
spacing~\cite{Marzari1997,Mostofi2008}. The direct real-space mesh
integration method mentioned below Eq.~\eqref{eq:H-R} should be free
of such errors, but it is not as practical in the context of $k$-space
Wannierization schemes.  It is therefore desirable to develop
improved discretized $k$-space formulas for
$\me{{\bf 0}i}{\hat\rr}{\RR j}$ and related matrix elements.  A
higher-order generalization of the discretization scheme of~\textcite{Marzari1997} and~\textcite{Mostofi2008} was recently
introduced~\cite{Cistaro2023}, and further improvements are currently
under way~\cite{JML_Wannier2022,MG_Wannier2022}.

In empirical tight-binding, it is customary to approximate the
position matrix elements by dropping the second term in
Eq.~\eqref{eq:r-matrix}~\cite{Vanderbilt2018,foreman-prb02},
\begin{equation}
\me{{\bf 0}i}{\hat\rr}{\RR j}\approx
\delta_{\RR,{\bf 0}}\delta_{ij}\ttau_j\,.
\label{eq:r-matrix-diag}
\end{equation}
In this approximation, and when the modified phase
convention~\eqref{eq:phase-convention} is used, the matrix
${\bf A}^\text{W}_{\kk'}$ in Eq.~\eqref{eq:A-W} vanishes;
Eq.~\eqref{eq:A-H} for ${\bf A}^\text{H}_{\kk'}$ then reduces to its
first term, which can be interpreted as the ``internal'' Berry
connection of the tight-binding eigenvectors; this is how
tight-binding codes such as \code{PythTB} evaluate Berry phases and
curvatures~\cite{pythtb_notes}, which is quite natural in the context
of toy-model calculations. 

The above approximation is harder to justify when using \textit{ab
  initio} \glspl{WF}, given that the discarded ${\bf d}_{ij}(\RR)$ matrix
elements are readily available, as mentioned above; even so, that
approximation is made by some codes, including
\code{WannierTools}~\cite{Wu2018} and
\code{linres}~\cite{Zelezny-rinres-code}. The role of intra-atomic
${\bf d}_{ij}(\RR)$ matrix elements in tight-binding calculations of
the linear dielectric function was studied in~\textcite{pedersen-prb01}; in
~\textcite{azpiroz-scipost22}, that analysis was extended to
inter-atomic matrix elements and to non-linear optical responses, using
Wannier interpolation. The importance of the ${\bf d}_{ij}(\RR)$ off-diagonal elements for gauge invariance has been discussed also in the context of time-dependent dynamics~\cite{schuler_prb_2021}.
\subsubsection{Other Berry-type quantities}

The interpolation of the Berry curvature ${\boldsymbol\Omega}_\kk$
proceeds along similar lines, allowing the computation of \gls{AHC} from
Eq.~\eqref{eq:ahc}~\cite{Wang-AHC-PhysRevB.74.195118}, and the
procedure can be extended to \gls{SHC}~\cite{Qiao-SHC-PhysRevB.98.214402,Ryoo-SHC-PhysRevB.99.235113}
or non-linear responses. For example, non-linear
optical and \glspl{AHC} involve $k$-derivatives of
${\bf A}_\kk$ and ${\boldsymbol\Omega}_\kk$,
respectively~\cite{Aversa1995,sodemann-prl15}; in the same way as band
derivatives (see Sec.~\ref{sec:band-der}), both are conveniently
evaluated by Wannier
interpolation~\cite{wang-prb17,Ibanez-shift-current-PhysRevB.97.245143,liu2023covariant}.
To interpolate ${\bf A}_\kk$, ${\boldsymbol\Omega}_\kk$, and their
$k$-derivatives, only $\me{{\bf 0}i}{\hat{H}}{\RR j}$ and
$\me{{\bf 0}i}{\hat\rr}{\RR j}$ are needed, but other quantities
require additional matrix elements
\footnote{By explicitly plugging the Bloch sum of Eq.~\eqref{eq:psi-W} into the RHS of Eq.~\eqref{eq:curv}, one may think that a matrix element 
$\me{{\bf 0}i}{\hat{r}_a\hat{r}_b}{\RR j}$ might be needed. However, 
this term is symmetric under the exchange of $a$ and $b$, therefore it does not contribute to the cross-product, as shown in Eq.~(40) of Ref.~\textcite{Wang2007}.}.
For example, the orbital moment in
Eq.~\eqref{eq:orbital-moment} requires
$\me{{\bf 0}i}{\hat H(\hat\rr-\RR)}{\RR j}$ and
$\me{{\bf 0}i}{\hat r_a\hat H(\hat r-R)_b}{\RR
  j}$~\cite{Lopez-orbital-PhysRevB.85.014435}; and while the former
can be evaluated on the \textit{ab initio} grid using the same
ingredients entering Eqs.~\eqref{eq:H-R} and~\eqref{eq:r} for
$\me{{\bf 0}i}{\hat{H}}{\RR j}$ and $\me{{\bf 0}i}{\hat\rr}{\RR j}$,
the latter also requires
$\me{u_{m\kk+\bb_1}}{\hat{H}_{\kk}}{u_{n\kk+\mathbf{b}_2}}$, which
must be calculated separately~\cite{Lopez-orbital-PhysRevB.85.014435}.
As in the case of Eq.~\eqref{eq:A-H} for ${\bf A}_{\kk'}$, the
resulting interpolation formula contains an ``internal'' term
analogous to Eq.~\eqref{eq:orbital-moment} itself, but expressed in
terms of the \gls{TB} eigenvectors and Hamiltonian. That term can be
evaluated from $\me{{\bf 0}i}{\hat{H}}{\RR j}$ and $\ttau_j$ alone,
but it leaves out important atomic-like contributions that are encoded
in the other matrix elements needed for a full
calculation~\cite{Lopez-orbital-PhysRevB.85.014435,nikolaev-prb14}.
\begin{table*}[!htb]
    \centering
    \begin{tabular}{c|c|c|c|c}
         Matrix element & Wannier90 file & Needed for
         & Implemented in & Computed by \code{mmn2uHu} from  \\
         \hline 
          $\me{u_{m\kk}}{\hat{H}_{\kk}}{u_{n\kk}}= \varepsilon_{n\kk}\delta_{mn}$ & \texttt{*.eig} &  Transport, optics, ${\bf m}^\text{orb}_\kk$ & All &\\
          $\me{u_{m\kk}}{\hat{\boldsymbol\sigma}}{u_{n\kk}}$ & \texttt{*.spn} &  % Spin-related properties
          ${\bf m}^\text{spin}_\kk$  & QE,VASP & \\
         $\ip{u_{m\kk}}{u_{n\kk+\mathbf{b}}}$  & \texttt{*.mmn} & ${\bf A}_\kk$, $\boldsymbol{\Omega}_\kk$, ${\bf m}^\text{orb}_\kk$,  % optical effects,
         all  below & All &  \\
         $\me{u_{m\kk+\bb_1}}{\hat{H}_{\kk}}{u_{n\kk+\mathbf{b}_2}}$  & \texttt{*.uHu} &  % Orbital moment
         ${\bf m}^\text{orb}_\kk$ & QE & \texttt{*.eig}, \texttt{*.mmn} \\
         $\ip{u_{m\kk+\bb_1}}{u_{n\kk+\mathbf{b}_2}}$ & \texttt{*.uIu} &
         % Spatial dispersion
% Optical activity
$q^{ab}_\kk$
         & QE & \texttt{*.mmn}\\
          $\me{u_{m\kk}}{\hat{\boldsymbol\sigma} \hat{H}_{\kk}}{u_{n\kk+\bb}}$ & \texttt{*.sHu} &  \multirow{2}{*}{$\Bigr\}$ SHC in~\textcite{Ryoo-SHC-PhysRevB.99.235113}} & QE & \texttt{*.eig}, \texttt{*.mmn}, \texttt{*.spn} \\
           $\me{u_{m\kk}}{\boldsymbol{\sigma}}{u_{n\kk+\bb}}$& \texttt{*.sIu} &   & QE &  \texttt{*.mmn}, \texttt{*.spn} \\
            \end{tabular}
            \caption{\textit{Ab initio} matrix elements  that are used \textit{explicitly} in setting up
              the WF matrix elements needed to perform common
              interpolation tasks, SHC stands for spin Hall conductivity.
               * denotes the seedname specified in
              the input file of \code{Wannier90}. QE~=~\code{Quantum ESPRESSO},
              ${\bf A}_\kk$ is the Berry connection or electric dipole
              matrix~\eqref{eq:connection}, $\boldsymbol{\Omega}_\kk$
              is the Berry curvature~\eqref{eq:curv},
              ${\bf m}^\text{orb}_\kk$ is the intrinsic orbital
              magnetic dipole~\eqref{eq:orbital-moment}, and
              $q^{ab}_\kk$ is the intrinsic electric quadrupole (for
              the matrix definitions of ${\bf m}^\text{orb}_\kk$ and
              $q^{ab}_\kk$, see~\textcite{pozo2023multipole}). The
              optical conductivity~\eqref{eq:optical-conductivity} in
              the electric-dipole approximation involves
              $\varepsilon_{n\kk}$ and ${\bf A}_\kk$, the \gls{AHC}~\eqref{eq:ahc} involves
              $\varepsilon_{n\kk}$ and ${\boldsymbol\Omega}_\kk$, and
              the orbital
              magnetization~\eqref{eq:orbital-magnetization} involves
              $\varepsilon_{n\kk}$, ${\boldsymbol\Omega}_\kk$ and
              ${\bf m}^\text{orb}_\kk$.  Spatially dispersive
              responses such as natural optical activity depend on
              $\varepsilon_{n\kk}$, ${\bf A}_\kk$,
              ${\bf m}^\text{orb}_\kk$, ${\bf m}^\text{spin}_\kk$, and
              $q^{ab}_\kk$~\cite{pozo2023multipole}. The \texttt{mmn} (and \texttt{eig})
              matrix elements are needed for constructing the
              (disentangled) WFs, and hence they are used
              \textit{implicitly} when interpolating any physical
              quantity.}
    \label{tab:matrix_elements-revised}
\end{table*}

The \textit{ab initio} matrix elements needed for various
interpolation tasks are listed in
Table~\ref{tab:matrix_elements-revised}.
The files \texttt{*.eig} and
\texttt{*.mmn} are required already for constructing the \glspl{WF},
and therefore they are provided by the interface code of every
\textit{ab initio} code that is compatible with \code{Wannier90}. The file
\texttt{*.spn} is provided by the interface of both \code{Quantum
  ESPRESSO} and \code{VASP}. As for the other matrix elements listed
in Table~\ref{tab:matrix_elements-revised}, at present they are only
implemented in \code{pw2wannier90.x}, the interface of \code{Quantum
  ESPRESSO}.

As a workaround for obtaining these quantities from the output of
  other \textit{ab initio} engines, one can resort to a
  sum-over-states procedure.
For example, the \texttt{uHu} matrix elements may be expressed as
\begin{equation}
    \me{u_{m\kk+\mathbf{b}_1}}{\hat{H}_{\kk}}{u_{n\kk+\mathbf{b}_2}} \approx\sum_l^{l_{\rm max}}  \ip{u_{m\kk+\mathbf{b_1}}}{u_{l\kk}} \varepsilon_{l\kk}   \ip{u_{l\kk}}{u_{n\kk+\mathbf{b_2}}}
    \label{eq:Cmnq-sum}
\end{equation}
in terms of the energy eigenvalues and overlap
matrices, and the relevant matrices for spin Hall
conductivity can be obtained similarly~\cite{Qiao-SHC-PhysRevB.98.214402}.  Since the
summation is done before Wannierization, the number $l_{\rm max}$ of
states included in the non-self-consistent \textit{ab initio}
calculation can be systematically increased until the desired level of
convergence is reached.

The above procedure is implemented for \texttt{uHu}, \texttt{uIu},
 \texttt{sHu} and \texttt{sIu} in the
utility \code{mmn2uHu}~\cite{mmn2uHu_webpage} provided with the
\code{WannierBerri} code package \cite{Tsirkin2021-WannierBerri}.
Besides its use as a workaround, it can serve as a
  benchmark for testing future implementations of those matrix elements
  in various interface codes between \textit{ab initio} and Wannier
  engines.

\subsection{\label{subsec:topoinv}Topological invariants and related properties}

The topological aspects of band theory have been studied intensively
over the past two decades~\cite{Vanderbilt2018,Hasan2010}, and
\textit{ab initio} calculations have been central to that
effort~\cite{Zhang_natmat_2017}: they are used for identifying candidate
topological materials, to determine topological invariants, and to
calculate surface bands that can be compared with angle-resolved
photoemission measurements.

\glspl{WF} feature prominently in topological band theory.  For
example, quantum anomalous Hall insulators~\cite{Haldane1988}
(a.k.a. Chern insulators) can be defined as 2D systems where it is
not possible~\footnote{Very recent work~\cite{barnett_PRR_2024} suggests that it might be possible to construct optimally localized WFs even in presence of non-vanishing Chern numbers, at least for the simple case of one isolated band.} to construct a set of
exponentially localized \glspl{WF}~\cite{Brouder2007,timo_prb_2006,Vanderbilt2018}: this is known as a
``topological obstruction''.  More generally, symmetry-protected
topological insulators can typically be defined as insulators for
which it is not possible to construct a set of \glspl{WF} spanning the
valence bands without breaking the protecting symmetry in the choice
of gauge~\cite{Soluyanov2011,Bradlyn2017}. Prominent examples include
2D quantum spin Hall
insulators~\cite{KM2005a,KM2005b,bernevig_prl_2006} and 3D
$\mathbb{Z}_2$ topological insulators~\cite{fu_prl_2007}, where the
protecting symmetry is time reversal, and topological crystalline
insulators, which are instead protected by crystalline
symmetries~\cite{fu_prl_2011}.

A second example is the ``Wannier spectrum'' defined
  by the centers of hybrid (a.k.a. hermaphrodite) orbitals~\cite{Sgiarovello2001} that are Wannier-like along
  $\hat{\bf z}$ and Bloch-like along $\hat{\bf x}$ and~$\hat{\bf
    y}$. The surface energy spectrum $\varepsilon_n(k_x,k_y)$ can be
  continuously deformed into the bulk Wannier spectrum $z_n(k_x,k_y)$
  obtained by Wannierizing along the surface
  normal~\cite{Fidkowski2011,Neupert2018}, allowing to infer the
  topological flow of the surface energy bands from that of the bulk
  Wannier bands~\cite{Taherinejad2014}. In some cases, the topological
  indices can be deduced from the Wannier band
  structure~\cite{Gresch2017,Varnava2020}.
  
\glspl{WF} also play a more practical role in the study of
topological materials, as several of the electronic-structure packages
that are commonly used to characterize them rely on a Wannier/TB
representation (despite the obstruction mentioned
above~\cite{Soluyanov2011,Bradlyn2017}, topological insulators still
afford a Wannier representation, provided that the \glspl{WF} span a few
low-lying conduction states along with the valence bands).

In particular, both \code{PythTB}~\cite{pythtb} and
\code{WannierTools}~\cite{Wu2018} work with orthogonal \gls{TB}
models, and have the option to import \gls{TB} Hamiltonians generated
by a Wannier engine using either the $*${\_}{\tt hr.dat} or the
$*${\_}{\tt tb.dat} file format (as mentioned above, the latter also includes the matrix
elements of the position operator in the Wannier basis, and the
coordinates of the lattice vectors). The key point is that the
Wannierized Hamiltonian preserves the topological features of the
original first-principles electronic structure; the identification and
characterization of those features can therefore be carried out
entirely in the Wannier representation, which is often more convenient
and/or efficient than proceeding directly from the \textit{ab initio}
Bloch states.

The simplest example of a topological band-structure feature is an
isolated touching between a pair of bands, known as a ``Weyl
point''~\cite{Vanderbilt2018,Armitage2018}. Weyl points are
fundamentally different from weak avoided crossings, but most band
interpolation schemes are unable to tell them apart; instead, Wannier
interpolation correctly distinguishes between the two. The
distinction is rooted in the fact that a Weyl node acts as a monopole
source or sink of Berry curvature in ${\bf k}$ space, allowing to
associate with it a topological invariant known as the ``chiral
charge.''

The chiral charge $\chi$ (typically $\pm 1$, but sometimes $\pm 2$ or
$\pm 3$~\cite{Fang2012,Tsirkin2017}) can be determined in two
different ways: (i) from the quantized Berry-curvature flux
through a small surface ${\mathcal S}$ enclosing the Weyl
point~\cite{Vanderbilt2018,Gosalbez2015},
\begin{equation}
\int_{\mathcal S}{\boldsymbol\Omega}_{n\kk}\cdot\hat{\bf n}=-2\pi\chi\,,
\end{equation}
where $\hat{\bf n}$ is a unit vector in the direction of
${\boldsymbol\nabla}_\kk \varepsilon_{n\kk}$; (ii) by evaluating the
Berry phase
\begin{equation}
\phi_n({\mathcal C})=\oint_{\mathcal C}{\bf A}_{nn\bf k}\cdot d\kk
\label{eq:berry-phase}
\end{equation}
around contours ${\mathcal C}$ at fixed latitude on an enclosing
spherical surface, and then tracking its evolution from zero at the
south pole to $2\pi\chi$ at the north pole~\cite{Gresch2017}.

The latter procedure is implemented in both \code{Z2Pack}~\cite{z2pack_code} and
\code{WannierTools}. All that is required is the \gls{TB} Hamiltonian
$\me{{\bf 0}i}{\hat{H}}{\RR j}$, from which one obtains the
eigenvectors on a discrete mesh $\{\kk_j\}$ of points along each
contour; the Berry phase is then evaluated by finite differences from
the overlaps between \gls{TB} eigenvectors on consecutive points along
${\mathcal C}$ as follows~\cite{Vanderbilt2018},
\begin{equation}
\phi_n^{\text{(int)}}({\mathcal C})=-\text{Im ln}\,\Pi_j
\lip{u_{n\kk_j}}{u_{n\kk_{j+1}}}\,,
\label{eq:phi-int}
\end{equation}
where $\lket{u_{n\kk_j}}$ denotes a column vector of the matrix
${\mathcal U}_{\kk_j}$ defined by Eq.~\eqref{eq:eig-int}. The above
expression corresponds, in the language of Sec.~\ref{subsec:Berry}, to
the internal part of the Berry phase~\eqref{eq:berry-phase}, which
also contains an external part
\begin{equation}
\phi_n^{\text{(ext)}}({\mathcal C})=\sum_j\,
\lme{u_{n\kk_j}}{{\bf A}^\text{W}_\kk}{u_{n\kk_j}}\cdot\Delta\kk\,,
\label{eq:phi-ext}
\end{equation}
where $\Delta\kk=(\kk_{j+1}-\kk_{j-1})/2$~\cite{Wang2007}. The two
parts arise from discretizing the integral along ${\mathcal C}$ of the
two terms in Eq.~\eqref{eq:A-H} for the interpolated Berry connection;
the internal term only depends on $\me{{\bf 0}i}{\hat{H}}{\RR j}$,
while the external one also requires $\me{{\bf 0}i}{\hat\rr}{\RR
  j}$. The $2\pi$ indeterminacy in the Berry phase comes from the
former, while the latter is single-valued and hence it does not
contribute to the quantized change in Berry phase from the south to
the north pole of a spherical surface; this is why $\chi$ can be
determined from the \gls{TB} Hamiltonian alone.

\begin{figure}[tb]
  \centering\includegraphics[width=0.45\textwidth]{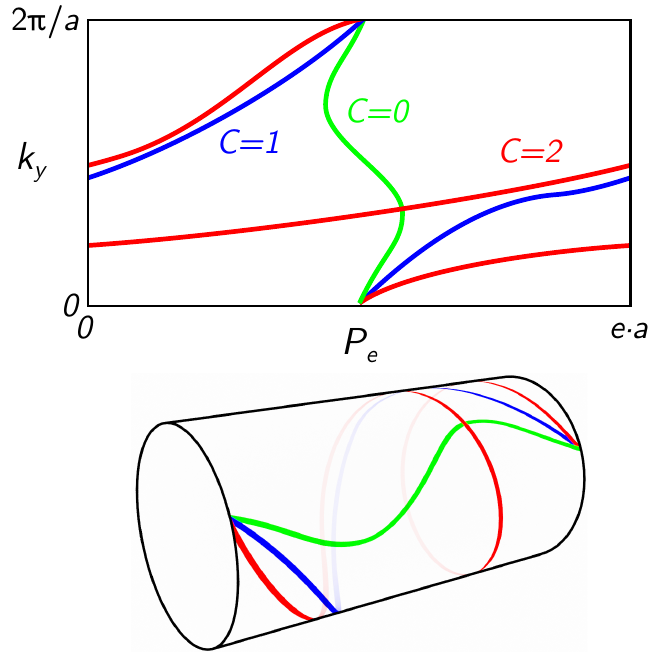}
  \caption{\label{fig:hwcc}Sketch of some possible evolutions of
    hybrid polarization $P_e(k_y)$, i.e. the sum of hybrid Wannier
    charge centers, across the \gls{BZ}. Chern numbers $C$ correspond
    to different winding numbers. See~\textcite{Gresch2017} for an in-depth discussion.}
\end{figure}

Weyl crossings can occur at arbitrary points in the \gls{BZ}, which makes it
difficult to spot them in the band structure. By allowing to quickly
evaluate energy eigenvalues and band velocities at arbitrary $k$ points, Wannier interpolation provides a practical solution to this
problem~\cite{Gosalbez2015}: to locate the degeneracies between bands
$n$ and $n+1$, define a gap function
$\varepsilon_{n+1\kk}-\varepsilon_{n\kk}$, and search for its minima
using a minimization method such as conjugate-gradient, starting from
a sufficiently dense grid of $k$ points. After discarding local minima
where the gap function is above some numerical threshold, one is left
with candidate degeneracies that can be further characterized; they
include not only point nodes such as Weyl and Dirac
nodes~\cite{Armitage2018}, but also nodal
lines~\cite{Fang2016,Yang2018}.  This procedure is implemented in
\code{WannierTools}.

Topological materials feature characteristic boundary modes that
reflect the bulk topology~\cite{Vanderbilt2018,Hasan2010}. In the case
of Weyl semimetals, those modes take the form of ``Fermi arcs''
connecting the projections of bulk Weyl nodes of opposite chirality
onto the surface \gls{BZ}~\cite{Vanderbilt2018,Armitage2018}.  \code{PythTB} and
\code{WannierTools} allow to terminate a bulk \gls{TB} model along specified
directions, creating ribbons or slabs whose boundary modes can then be
inspected by plotting the energy bands. \code{WannierTools} also has the
option to obtain a surface spectral function from the Green's function
calculated for a semi-infinite system. We remark that, in general, these methods yield boundaries that are the result of a truncation of the Wannier Hamiltonian, where the on-site energies and hoppings between surface (edge) atoms are left unchanged to their bulk value. More broadly, these methods completely neglect any relaxation, charge redistribution and reconstruction of the surface (edge). Hence, these are very crude and unrealistic approximations, and the corresponding calculations do not provide a first-principles description of the boundary electronic structure, even if they derive from a bulk Wannier Hamiltonian that was calculated with first-principles methods. Still, they provide valuable insights for the prediction of purely topological properties, in particular the existence of topologically protected surface (edge) states such as chiral or helical edge states, surface Dirac cones, or Fermi arcs. While these calculations can support the \emph{existence} of these boundary states, their precise \emph{band dispersion} requires to treat more explicitly the boundary electronic structure, typically through supercell slab (ribbon) simulations with structural optimization.

As already mentioned, another very useful tool for diagnosing
topological behaviors is a hybrid representation of the electronic
structure in terms of orbitals that are localized in one spatial
direction only, remaining extended in the
others~\cite{Sgiarovello2001}.
To define these \glspl{HWF} for lattices of
arbitrary symmetry, it is convenient to work in reduced
coordinates. Consider a 2D crystal, and let
$\kk=k_1{\bf b}_1+k_2{\bf b}_2$ and
$\hat{\bf r}=\hat x_1{\bf a}_1+\hat x_2{\bf a}_2$. Choosing
${\bf b}_2$ as the localization direction, the \glspl{HWF} are defined as
\begin{equation}
\ket{h_{k_1ln}}=\frac{1}{N_2}\sum_{k_2}\,e^{-i2\pi k_2l}\ket{\psi_{n \kk}}\,,
\end{equation}
where $N_2$ is the number of distinct values of $k_2$ in the \gls{BZ}, and
$l$ labels cells along ${\bf a}_2$.
The topological indices can be determined
from the winding of the \gls{HWF} centers
\begin{equation}
x_{2,k_1ln}=\me{h_{k_1ln}}{\hat x_2}{h_{k_1ln}}\,;
\label{eq:hwf-center}
\end{equation}
for bulk materials, the analysis is carried out on high-symmetry
\gls{BZ} planes. 

A different physical perspective on the \gls{HWF} centers is provided by the Wilson loop, which is calculated over a closed curve $\mathcal{C}$ in $k$ space and discretized in $L$ points as
\begin{equation}
\label{eq:wilson}\mathcal{W}(\mathcal{C})= \prod_{i=0}^{L-1}P_{\mathbf{k}_i}^{\rm occ}
\end{equation}
and is a $J\times J$ matrix~\cite{rui_prb_2011} obtained from the product of ground-state projectors $P_{\mathbf{k}_i}^{\rm occ}
$. The Wilson-loop approach was first developed for TR-symmetric systems and later generalized to other topological phases~\cite{alex_prb_2014,alex_prl_2014,alex_prb_2016,david_prb_2014}.

The two approaches are essentially equivalent~\cite{Gresch2017}: the logarithm of the eigenvalues of the Wilson loop at a given $k$ point correspond to a gauge-invariant set of \glspl{HWF} centers, which coincide with those obtained from maximal localization~\cite{Soluyanov2011a}. Indeed, while the original implementation based on \glspl{HWF} enforced parallel transport by performing singular value decomposition on each overlap matrix along the line in $k$ space, in the Wilson loop formalism the full gauge-invariant loop $\mathcal{W}$ is diagonalized; the second approach has been found to converge a bit faster and allows studying the expectation value of the Wilson-loop eigenstates over symmetry operators~\cite{Gresch2017,z2pack_code}. 

This hybrid-Wannier (or ``Wilson-loop'') scheme is
implemented in \code{Z2Pack}, and a detailed description of the
methodology can be found in~\textcite{Gresch2017}.  In \code{Z2Pack},
the hybrid Wannier centers~\eqref{eq:hwf-center} are obtained from a
parallel-transport construction, starting from the overlap
matrices~\eqref{eq:Mkb}~\cite{Taherinejad2014}. The same procedure is
implemented in \code{PythTB} and \code{WannierTools} for Wannier/TB
Hamiltonians; the required overlaps are then taken between TB
eigenstates, as in Eq.~\eqref{eq:phi-int} above (\code{Z2Pack} can
also operate in this mode). We remark that
\code{Z2Pack}~\cite{Gresch2017} works with both Wannier/TB
Hamiltonians, and directly with first-principles engines such as
\code{Quantum ESPRESSO} and \code{VASP}.

In closing, it is
worth bearing in mind the different design philosophies of the three
packages surveyed in this section. As already mentioned in 
Sec.~\ref{subsec:tb}, \code{PythTB}~\cite{pythtb} was designed with \gls{TB} ``toy
models'' in mind, and to be used as a pedagogical tool;
it enables the computation of several geometric and topological quantities
(Berry phases and curvatures, Chern numbers, hybrid Wannier centers) as well as to generate ribbons and slabs to expose their boundary modes.
Although \code{PythTB} can also import large Wannier models (which can be
truncated internally), the code is not optimized for speed; however, a
high-performance Numba~\cite{numba} implementation of \code{PythTB} better suited for that
purpose is available~\cite{yeet-pythtb}. \code{WannierTools}~\cite{Wu2018}, on the other
hand, is primarily designed to work with large Wannier models, and is
parallelized using MPI. Its distinctive features include searching for
band degeneracies, and plotting surface spectral functions. Finally,
\code{Z2Pack} is focused on the \gls{HWF} scheme; it is not primarily a
``post-Wannier'' code, since it can circumvent the need to use a
Wannier engine by directly reading the \textit{ab initio} overlap
matrices, which can help streamline high-throughput calculations~\cite{MarrazzoNano2019,Grassano_2023}.

While topological invariants are generally introduced for crystalline
periodic systems in \glspl{PBC}, there are a few scenarios that
require either the use of open boundary conditions (OBCs) or the
adoption of large supercells with $\Gamma$-only sampling; in both
cases, standard approaches are of no avail. Relevant examples include
the study of Anderson~\cite{Li2009,Groth2009,Jiang2009} and amorphous
topological insulators~\cite{Corbae2023}, heterogeneous systems such
as trivial/topological junctions~\cite{Bianco2011}, molecular dynamics
simulations, and any other use case that does not fit a small
primitive cell with \gls{BZ} sampling. Among many other approaches
(see \textcite{Corbae2023} for a dedicated overview), single-point
sampling~\cite{Ceresoli2007,Favata2023} and local
markers~\cite{Bianco2011,bau_prb_2024,bau_arxiv_2024} have been introduced to study topology for
non-crystalline systems.

Thanks to the use of \glspl{WF} as a basis set, these techniques can
be implemented seamlessly in the same framework both for model
\gls{TB} (like the Haldane~\cite{Haldane1988} or the Kane--Mele
models~\cite{KM2005a,KM2005b}) and {\it ab initio} \gls{TB}, where the
latter is obtained by constructing \glspl{WF} on top of any electronic
structure calculations including \gls{DFT}, GW or \gls{DMFT} (as
discussed in Sec.~\ref{subsec:tb}). This strategy has been adopted by
the \code{StraWBerryPy} code~\cite{Favata2023,bau_prb_2024,bau_arxiv_2024,strawberrypy}, where
model or {\it ab initio} \gls{TB} Hamiltonians are read and
manipulated either through \code{PythTB} or \code{TBmodels}. This code can then calculate a few single-point and local topological invariants such
as the Chern number and the $\mathbb{Z}_2$ invariant, as well as other
quantum-geometrical quantities of the electronic structure that are relevant
for topological materials.

\subsection{\label{subsec:elph}Electron-phonon interactions}

\subsubsection{Methodology}

In the past decade, we have witnessed a community-wide effort to
develop advanced computational approaches and simulation tools for
atomistic modeling of function-defining properties of materials. A
primary focus of this ongoing research has been the accurate
description of \gls{e-ph} interactions from
first-principles~\cite{Giustino2017}, as they determine many materials
properties of technological interest such as electrical and thermal
transport~\cite{Li2015,Gunst2016,Ponce2018,Ponce2020,Ponce2021,Bernardi2018a,Bernardi2020b,Bernardi2021,Bernardi2021b,Brunin2020a,Brunin2020b,Protik2020a,Protik2020b,Liu2020,Chaves2020,Macheda2018,Macheda2022,Ponce2023a},
phonon-assisted light absorption~\cite{Noffsinger2012,Bushick2022,Kioupakis2022},
phonon-mediated
superconductivity~\cite{Calandra2010,Margine2013,Eiguren2020a,Errea2020,Roadmap2022,Tomassetti2024,Lucrezi2024,Mori2024}, polaron
formation~\cite{Sio2019,Lafuente2022,Falletta2022,Zhou2019,Lee2021,Sio2023}, and excitonic effects~\cite{Chen2021,Paleari2022,Antonius2022,Haber2023,Dai2024a,Dai2024b}. This list of references is by no means exhaustive, and it is only intended  to serve as a starting point for the respective topics. 

A comprehensive review of the theory of \gls{e-ph} interactions in solids from the point of view of {\it ab initio} calculations is given
by \textcite{Giustino2017}. An important contribution to the \gls{e-ph}
problem has been recently made by \textcite{Stefanucci2023} who have developed an {\it ab initio} many-body quantum theory of electrons and phonons in and out-of-equilibrium.

There is a well established formalism for computing the \gls{e-ph}
matrix elements from first-principles using
\gls{DFPT}~\cite{Baroni1987,Savrasov1992,Gonze1997,Baroni2001}. 
Furthermore, a linear-response approach employing the GW method, 
named \gls{GWPT}, has recently been proposed 
to improve the accuracy of the DFPT \gls{e-ph} matrix elements~\cite{Li2019,Li2024}. 
However, all aforementioned materials properties are notoriously difficult to evaluate with desired accuracy using \gls{DFPT} or \gls{GWPT} calculations directly due to the prohibitive computational cost. To achieve numerical convergence, the \gls{e-ph} matrix elements need to be computed on ultra-dense electron ($\bk$) and phonon ($\bq$) \gls{BZ} grids with $10^6-10^7$ points. 

Specialized numerical techniques based on Fourier interpolation of the perturbed potential~\cite{Eiguren2008,Gonze2020}, linear interpolation~\cite{Li2015}, Wannier interpolation~\cite{Giustino2007a,Giustino2007b,Calandra2010,Bernardi2018b} or atomic orbital interpolation~\cite{Gunst2016,Bernardi2018b} of the \gls{e-ph} matrix elements, and Fermi-surface harmonics representation of the \gls{e-ph} matrix elements~\cite{Allen1976,Eiguren2014,Eiguren2020b} have been
developed to address this convergence problem. In particular, the interpolation of the \gls{e-ph} matrix elements using \glspl{MLWF}~\cite{Marzari2012} introduced by \textcite{Giustino2007b} has
proven very successful for enabling highly accurate and efficient
calculations of \gls{e-ph} interactions, and the approach has been
implemented in a number of
codes~\cite{Noffsinger2010,EPW2016,EPW2023,PERTURBO2021,Phoebe2022,EPIq2023}. WFs have also been used in the context of downfolding methods to calculate the Coulomb interaction via the constrained random phase approximation (cRPA)~\cite{Aryasetiawan_et_al:2004} and the \gls{e-ph} interaction via constrained  density-functional perturbation theory (cDFPT)~\cite{Nomura_et_al:2015,Berge2023,Giovannetti2014}, respectively.  

We note that an alternative to the computation of the \gls{e-ph} interaction with DFPT is offered by the finite displacement scheme in real space~\cite{Gunst2016,Chaput2020,Engel2020,Engel2022}. While this approach requires large supercells to reach convergence and is adiabatic in nature~\cite{Chaput2020,Engel2020,Engel2022,Ponce2015}, 
it has the advantage of being universally applicable to any
functional, including hybrid or meta-GGA functionals, as well as
more complicated exchange-correlation potentials, where higher-order
derivatives of the functional are not readily available.

Below, we focus on the DFPT approach and outline the interpolation procedure 
to compute the \gls{e-ph} matrix elements on ultra-dense meshes using \glspl{WF}. 
One first determines the \gls{e-ph} matrix elements in the Bloch representation 
using the electronic states computed with DFT on a coarse $k$-point grid, and 
the deformation potentials computed with DFPT on a coarse $q$-point grid. 
The \gls{e-ph} matrix element is defined as:
\begin{equation} \label{eq:ephc}
\gmnvkqc = \braket{ \pmkqc|\pVc|\pnkc },  
\end{equation}
where $\pnkc$ and $\pmkqc$ are the \gls{KS} wavefunctions of the initial and 
final Bloch states (with $\bk$ being the electron wavevector, and $n$ being 
the band index), and $\pVc$ is the derivative of the self-consistent potential 
associated with a phonon with momentum $\bq$ and branch index $\nu$. 
The latter quantity can be obtained as:
\begin{equation} \label{eq:dV}
\pVc = \sum_{\k \a p} \left( \frac{\hbar}{2 M_\kappa \omega_{\mathbf{q}\nu}} \right)^{1/2} e^{i \bq \cdot \bRp} \frac{\D V^{\rm KS}}{\D \tau_{\k \a p}} e_{\k\a,\bq\nu},
\end{equation}
where $\bRp$ is the lattice vector identifying the unit cell $p$, $\tau_{\k \a p}$ 
is the position of atom $\k$ in unit cell $p$ in the Cartesian direction $\a$, 
$M_\k$ is the mass of  atom $\k$, $\w_{\bq \nu}$ is the phonon frequency, and 
$e_{\k\a,\bq\nu}$ is the eigendisplacement vector corresponding to atom $\k$ 
in the Cartesian direction $\a$ for a collective phonon mode $\bq\nu$. 
When choosing the electron and phonon grids in Eq.~\eqref{eq:ephc}, 
it is necessary that the $q$-point grid for phonons is commensurate with and 
smaller than (or equal to) the $k$-point grid for electrons in order to
map the wavefunctions $\pmkqc$ onto $\pmkGc$, where $\bk''$ is on the coarse $\bk$-grid 
and $\bG$ is a reciprocal lattice vector.

Next, one finds the \gls{e-ph} matrix elements in the Wannier representation:
\begin{eqnarray} \label{eq:ephw1}
g_{ij\k\a}(\bRe,\bRp) = \Braket{\boe i| \frac{\D V^{\rm KS}}{\D \tau_{\k \a p}} | \bRe j} ,
\end{eqnarray}
where $\bRe$ and $\bRp$ are the Bravais lattice vectors associated to
the electron and phonon \gls{WS} supercells, and $|\bRe j\rangle$ are the
\glspl{MLWF} with index $j$ and centered in the cell at $\bRe$. This is
done by transforming the \gls{e-ph} matrix elements from the coarse
\gls{BZ} $(\bk, \bq)$ grids into the corresponding
real-space supercells $(\bRe, \bRp)$ as:
\begin{eqnarray} \label{eq:ephw2}
g_{ij\k\a}(\bRe,\bRp) = \frac{1}{N_e N_p} \sum_{\bk,\bq}
e^{-i(\bk \cdot \bRe + \bq \cdot \bRp)}\nonumber\\ 
\times \sum_{m n \nu} \left( \frac{\hbar}{2 M_\kappa \omega_{\mathbf{q}\nu}} \right)^{1/2} V^{\dagger}_{\bk+\bq, i m} \gmnvkqcp V_{\bk, n j} e^{*}_{\k \a, \bq \nu}. \nonumber \\
\end{eqnarray}
In Eq.~(\ref{eq:ephw2}), $N_e$ and $N_p$ are the number of unit cells in the 
periodic \gls{BvK} supercells corresponding to the number of $k$ and 
$q$ points on the coarse electron and phonon grids, respectively, and 
$V_{\bk}$ is the Wannierization matrix introduced in 
Eq.~\eqref{eq:wannier-psi-dis}. That matrix is provided by the Wannier engine, 
while $e_{\k \a, \bq \nu}$ is obtained by diagonalizing the dynamical matrix 
at wavevector $\bq$. In the same spirit as the interpolation of the Hamiltonian discussed in Sec.~\ref{sec:band-inter}, a \gls{WS} construction can be used for the interpolation of the electron--phonon matrix elements~\cite{Ponce2021}. In this case, the construction is based on three quantities (two Wannier centers and one atomic position).

Finally, performing the inverse Fourier transform of Eq.~(\ref{eq:ephw2}), 
the \gls{e-ph} matrix elements on very fine $(\bk', \bq')$ \gls{BZ} 
grids are given by:
\begin{eqnarray} \label{eq:ephf}
\gmnvkqf  &=& \sum_{e p} e^{i(\bk' \cdot \bRe + \bq' \cdot \bRp)}
\sum_{i j \k \a} \left( \frac{\hbar}{2 M_\kappa \omega_{\bq' \nu}} \right) \nonumber\\ 
&\times& {\mathcal U}_{\bk'+\bq', m i} g_{i j \k\a}(\bRe,\bRp) 
{\mathcal U}^{\dagger}_{\bk', jn} e_{\k \a, \bq' \nu}. \nonumber \\
\end{eqnarray}
In this step, it is assumed that the \gls{e-ph} matrix elements outside of
the \gls{WS} supercells defined by the initial coarse grids can be
neglected. Prior to computing Eq.~(\ref{eq:ephf}), the transformation
matrices ${\mathcal U}_{\bk', n j}$ given by Eq.~\eqref{eq:eig-int} and the phonon eigenvectors
$e_{\k \a, \bq' \nu}$ for the new set of points $(\bk', \bq')$ must be
found as described in~\textcite{Giustino2007b}.

The accuracy of the Wannier--Fourier interpolation approach depends on
the spatial localization of the $g_{ij\k\a}(\bRe,\bRp)$ matrix
elements. Eq.~(\ref{eq:ephw1}) can be seen as a hopping integral between
two localized \glspl{WF}, one at $\boe$ and one at $\bRe$, due to a
perturbation caused by the displacement of the atom at $\bt_{\k
  p}$. If the \gls{e-ph} interactions are short-ranged in real space,
the quantity $g_{ij\k\a}(\bRe,\bRp)$ decays rapidly with $|\bRe|$ and
$|\bRp|$, and it is sufficient to only compute the matrix elements on
a small set of $(\bRe,\bRp)$ lattice vectors to fully capture the
coupling between electrons and phonons. As discussed 
in~\textcite{Giustino2007b,Giustino2017}, the spatial decay is bound by 
the limiting cases $g_{ij\k\a}(\bRe, \bop)$ and
$g_{ij\k\a}(\boe,\bRp)$.
In the first case, the matrix element decays in the variable $\bRe$ at
least as fast as the \glspl{MLWF}. In the second case, the matrix
element decays with the variable $\bRp$ at the same rate as the
screened Coulomb potential generated by the atomic
displacements. Thus, the localization of $g_{ij\k\a}(\bRe, \bRp)$
depends strongly on the dielectric properties of the system. In metals
and nonpolar semiconductors and semiconductors, the screening properties 
are dictated by Friedel oscillations~\cite{Fetter} $|\bRp|^{-4}$ and 
quadrupole behavior~\cite{Pick1970} $|\bRp|^{-3}$, respectively. In polar
materials (i.e., materials exhibiting nonzero Born effective charges), 
the dominant contribution to the potential is the dipole Fr\"{o}hlich
term~\cite{Vogl1976}, which is long-ranged and decays as
$|\bRp|^{-2}$. Therefore, in the case of semiconductors and insulators, 
the long-range electrostatic fields arising from the nonanalytic behavior
of the Coulomb potential in the long-wavelength limit
($\bq \rightarrow 0$), and the \gls{e-ph} matrix elements cannot be 
directly interpolated from a coarse to a fine grid using the Wannier-based 
interpolation approach. 

To address this problem, the \gls{e-ph} matrix elements are separated into 
short- ($\mathcal{S}$) and long- ($\mathcal{L}$) range contributions, as follows: 
\begin{equation} \label{eq:g_full}
    g_{mn\nu}(\mathbf{k},\mathbf{q}) = 
    g^{\mathcal{S}}_{mn\nu}(\mathbf{k},\mathbf{q}) +
    g^{\mathcal{L},D}_{mn\nu}(\mathbf{k},\mathbf{q})+
    g^{\mathcal{L},Q}_{mn\nu}(\mathbf{k},\mathbf{q}),
\end{equation}
where the terms on the right-hand side are the short-range, the dipole, and the quadrupole components. This strategy allows the short-range component to be treated  using the Wannier-Fourier interpolation approach described above, once the long-range components have been subtracted from the total matrix elements $g_{mn\nu}(\mathbf{k},\mathbf{q})$. A data-driven compression technique based on a singular value decomposition of the short-range e-ph matrix elements in the Wannier basis has been recently developed and shown to significantly accelerate e-ph calculations while preserving quantitative accuracy~\cite{Luo2024}.

\textcite{Verdi2015} and \textcite{Sjakste2015} derived the analytic expression for the dipole \gls{e-ph} matrix, which takes the form:
\begin{align} \label{eq:frohlich}
    g^{\mathcal{L},D}_{mn\nu}(\mathbf{k},\mathbf{q}) 
    &= i \, \frac{4\pi}{V_\text{cell}} \frac{e^{2}}{4\pi\epsilon_{0}}
    \sum_{\kappa} \left( \frac{\hbar}{2 M_\kappa \omega_{\mathbf{q}\nu}} \right)^{1/2}  \nonumber \\
    \times & \sum_{\mathbf{G}\neq -\mathbf{q}} e^{-i (\mathbf{q}+\mathbf{G}) \cdot \boldsymbol{\tau}_{\kappa}}
    \frac{(\mathbf{q}+\mathbf{G}) \cdot \mathbf{Z_{\kappa}^{*}} \cdot\mathbf{e}_{\kappa,\mathbf{q}\nu}}      
    {(\mathbf{q}+\mathbf{G}) \cdot \boldsymbol{\epsilon}^{\infty} \cdot (\mathbf{q}+\mathbf{G})} \nonumber \\
    \times & \sum_i V_{\mathbf{k+q+G}, mi}  V_{\mathbf{k}, i n}^{\dagger}.
\end{align}
This term is of the order $1/\mathbf{q}$ and diverges as $\mathbf{q}$ approaches the zone center. In Eq.~\eqref{eq:frohlich}, $\epsilon_{0}$ is the vacuum permittivity, $\boldsymbol{\epsilon}^{\infty}$ is the high-frequency dielectric tensor of the material, $\mathbf{G}$ is a reciprocal lattice vector, and $\mathbf{Z_{\kappa}^{*}}$ is the Born effective charge tensor of the atom $\kappa$. The unitary matrix $V_{\mathbf{k}}$ is the Wannierization matrix introduced in Eq.~\eqref{eq:wannier-psi-dis} and comes from the overlap integral between the KS wavefunctions $\langle \psi_{m\mathbf{k+q}} | e^{i \mathbf{q} \cdot \mathbf{r}} | \psi_{n\mathbf{k}} \rangle = \sum_i V_{\mathbf{k+q}, m i}  V_{\mathbf{k}, i n}^{\dagger}$ in the $\mathbf{q+G}\rightarrow 0$ limit.

The quadrupole contribution is of the order of $|\bq|^0$, and the corresponding $g^{\mathcal{L},Q}_{mn\nu}(\mathbf{k},\mathbf{q})$ expression was derived in~\textcite{Brunin2020a,Brunin2020b,Bernardi2020a,Bernardi2020b,Ponce2023b}. In the maximally localized Wannier gauge, $g^{\mathcal{L},Q}_{mn\nu}(\mathbf{k},\mathbf{q})$ can be written as:
\begin{align} \label{eq:quadrupole}
     g^{\mathcal{L},Q}_{mn\nu}(\mathbf{k},\mathbf{q}) 
    & =
    \frac{4\pi}{V_\text{cell}} \frac{e^{2}}{4\pi\epsilon_{0}}
    \sum_{\kappa\alpha} \left( \frac{\hbar}{2 M_\kappa \omega_{\mathbf{q}\nu}} \right)^{1/2}
    \nonumber \\
    \times & \sum_{\mathbf{G}\neq -\mathbf{q}} \frac{e^{-i (\mathbf{q}+\mathbf{G}) \cdot \boldsymbol{\tau}_{\kappa}}e_{\kappa\alpha,\mathbf{q}\nu} }{(\mathbf{q}+\mathbf{G}) \cdot \boldsymbol{\epsilon}^{\infty} \cdot (\mathbf{q}+\mathbf{G})} \nonumber \\
    \times & \sum_{\beta\gamma} \bigg[ \frac{1}{2}  Q_{\kappa,\alpha\beta\gamma} (q_\beta+G_\beta) (q_\gamma+G_\gamma)\, \nonumber \\
    \times & \sum_i V_{\mathbf{k+q+G}, m i}  V_{\mathbf{k}, i n}^{\dagger}  \nonumber \\
     - &  Z_{\kappa,\alpha\beta}^* (q_\beta+G_\beta)
    \sum_{i j} V_{\mathbf{k+q+G}, m i} (q_\gamma + G_\gamma)  \nonumber \\
    \times & \Big( A_{\mathbf{k},i j}^{W, \gamma}  
    + \langle u_{i \mathbf{k}}^{\text{W}} | V^{\text{Hxc},\mathcal{E}_\gamma}| 
    u_{j \mathbf{k}}^{\text{W}} \rangle  \Big)  V_{\mathbf{k}, j n}^{\dagger}
    \bigg].
\end{align}
where $\mathbf{Q}_{\kappa}$ is the dynamical quadrupole tensor that can be computed using DFPT~\cite{Royo2019},  $V^{\text{Hxc},\mathcal{E}_\gamma}(\mathbf{r})$ is the self-consistent potential induced by a uniform electric field $\mathcal{E}_\gamma$ along the Cartesian direction $\gamma$~\cite{Brunin2020a,Ponce2023b}, 
${\bf A}_{\mathbf{k}, i j}^{\text{W}}$ is the Berry connection introduced in Eq.~\eqref{eq:A-W}, and $u_{j\mathbf{k}}^{\text{W}}$ is the smooth cell-periodic part of the Bloch wavefunction in the Wannier gauge ($\ket{u_{j\mathbf{k}}^{\text{W}}}=\sum_n V_{\mathbf{k}, j n} \ket{u_{n\mathbf{k}}} $)~\cite{Ponce2023b}. 

Equations~\eqref{eq:frohlich} and \eqref{eq:quadrupole} are for the long-range dipole and quadrupole components of the \gls{e-ph} matrix elements in 3D bulk crystals. During the past few years, several formalisms have been proposed to treat the long-range contributions in 2D materials~\cite{Sohier2016, Deng2021,Zhang2022,Sio2022,Ponce2023a,Ponce2023b,Sio2023}. For example, a unified description of polar \gls{e-ph} interactions that allows a smooth transition from 3D to 2D was developed by \textcite{Sio2022}. Their formalism reduces to the 3D approach of \textcite{Verdi2015,Sjakste2015}, and to the 2D approach of \textcite{Sohier2016,Deng2021}. Another strategy was followed by \textcite{Zhang2022} and \textcite{Ponce2023a,Ponce2023b} who built on the general formalism for treating long-range electrostatic interactions in 2D crystals developed by \textcite{Royo2021}. Importantly, \textcite{Ponce2023a,Ponce2023b} showed that the long-range \gls{e-ph} matrix elements have a spurious dependence on the Wannier gauge that can be eliminated by including the contribution associated with the Berry connection in Eq.~\eqref{eq:quadrupole}. Therefore, in order to restore the gauge covariance in the long-wavelength limit any beyond-Fr\"{o}hlich Wannier approach should incorporate this term. The contribution from $V^{\text{Hxc},\boldsymbol{\mathcal{E}}}$ term, on the other hand, was found to represent less that 0.1\% of the total quadrupole correction~\cite{Brunin2020a,Ponce2023b}.

\subsubsection{Codes}

\begin{figure*}[t]
	\centering
	\includegraphics[width=0.99\textwidth]{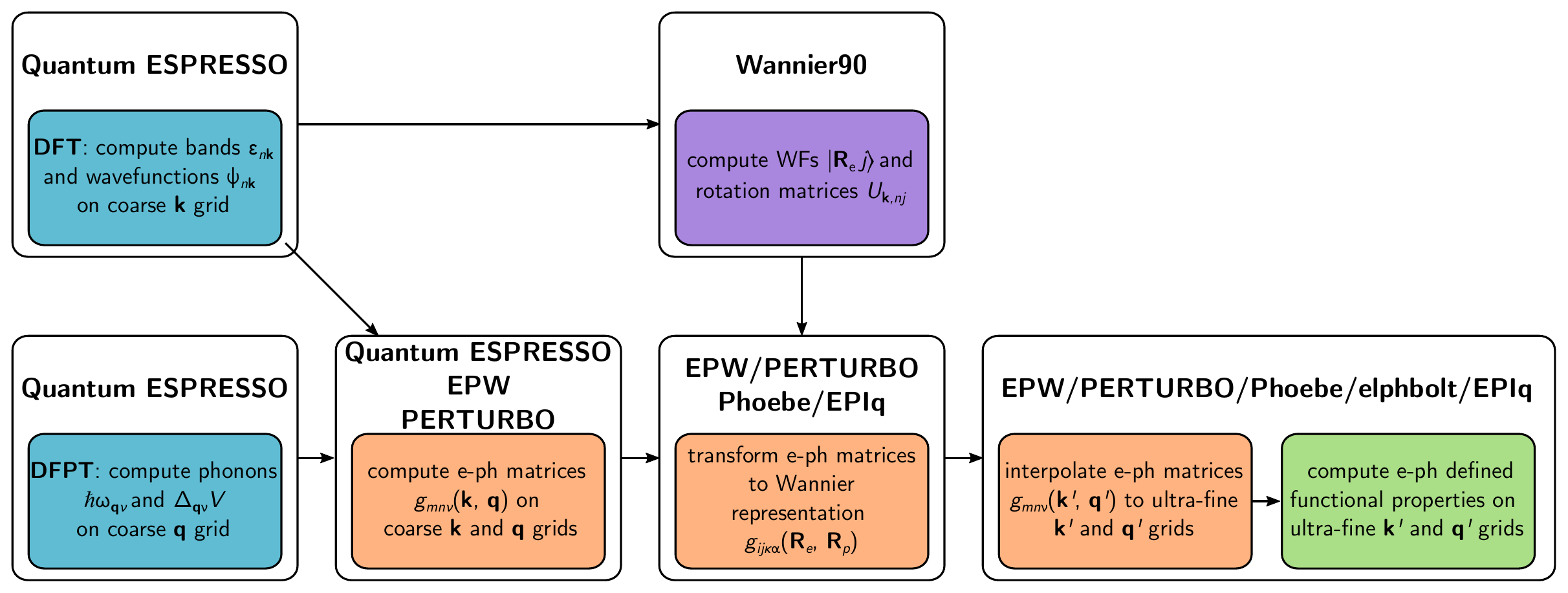}
	\caption{Workflow to compute \gls{e-ph} interactions on ultra-fine grids using \glspl{WF}. The user performs \gls{DFT} and \gls{DFPT} calculations on coarse $k$ and $q$-point grids with \code{Quantum ESPRESSO}. Next, the user computes the \gls{e-ph} matrix elements on the coarse grids and transforms them in localized Wannier basis using the rotation matrices obtained with \code{Wannier90}. Note that \code{Phoebe} and \code{EPIq} read the \gls{e-ph} matrix elements on the coarse grids computed with \code{Quantum ESPRESSO}, while \code{EPW} and \code{PERTURBO} compute them internally. Finally, \code{EPW}/\code{PERTURBO}/\code{Phoebe}/\code{EPIq} interpolate the band structure, phonon dispersion, and \gls{e-ph} matrix elements on ultra-fine $k'$- and $q'$-point grids, and perform calculations of various materials properties.} 
	\label{fig-eph}
\end{figure*}

\code{EPW}~\cite{Noffsinger2010,EPW2016,EPW2023,EPW-web}, the first
open-source code for the study of \gls{e-ph} interaction using
\glspl{MLWF}, was publicly released in 2010 and has been distributed
within the \code{Quantum ESPRESSO} suite~\cite{Giannozzi2017,QEF}
since 2016. Several Wannier-based open-source codes exist today to
compute physical properties related to e-ph interactions such as
\code{Perturbo}~\cite{PERTURBO2021,PERTURBO-web},
\code{Phoebe}~\cite{Phoebe2022,Phoebe-web},
\code{elphbolt}~\cite{Elphbolt2022, Elphbolt-web}, and
\code{EPIq}~\cite{EPIq2023, EPIq-web}. At present, \code{EPW},
\code{Perturbo}, \code{Phoebe}, and \code{EPIq} are all interfaced
with \code{Quantum ESPRESSO}~\cite{Giannozzi2017} to generate the
relevant first-principles input data, and use
\code{Wannier90}~\cite{Pizzi2020} in standalone or library mode to
compute the required quantities in the Wannier
representation. \code{elphbolt}, on the other hand, relies on \code{EPW}
to generate the required Wannier space information. All codes follow
an overall similar workflow to compute e-ph matrix elements on fine
grids, as outlined below and summarized in Fig.~\ref{fig-eph}.

1. The initial step is to perform \gls{DFT} calculations with \code{Quantum ESPRESSO} on a uniform coarse electronic $\bk$ grid to obtain the band energies and Bloch wavefunctions. In addition, \gls{DFPT} calculations with \code{Quantum ESPRESSO} are carried out on an irreducible coarse $\bq$ grid to obtain the dynamical matrices and the derivatives of the self-consistent potential with respect to the phonon perturbations. \code{Phoebe} and \code{EPIq} require that the \gls{e-ph} matrix elements on the coarse electron and phonon grids are also computed with \code{Quantum ESPRESSO} since these quantities are later passed to the two codes. \code{EPW} and \code{PERTURBO}, on the other hand, compute the \gls{e-ph} matrix elements on the coarse $\bk$ and $\bq$ grids internally by reading the files generated from the \gls{DFT} and \gls{DFPT} calculations.

2. Next, one must perform a precise Wannierization of the system using
\code{Wannier90} in standalone mode with \code{PERTURBO},
\code{Phoebe}, and \code{EPIq}, or in library mode with \code{EPW}.
This step produces the \glspl{MLWF} and the
Wannierization  matrices $V_\kk$ that transform the \gls{DFT} Bloch wavefunctions
  into \glspl{MLWF}.

3. The next step is to compute the \gls{e-ph} matrix elements on the coarse
$\bk$ and $\bq$ grids in the Bloch representation and transform them,
along with the electronic Hamiltonian and the dynamical matrix, from
the Bloch to the Wannier representation. As noticed at point 2,
\code{Phoebe} and \code{EPIq} use the \gls{e-ph} matrix elements on the
coarse electron and phonon grids computed directly with
\code{Quantum ESPRESSO}.

4. The final step is to perform an inverse Fourier transform of the electronic Hamiltonian, dynamical matrix, and \gls{e-ph} matrix elements from the Wannier to the Bloch representation. At this stage, the electronic eigenvalues, phonon frequencies, and \gls{e-ph} matrix elements can be efficiently computed on ultra-dense $\bk'$ and $\bq'$ grids and further used to carry out calculations of various materials properties.

To extend the Wannier-based interpolation scheme to systems with long-range \gls{e-ph} contributions, the strategy is as follows. First, the total matrix elements $g_{mn\nu}(\mathbf{k},\mathbf{q})$ are calculated on the coarse $\bk$ and $\bq$ grids.  Second, the long-range contributions $g^{\mathcal{L},D}_{mn\nu}(\mathbf{k},\mathbf{q})$ and $g^{\mathcal{L},Q}_{mn\nu}(\mathbf{k},\mathbf{q})$ are evaluated on the same coarse grids using Eqs.~\eqref{eq:frohlich} and \eqref{eq:quadrupole} and subtracted from $g_{mn\nu}(\mathbf{k},\mathbf{q})$, leaving out the short-range component $g^{\mathcal{S}}_{mn\nu}(\mathbf{k},\mathbf{q})$. Third, the short-range \gls{e-ph} matrix elements are interpolated to ultra-dense $\bk'$ and $\bq'$ grids using the standard approach based on \glspl{MLWF}. Fourth, the long-range contributions are computed using Eqs.~\eqref{eq:frohlich} and \eqref{eq:quadrupole} on the fine $\bk'$ and $\bq'$ grids and added back to the short-range component to recover the total $g_{mn\nu}(\bk',\bq')$ matrix elements.

A wide range of properties can be currently computed with \code{EPW},
\code{Perturbo}, \code{Phoebe}, \code{elphbolt}, and \code{EPIq}. A
full list of the capabilities for the released version of each code
can be found on the respective website. In particular, \code{EPW} computes 
charge carrier mobility under electric and magnetic fields
using \gls{BTE}, phonon-mediated superconductivity using the anisotropic 
Migdal-Eliashberg formalism, phonon-assisted direct and indirect optical absorption using quasi-degenerate perturbation theory, 
small and large polarons without supercells, and zero-point renormalization 
and temperature dependence of band structures using \gls{WFPT}~\cite{EPW2023,EPW-web}. \code{EPW} also comes with the \code{ZG} toolkit for calculations of
band structure renormalization, temperature-dependent optical spectra, 
temperature-dependent anharmonic phonon dispersions, and anharmonic  
\gls{e-ph} couplings via the special displacement
method~\cite{ZG2016,ZG2020,ZG2023a,ZG2023b}.  \code{Perturbo}
calculates phonon-limited transport properties using \gls{BTE},
ultrafast carrier dynamics, magnetotransport, and high-field
electron transport~\cite{PERTURBO2021,PERTURBO-web}. \code{Perturbo} is
interfaced with the \code{TDEP} package~\cite{TDEP-web} that 
computes temperature-dependent anharmonic phonons. 
\code{Phoebe} provides various tools to predict electron 
and phonon transport properties at different levels of theory 
and accuracy, including full scattering matrix \gls{BTE} solutions 
such as the relaxons method and models based on the Wigner 
distribution~\cite{Phoebe2022,Phoebe-web}. \code{elphbolt} solves the
coupled electron and phonon \glspl{BTE}, and the effect of the mutual
e-ph drag on the electrical and thermal transport
coefficients~\cite{Elphbolt2022,Elphbolt-web}. \code{EPIq} computes 
phonon-mediated superconducting properties based on the 
Migdal-Eliashberg formalism, adiabatic and non-adiabatic 
phonon frequencies, double-resonant Raman intensities, and excited carriers 
lifetimes~\cite{EPIq2023,EPIq-web}. \code{EPIq} is interfaced 
with the stochastic self-consistent harmonic approximation 
(SSCHA)~\cite{SSCHA-web,Errea2013,Errea2014,Monacelli2021} 
in order to calculate \gls{e-ph} interactions in the 
presence of strong quantum anharmonicity.

\subsection{\label{subsec:dmft}Beyond DFT with localized orbitals}

While in most cases \gls{DFT} is the method of choice for electronic ground
state calculations, certain excited state properties, and even the ground state properties of certain materials, 
may require a ``beyond-\gls{DFT}'' treatment for accurate first-principles
predictions. Two examples are finite-temperature and spectroscopic properties, 
as observed in direct and inverse photoemission experiments, which cannot be 
addressed adequately within conventional \gls{DFT}.
Similarly, the complex physics arising from strong local Coulomb
interaction in partially filled orbitals is beyond the scope of a
single-particle picture, which can manifest itself in an inaccurate
description of the material. In such cases, more advanced methods are
needed. One class of approaches is based on diagrammatic many-body perturbation theory; examples include the GW approximation or \gls{DMFT} (see Sec.~\ref{subsubsec:DMFT}).
Since such methods are often
computationally costly and complex, it may be necessary to extract
accurate low-energy effective Hamiltonians that are treated using
these methods in a post-processing step.  An efficient alternative is
to retain the functional character of \gls{DFT} and apply physically
motivated corrections, as in hybrid or Koopmans-compliant functionals
(see Sec.~\ref{subsubsec:KF}). In this case the theory is no longer based on a pure functional of the density, but the orbitals
themselves or their orbital densities become the key variables.
Common to both approaches is the importance of improving the
description of local, orbital-dependent physics.  This is where
\glspl{WF} come in, providing a useful basis for such applications and
supporting the physical understanding with chemical intuition.

\subsubsection{\label{subsubsec:DMFT}Dynamical mean-field theory and embedding}
Strongly correlated electron systems host a wide variety of physical
phenomena, ranging from Mott physics to high-temperature
superconductivity to exotic ordered
phases~\cite{Tokura/Kawasaki/Nagaosa:2017}.  Fundamental to these
phenomena is the competition of itinerant versus
localized character of electrons, which requires computational methods
beyond the single-particle picture. While the representation of electronic 
states in reciprocal space can be beneficial, the theory of strong local correlations
is most naturally formulated in a real-space basis.  A key aspect of many
beyond-\gls{DFT} schemes is therefore the combination of itinerant
Bloch states with localized molecular
orbitals, which can be elegantly formalized with
\glspl{WF}~\cite{Lechermann_et_al:2006}.

Beyond-\gls{DFT} schemes are typically computationally demanding and
require a compromise in terms of the number of orbitals that can be
treated.  Starting from an \emph{ab initio} description of a large number of bands, it is common
practice to select a subset of correlated orbitals, such as those
describing electronic states in the vicinity of the Fermi level, using
projector functions.  The result is a so-called downfolded model,
which contains only the relevant degrees of freedom~\cite{Kotliar_et_al:2006}.
It divides the total system into a subspace of localized orbitals, for
which a higher-level method is used, and the
remaining \gls{KS} states, for which the single-particle description
within \gls{DFT} accurately reflects the physics.  The embedding
\emph{ansatz} is further justified by the observation that correlated
physics phenomena typically occur on energy scales of meV to a few
eV~\cite{Chen_et_al:2022}. For the remainder of this subsection we focus 
on \gls{DMFT} as the higher-level method to solve the downfolded many-body problem,
but the concepts can be similarly applied to its extensions (see below) and conceptual other
approaches (see, e.g.,~\textcite{Zgid/Gull:2017,Eskridge_et_al:2019,Sheng_et_al:2022,Muechler_et_al:2022}).

While the multiband Hubbard model studied with \gls{DMFT} is naturally
related to \gls{TB} models, a rigorous
formalism based on energy functionals allows the combined approach
with \emph{ab initio} methods, as in \gls{DFT}+\gls{DMFT}~\cite{Kotliar_et_al:2006}.
The development of \gls{DFT}+\gls{DMFT}
methods and software for strongly correlated materials has seen a
tremendous surge in the last few decades~\cite{Held:2007,Pavarini_et_al:2011,Paul/Birol:2019}.  Routine calculations allow
to compute single-particle spectra~\cite{Pavarini_et_al:2004,Anisimov_et_al:2005,Nekrasov_et_al:2006},
optical conductivity~\cite{Haule_et_al:2005,Wissgott_et_al:2012}, transport~\cite{Oudovenko_et_al:2006,Zingl_et_al:2019}
and thermoelectric properties~\cite{Arita_et_al:2008,Wissgott_et_al:2010,Tomczak/Haule/Kotliar:2012}, electronic Raman~\cite{blesio_prr_2024},
and two-particle correlation
functions (susceptibilities)~\cite{Kunes:2011,Boehnke_et_al:2011,Park/Haule/Kotliar:2011}.
Recent advances include interactions with core holes as in x-ray absorption and
photoemission spectroscopy, and
resonant inelastic x-ray scattering~\cite{Haverkort_et_al:2012,Lueder_et_al:2017,Hariki/Uozumi/Kunes:2017,
  Hariki/Winder/Kunes:2018}.  Furthermore, routines for lattice
optimization~\cite{Leonov/Anisimov/Vollhardt:2014,Haule/Pascut:2016,Plekhanov/Bonini/Weber:2021}
and for computing phonon spectra~\cite{Kocer_et_al:2020} have been
formalized. This list of references is by no means exhaustive, but is intended to serve as a
starting point for the respective topics.

In beyond-\gls{DFT} methods, it is convenient to describe the total system of interacting electrons in a periodic solid in terms of the momentum- and frequency-dependent retarded single-particle Green's function 
\begin{align}\label{eq:lattice_gf}
    \hat G(\ke,\omega) =& \left[ (\omega+\mu)\mathds{1}-\hat H(\ke)-\hat\Sigma(\ke,\omega) + i\eta \right]^{-1} \,,
\end{align}
where $\mu$ is the chemical potential and $\hat H(\ke)$ represents the
non-interacting Hamiltonian.
The frequency- and momentum-dependent electron self-energy is given by
$\hat \Sigma(\ke,\omega)$, and $\eta$ is an infinitesimal positive parameter to ensure physical correctness. For clarity, we have omitted the
double-counting correction here, and we refer to~\textcite{Karolak_et_al:2010} for an overview.  Starting from a
\gls{DFT}-derived downfolded Hamiltonian, the challenge is to compute
the corresponding self-energy correction that accounts for dynamical
interaction effects.  Various approaches can be formalized, but for
the purpose of this review we outline the workflow of single-site
\gls{DMFT}~\cite{Georges_et_al:1996}.  In \gls{DMFT}, the self-energy
becomes a site-local quantity for a given atomic site $\mathcal{R}$
within the unit cell when expressed in a localized orbital basis.
This approximation is conceptually similar to
\gls{DFT}+$U$~\cite{Anisimov/Aryasetiawan/Lichtenstein:1997}, but in
\gls{DMFT} the full frequency dependence of the interaction is taken
into account.  Following this approach, in the \gls{DMFT}
self-consistency cycle, the local lattice self-energy
$\Sigma^{\mathcal{R}}(\omega)$
is approximated by that of an
auxiliary quantum impurity problem.  The most computationally
challenging step of the \gls{DMFT} loop is typically to find the
solution to the impurity problem, which allows a user to infer the impurity
self-energy via the Dyson equation.  The self-energy is embedded into
the Hilbert space of the effective Hamiltonian as
\begin{align}\label{eq:upfolding}
    \Sigma_{mn}(\ke,\omega) = \sum_{\mathcal{R},{ij}} P_{m i}^{\mathcal{R}*}(\ke) \Sigma^{\mathcal{R}}_{ij}(\omega) P^{\mathcal{R}}_{jn}(\ke)\,.
\end{align}
Approximating the lattice self-energy in  Eq.~\eqref{eq:lattice_gf} by the upfolded impurity self-energy becomes exact for infinite connectivity of the lattice~\cite{Metzner/Vollhardt:1989,Georges/Kotliar:1992}.
The projector functions $P^{\mathcal{R}}_{jn}(\ke)$ in \eqref{eq:upfolding} encode the basis transformation from band to orbital basis, i.e., from $\ket{\psi_{n\kk}}$, with band index $n$ and wavevector $\ke{}$, to $\ket{\psi^\text{W}_{\mathcal{R}_j{\bf k}}}$, with orbital index $j$ at site $\mathcal{R}$ (i.e., $j$ is an intrasite index here):
\begin{equation}\label{eq:projector}
    P^{\mathcal{R}}_{jn}(\ke) = \braket{\psi^\text{W}_{\mathcal{R}_j{\bf k}}|\psi_{n\kk}}\,.
\end{equation}
The local Green's function is then computed as
$G^{\text{loc},\mathcal{R}}_{ij}(\omega) =
\frac{1}{N}\sum_{\ke{},mn} P_{im}^{\mathcal{R}}(\ke)
G_{mn}(\ke,\omega) P_{n j}^{\mathcal{R}*}(\ke)$, where $N$ is the total number of $k$ points of the grid.
To determine a
suitable localized basis set, some \gls{DFT}+\gls{DMFT} codes use
projections on atomic orbitals, others rely on \code{Wannier90}
directly for a simple and user-friendly interface.
While the two approaches are conceptually similar (for a more detailed
overview see~\textcite{Chen_et_al:2022}), the choice of projectors
may affect the results and therefore needs to be carefully
analyzed~\cite{Karp/Hampel/Millis:2021}.
Note that Eq.~\eqref{eq:projector} assumes that the DFT+DMFT calculation is performed in the band basis in a charge self-consistent mode (i.e., $P(\kk)$ corresponds to $V_{\kk}$ in Eq.~\eqref{eq:psi-smooth}).
However, for one-shot calculations, the equations simplify~\cite{Beck_et_al:2022}.
Wannier interpolation can be used in the DMFT self-consistent loop for an isolated set of bands or at the TB level, which is crucial for accurately resolving low-energy physics~\cite{Kaye2023}.

Multiple schemes go beyond standard \gls{DMFT}, but the discussion above carries over directly to these extensions.
Examples include cluster-\gls{DMFT} approaches (either in real or reciprocal space)~\cite{Kotliar_et_al:2006} or diagrammatic extensions of the self-energy~\cite{Rohringer_et_al:2018}, as well as non-equilibrium \gls{DMFT}~\cite{Aoki_et_al:2014}.
To improve some of the shortcomings of the \gls{DFT}+\gls{DMFT} method, a better starting point than \gls{DFT} may be GW. The combination of GW with \gls{DMFT} provides a route to include non-local effects beyond \gls{DFT} as well as to formalize the double-counting correction term~\cite{Biermann:2014}.

\paragraph{DFT+DMFT codes}
From the previous section, it becomes clear that a fully integrated \gls{DFT}+\gls{DMFT} software suite requires three main components: 1) a \gls{DFT} implementation (\emph{ab initio} engine) and a routine to construct the localized basis set (e.g., a Wannier engine), 2) a Green's function formalism to implement the \gls{DMFT} equations, and 3) an impurity solver.
At present, there are several open-source implementations that meet these requirements to varying degrees.
On the side of more monolithic, publicly available beyond-\gls{DFT} codes, there are implementations in, for example, \code{CASTEP}~\cite{Plekhanov_et_al:2018}, \code{Abinit}~\cite{Romero_et_al:2020}, \code{RSPt}~\cite{Grechnev_et_al:2007,DiMarco_et_al:2009,Thunstrom_et_al:2009}, \code{Amulet}~\cite{amulet}, and \code{eDMFT}~\cite{Haule:2010}, which include an implementation of \gls{DFT} and a downfolding routine, as well as choices of internal and externally linked impurity solvers.
\code{ComDMFT}~\cite{Choi_et_al:2019}, on the other hand, interfaces directly with \code{Wannier90} for the downfolding procedure.
All codes support charge self-consistency.

An alternative philosophy is a more modular library approach, focusing on providing the framework for performing \gls{DMFT} calculations based on input from a \gls{DFT} calculation.
For this purpose, most codes directly rely on \code{Wannier90} to benefit from a generic, robust, and flexible interface independent of the flavor of \gls{DFT}.
Examples include \code{w2dynamics}~\cite{Wallerberger_et_al:2019} and \code{DCORE}~\cite{Shinaoka_et_al:2021}, as well as \code{EDIpack}~\cite{Amaricci_et_al:2022}, \code{DMFTwDFT}~\cite{Singh_et_al:2021}, and \code{TRIQS}~\cite{Parcollet_et_al:2015,Aichhorn_et_al:2016,Merkel_et_al:2022}, with only the latter two packages supporting charge self-consistency at present~\cite{Singh_et_al:2021,Schueler_et_al:2018,James_et_al:2021,Beck_et_al:2022}.
All packages contain a number of internal and external impurity solvers.
A list of all currently available impurity solvers is beyond the scope of this article, but can be found on the respective websites.

\paragraph{Interaction-parameters codes}

As in \gls{DFT}+$U$, the local interaction parameters that enter the Hubbard model in \gls{DFT}+\gls{DMFT} must be chosen appropriately.
Starting from a local basis set, the Coulomb integrals can be evaluated, ideally taking into account screening processes and the symmetries of the system to simplify the parametrization of the interaction Hamiltonian~\cite{Chen_et_al:2022}.
The most widely used approach is the constrained random phase approximation (cRPA)~\cite{Aryasetiawan_et_al:2004}.
Currently, the method is implemented in \code{ABINIT}~\cite{Amadon_et_al:2014}, \code{SPEX}~\cite{Friedrich/Blugel/Schindlmayr:2010}, \code{VASP}~\cite{Merzuk:2015}, and \code{RESPACK}~\cite{Nakamura_et_al:2021}.
The latter three offer the possibility to use \glspl{WF}, which are constructed either using an internal Wannier engine or via an interface to \code{Wannier90} (\code{wan2respack}~\cite{Kurita_et_al:2023} in the case of \code{RESPACK}).
The usage of \glspl{WF} in cRPA has several benefits, e.g., a simplified interpretation of resulting interactions in terms of orbitals, a more compact representation, and the possibility to utilize Wannier interpolation on the Coulomb kernel (see, e.g., ~\cite{Roesner_et_al:2015}).
cRPA calculates the Coulomb interaction as a frequency-dependent response function that can be treated in extended \gls{DMFT}~\cite{Werner/Casula:2016}, and which can also be combined with the frequency-dependent \gls{e-ph} interaction~\cite{Nomura_et_al:2015}.
Alternatively, the interaction parameters are treated as free parameters that must be adjusted to match experimental observables.

\paragraph{Post-processing}

\begin{figure}
  \centering
  \includegraphics[width=8cm]{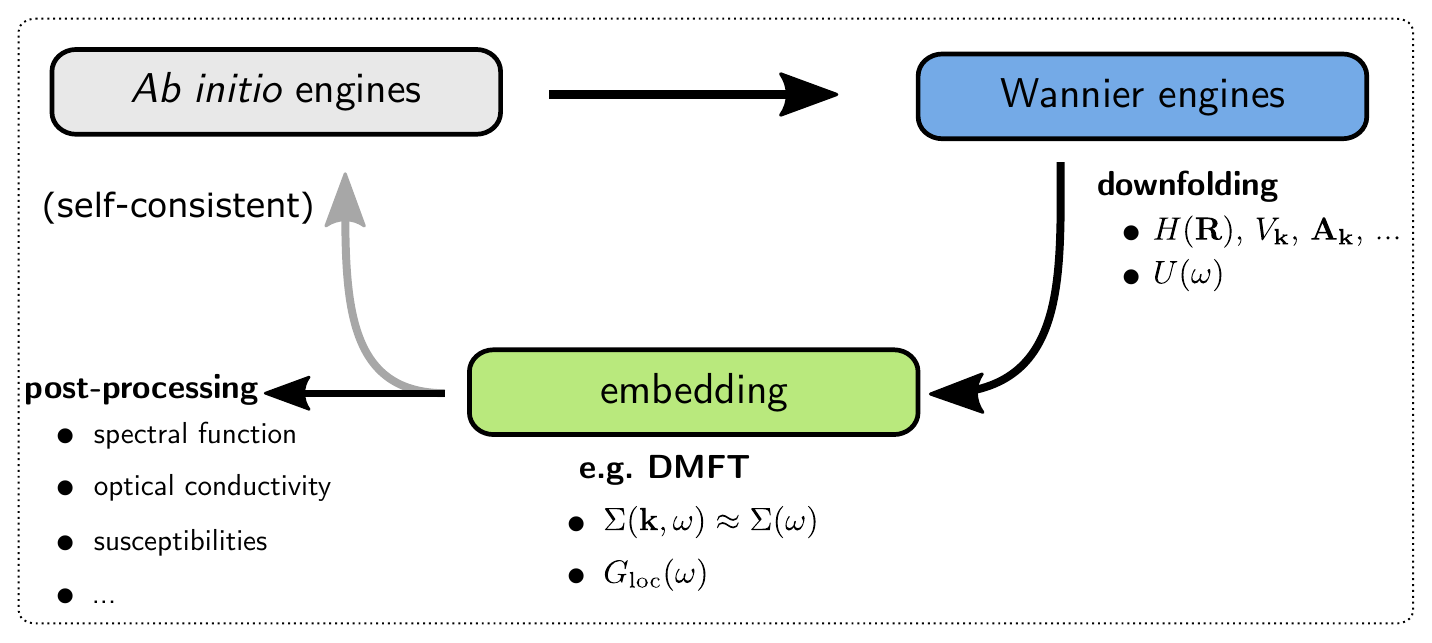}
  \caption{Typical workflow of the embedding formalism.  First, the
    user performs an \emph{ab initio} calculation from which a downfolded
    model is derived in the basis of localized orbitals.  The
    downfolded model is solved using an appropriate many-body method
    such as \gls{DMFT} (depicted). Physical observables can be computed in a
    post-processing step.  For full self-consistency the cycle is
    iterated until convergence.}
  \label{fig:dftdmft}
\end{figure}

Once the solution to the \gls{DFT}+\gls{DMFT} scheme is found (charge)
self-consistently (see Fig.~\ref{fig:dftdmft}), converged \gls{DMFT}
quantities such as the self-energy, the local Green's function, or the
hybridization function allow computing several physical observables
in post-processing applications.  Depending on the frequency domain in
which the impurity solver operates, it may be necessary to use
analytic continuation to obtain the self-energy in the real frequency
domain~\cite{Gubernatis_et_al:1991,Wang_et_al:2009}.  Such programs
are often included in the respective software packages.  Since the
post-processing step is performed only once after convergence, and
therefore does not significantly contribute to the overall
computational cost, Wannier interpolation (see Sec.~\ref{subsec:tb})
is particularly beneficial at this stage.  A standard observable is
the lattice or impurity spectral function, which is directly related
to photoemission and absorption spectra.  Most software packages provide tools for users to compute
such quantities routinely.  Transport tensors based on Kubo's linear
response theory~\cite{Kubo:1957}, such as optical and thermal conductivities, as well as Hall and Seebeck coefficients, can be computed using various tools, including \code{TRIQS}/\code{DFTTools}~\cite{Aichhorn_et_al:2016} and the
\code{woptic} package~\cite{Assmann_et_al:2016}.
Another option is \code{LinReTraCe}~\cite{Pickem/Maggio/Tomczak:2022}, which relies on a semi-analytical approach valid when a linear expansion of the self-energy is adequate.
Tools to evaluate core level spectroscopies within different levels of approximations
are available, for example, in \code{Quanty}~\cite{Haverkort:2016} and \code{EDRIXS}~\cite{Wang_et_al:2019}.

\subsubsection{\label{subsubsec:KF}Koopmans functionals} \gls{KC}
functionals~\cite{Dabo2010,Borghi2014,Colonna2018,Colonna2019,Linh2018,Elliott2019,DeGennaro2022,Colonna2022,linscott_koopmans_2023,Marrazzo_arxiv_2024}
are orbital-dependent functionals capable of delivering accurate
spectral properties for molecular and extended systems at low
computational cost. Remarkably, the \gls{KC} approach maintains a
simple functional formulation while being more accurate than $G_0W_0$
and comparable to
\gls{QSGtildeW}~\cite{Linh2018,Colonna2019,Colonna2022,Marrazzo_arxiv_2024}, at a cost
which is broadly comparable to standard \gls{DFT}. The simplicity and
accuracy of the \gls{KC} framework rests on three fundamental
concepts: linearization, screening, and localization. First, a
generalized linearization condition is imposed on each charged
excitation: the energy of any orbital must be independent of the
occupation of the orbital itself. This implies that the \gls{KC} total
energy functional is piecewise linear with respect to fractional
occupations, and essentially implements a generalized definition of
self-interaction-free orbital.  Second, electronic screening and
orbital relaxation (due to the electron addition/removal process) are
taken into account by orbital-dependent screening coefficients, which
can be calculated by finite differences~\cite{Linh2018} or
linear-response approaches~\cite{Colonna2018,Colonna2022}. Finally,
the Koopmans compliance condition is imposed on those
variational orbitals---i.e., those minimizing the \gls{KC} energy
functional---which are localized. For periodic systems, these
variational orbitals resemble
\glspl{MLWF}~\cite{Linh2018,Colonna2018,Colonna2019,Colonna2022}.

Using \glspl{WF} as a proxy for variational orbitals has allowed the
development of a Wannier-interpolation and unfolding scheme to
calculate the band structure from a supercell Koopmans-functional
calculation~\cite{DeGennaro2022}. In addition, \glspl{WF} have
fostered the development of a convenient Koopmans formulation that
operates fully under \glspl{PBC}, and is based on explicit \gls{BZ} sampling and \gls{DFPT}~\cite{Colonna2022,linscott_koopmans_2023}. This \emph{Koopmans-Wannier} implementation, which goes under the name of
\code{KCW}, is available in the
\code{Quantum ESPRESSO} distribution; it delivers improved scaling
with system size, and makes band-structure calculations with Koopmans
functionals straightforward~\cite{Colonna2022,linscott_koopmans_2023,Marrazzo_arxiv_2024}. \gls{KC} functionals resonate with other
efforts aimed at calculating excitation energies where \glspl{WF} and
localized orbitals are often a key
ingredient~\cite{Anisimov2005,Anisimov2007,Kronik2007,Galli2014,Weitao2015,Weitao2017,Kronik2021,Ma2016}.

\subsection{\label{sec:interoperability-in-ecosystem}Interoperability between codes in the ecosystem}
\subsubsection{Library mode for the Wannier engines}

\begin{figure}[tb]
\centering\includegraphics[width=8cm]{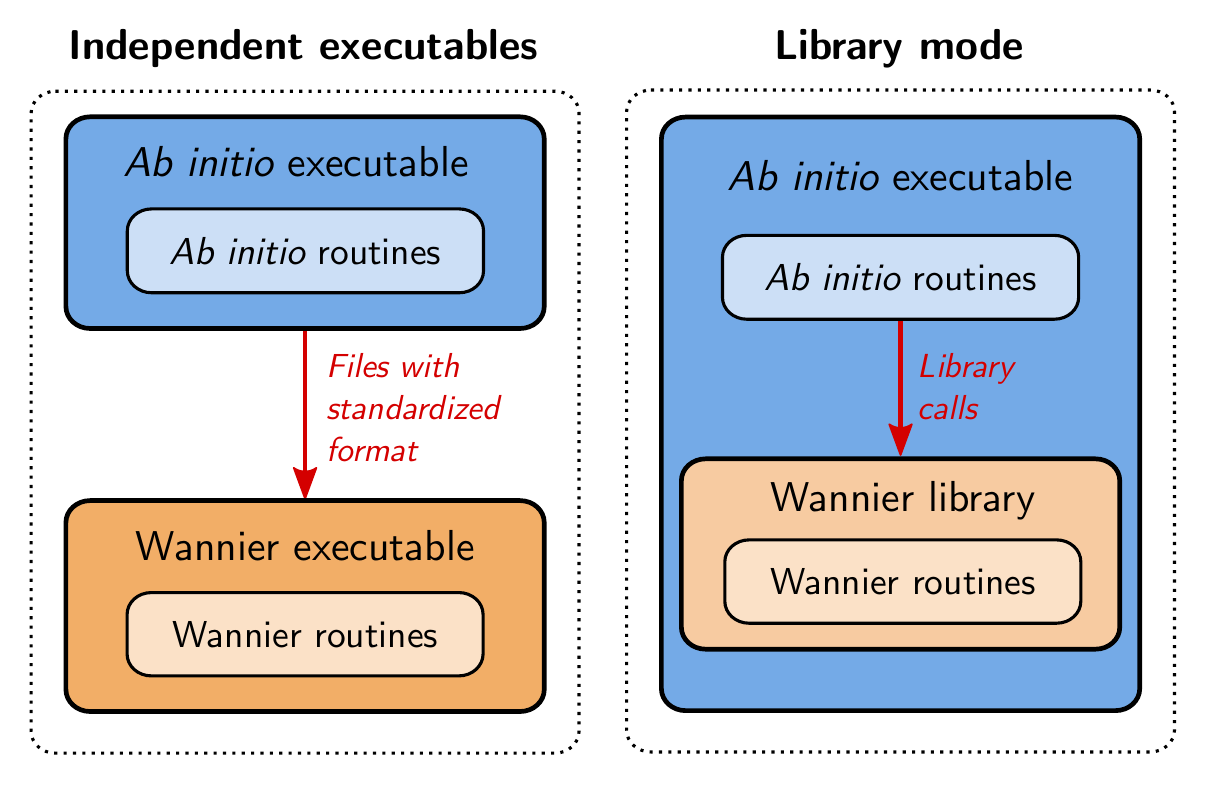}
\caption{\label{fig:library-vs-standalone}Different approaches for
  interaction of the \emph{ab initio} codes with the library
  routines. Left: the \emph{ab initio} codes and the Wannier engine
  are compiled independently in two different executables. Data
  exchange happens via files with
  standardized formats (see also discussion in
  Secs.~\ref{sec:ecosystem-concept} and~\ref{sec:iolib}). Right: the
  Wannier engine provides a library mode. The \emph{ab initio} code is
  linked at compile time to the Wannier library, and
  a single executable is created. The \emph{ab initio} routines call
  directly the routines from the Wannier engine via library calls.}
\end{figure}

When combining two (or more) codes, various approaches are
possible. While the most common approach has been to compile the
Wannier engine into a different executable than the \emph{ab initio}
code, with data being transferred
via files with standardized formats, an alternative approach is
 to expose the Wannier routines via a library interface, as
schematically depicted in Fig.~\ref{fig:library-vs-standalone}. In
this second approach, a single executable is created, and the main
\emph{ab initio} code is responsible for calling the appropriate
routines from the Wannier-engine library.

Making sure that the library interface is both easily usable and
covers all possible use cases, however, is a non-trivial task.  In the
specific case of \code{Wannier90}, for instance, its first release
contained a simple interface to allow it to be called as a library
from another Fortran program, with the necessary data being passed
by the calling program rather than by file. Over time, however, it
became apparent that this original library interface
did not provide the full functionality
needed by an ecosystem of codes, presenting three main issues. First,
the interface was not fully compatible with
parallel calling codes, because there  was no
means to distribute the data and make use of \code{Wannier90}'s
internal parallelism. Second, the use of global module variables
 meant that \code{Wannier90} was not thread-safe, i.e., a calling
program could not call more than one
instance of \code{Wannier90}. The third issue concerns error
handling. On detecting an error, \code{Wannier90}
would abort, causing the calling program
to crash. The desired behavior would be instead to return an error
code, allowing the calling
program to decide how to handle it (for instance,
exiting gracefully or retrying with different
parameters). Work to address all three of these
issues has been carried out by a team at the Scientific Computing
Department of STFC at Daresbury Laboratory, UK,
in collaboration with three of the present authors (AAM, JRY and
GP). A development version is available in the \code{Wannier90} GitHub
repository and will be merged into a future \code{Wannier90} release.

\subsubsection{\label{sec:iolib}File I/O generation and parsing}
In the first approach mentioned earlier, intermediate files (as listed in
Tab.~\ref{tab:matrix_elements-revised}) are used to decouple the
first-principles calculations from the Wannierization step.  As
already discussed earlier, the formats defined by \code{Wannier90}
became the \emph{de facto} standards for the codes in the
Wannier function software ecosystem.  For instance, many \gls{TB} codes can
read the so-called \texttt{\_tb.dat} or \texttt{\_hr.dat} files
(as already mentioned, these contain on-site energies, hopping terms, \ldots) to further
process the Wannier Hamiltonian.  In the past few years many software
packages have thus started to implement their own parsers for these
I/O files, often focusing only on a specific subset of the file
formats relevant for their own use case.  However, this leads to a
duplication of efforts, also considering that maintaining robust and
feature-complete parsers is a non-trivial task.  As many of these
codes use Python as the programming language of choice, such as
\code{TBmodels}~\cite{tbmodels}, \code{PythTB}~\cite{pythtb},
\code{WannierBerri}~\cite{Tsirkin2021-WannierBerri}, and
\code{AiiDA}~\cite{Pizzi2016,Huber2020,Uhrin2021}, a community effort
has been initiated by Jamal I. Mustafa and others, to implement a
centralized reference set of parsers for the \code{Wannier90}
input/output files in Python, hosted at
\url{https://github.com/wannier-developers/wannier90io-python/}.  The
goal of this project is not only to provide a parser library for
developers, but also to provide a convenient package directly for
users, allowing them to easily load and manipulate the
\code{Wannier90} input/output files for their own use case.
In addition, we note that part of the \code{Wannier90} code base already outputs
Python scripts for post-processing; 
  one example is the \texttt{berry} module, which outputs scripts
using the \code{matplotlib}~\cite{matplotlib} library to plot the
Berry curvature.
However, these scripts are hard-coded into the
\code{Wannier90} codebase as a series of \texttt{write} Fortran statements that output Python code, 
which makes them very difficult to update and maintain. 
The aforementioned Python library will also facilitate future efforts
on moving these post-processing functions into a dedicated Python
package, in the hope of smoothening the development experience, as
well as allowing users to easily postprocess and visualize the
calculation results.

The choice to support and push developments in the Python language is mostly driven by its current prevalence and adoption in the field.
Indeed, for the goal of interoperability, the choice of programming language is not crucial, while it is essential to have well defined APIs or file formats.
Nevertheless, providing reference implementations avoids duplication efforts in writing file writers and parsers, aiming at obtaining a robust library that can be easily reused, maintained, and where bugs can be quickly resolved. To efficiently address this goal, it is useful to select a popular language such as Python; this can also facilitate external contributions.
Nevertheless, we stress that the concept of a common parsing library is not limited to the Python language, but can also be applied to other emerging
languages. For example, the Julia  package \code{WannierIO.jl}~\cite{wannieriojl}
provides functions to read/write \code{Wannier90} file formats, and is
used by the \code{Wannier.jl}~\cite{wannierjl} and
\code{DFTK.jl}~\cite{dftk} packages as their I/O backend.

\subsection{\label{subsec:automation}Automation, workflows and high-throughput}
The diversity of the software ecosystem demonstrates the effectiveness
of \glspl{WF}. However, all methods depend on a robust Wannierization procedure. In the past this was not a very
straightforward process, since it
involves a series of \gls{DFT} calculations and
construction of \glspl{WF}, and more importantly it
also depends sensitively on various input parameters
(number of \glspl{WF}, initial projections for \glspl{MLWF},
energy windows, $k$-point sampling, etc.).  Their
selection often required experience and chemical intuition, and was often a major challenge not only for beginners, but even for
experienced researchers. Fully automated Wannierizations would make the procedure
straightforward, and, as a consequence,
allow any researcher to easily use all
capabilities of the whole ecosystem, while also
enabling high-throughput studies for accelerated materials discovery.
To this end, it became urgent and necessary to perform algorithmic developments on the Wannierization itself, and to implement robust workflows combining multiple software packages in the ecosystem. On the algorithmic side, Wannierization should provide well-localized WFs without
user input (for initial projections or energy windows, for instance);  on the
workflow side, one would like to orchestrate all different steps from the initial \gls{DFT}
calculations to the Wannier-engine executions to the post-processing
steps, while dynamically parsing the outputs and
generating new inputs. Moreover, the workflow engine should provide a set of well-tested convergence parameters, and it should be able to handle common errors, and  to automatically restart failed calculations.
In addition, an automated Wannierization workflow should be ideally modular and composable,
to allow better integration with the whole ecosystem: for instance, this has been exploited in the context of automated GW calculations with Wannier-interpolated band structures~\cite{Bonacci2023}.

Recent development of novel algorithms have largely solved the Wannierization challenge,
 starting from the ``selected columns of the density matrix''
\gls{SCDM}~\cite{Damle2015,Damle2017,Damle2018} algorithms, which generate
initial projections by decomposing the density matrix, to the ``projectability disentanglement''
\gls{PDWF}~\cite{Qiao2023}, that uses projectability thresholds on atomic orbitals, 
rather than energy windows, to select which states to drop, keep frozen, or throw in the disentanglement algorithm. Together with the ``manifold remixing'' \gls{MRWF}~\cite{Qiao2023a} using parallel transport~\cite{Gontier2019} these approaches remove what had been up to now a critical stumbling block. 

On the workflow-engine side, several software
packages are able to automate the electronic-structure calculations,
such as \code{pymatgen} \cite{Ong2013} and \code{FireWorks}
\cite{Jain2015}, \code{AFLOW$\pi$} \cite{Supka2017}, \code{mkite}
\cite{Schwalbekoda2023}, \code{ASE} \cite{Larsen2017} and \code{ASR}
\cite{Gjerding2021}, and \code{AiiDA}
\cite{Pizzi2016,Huber2020,Uhrin2021}; some of them, such as \code{ASE} and \code{AiiDA}, also
provide functionalities or workflows to compute \glspl{WF}.  Equipped
with automated Wannierization algorithms and robust workflow engines,
it has become now possible to create workflows for automated
Wannierizations. For instance,~\textcite{Gresch2018} implemented
\code{AiiDA} workflows and gathered Wannier \gls{TB} models for a
group of III-V semiconductor materials;~\textcite{Vitale2020} used
\gls{SCDM} algorithm together with \code{AiiDA} workflows, carefully
tested convergence parameters, and benchmarked Wannier interpolation
accuracy on a set of 200 structures for entangled bands and a set of
81 structures for isolated bands;~\textcite{Sakai2020} Wannierized
1419 ferromagnetic materials with spin-orbit coupling and computed anomalous
Hall/Nernst conductivities to identify high-performance
transverse-thermoelectric-conversion materials;
~\textcite{Garrity2021} created a
database of Wannier Hamiltonians for 1771 materials;
~\textcite{Fontana2021} implemented workflows in \code{ASE} and
Wannierized 30 inorganic monolayer materials using an automated
protocol; and finally~\textcite{Qiao2023} used \gls{PDWF} to automate the
Wannierization, obtaining over 1.3 million \gls{MLWF} for over 20,000 3D inorganics from the Materials Cloud~\cite{Talirz2020} MC3D database, then using manifold remixing~\cite{Qiao2023a} to separately Wannierize back these into the valence and
conduction bands of 77 insulators. These high-throughput studies can
not only expedite materials discoveries, but also help identify
challenging cases for the Wannierization algorithm, and promote
further development of robust and automated Wannierization approaches.

\section{Conclusions and perspectives\label{sec:conclusions}}
The Wannier function software ecosystem represents a positive model for
interoperability and decentralized code development in
electronic-structure simulations; a similar spirit is also found in CECAM's
Electronic Structure Library~\cite{Oliveira2020}. This was made
possible both by the nature of the scientific problem and physical
quantities involved, and by the design choices originally made in
\code{Wannier77} and \code{Wannier90}, planned early on as
Wannierization engines decoupled from the \emph{ab initio} codes used
to compute the electronic structure.  The availability of a
well-documented, maintained, and modular open-source Wannier engine
has pushed researchers to extend \code{Wannier90}'s functionalities or,
when deemed more practical or efficient, to develop novel packages
targeting specific materials properties. The growing availability of
post-processing features has ignited a positive loop which further
attracted developers from different electronic-structure domains to
work and use \glspl{WF}, strengthening the interest in
\gls{WF}-related methods and resulting in the current ecosystem of
interoperable software.  The ecosystem has been reinforced by the
organization of coding weeks and developer workshops, that have proven
to be crucial to keep the community engaged and synced, to avoid
duplication of efforts, and to collaborate on code maintenance.

While we could not cover here all the existing applications and codes
leveraging \glspl{WF}---notably, we did not discuss their use in the calculation of magnetic interaction parameters~\cite{
Nomoto2020,Yoon2020,He2021_TB2J,dfwannier.jl}---we have outlined some of the most popular
applications and summarized how they can be implemented in software
packages and workflows to calculate advanced materials properties.

Looking forward, we expect that the ongoing efforts in the redesigned
\code{Wannier90}'s library mode will be instrumental in smoothly
integrating automated Wannierization procedures within \emph{ab
  initio} and post-processing codes, with the benefit of reduced file
I/O and code maintenance.  As Wannierization becomes increasingly
automated, we expect researchers to focus on calculations of complex
properties, either through high-level programming of simulation
workflows, or through the development and extension of post-processing
packages.  As a result, even more materials properties will become
computationally accessible thanks to \glspl{WF}, and available to the
community through the release of dedicated functionalities, either in
existing or in new packages of the ecosystem.

Finally, it is worth commenting on two crucial features of an
ecosystem, being it biological or software: biodiversity and
resilience.  A certain level of biodiversity within a software
ecosystem, i.e., the existence of multiple software packages with
partially overlapping functionalities, can increase its robustness.
First, it enables cross-verification of different implementations,
increasing the reliability of the results and facilitating a rapid
identification of bugs.  Second, it can ensure that the ecosystem
capabilities are not lost if a package goes unmaintained or
disappears.  This aspect is connected to resilience, i.e., the
capability of the ecosystem to deliver functionalities---such as the
calculation of materials properties---under the loss of some of its
components. This is an especially relevant issue in a scientific community
where developers might not be able to guarantee long-term support for
their code.  We highlight that a software ecosystem might display the
same dynamics that can be seen in biological settings, including
competition and extinction. While a certain level of competition can
result in code improvements regarding feature coverage, efficiency,
and robustness, we caution that extreme competition might undermine
biodiversity.  It is thus important to sustain the work of individual
developers who contribute to the progress and maintenance of an
active, heterogeneous, and efficient ecosystem, encouraging measures
to ensure proper scientific recognition.  More broadly, the challenge
will be to support the software development work, which is crucial for the long-term maintenance and integration of heterogeneous software packages.

We believe that a diverse, resilient, and open Wannier function software ecosystem is a major asset for the electronic-structure community in its quest to understand, discover, and design materials.

\section{Acknowledgements\label{sec:acknowledgements}}
The authors acknowledge especially the generous and crucial support
(in alphabetical order) of CCP9, CECAM, E-CAM, ICTP, IYBSSD, MaX, NCCR
MARVEL, Psi-k, SISSA, STFC and University of Trieste in the
organization of summer schools and developer workshops related to the
Wannier function software ecosystem. Pivotal to the move to the
present community model were the 2016 San Sebasti\'an workshop,
which was funded by E-CAM,
NCCR MARVEL, and by the CECAM JCMaxwell node, and the
2022 ICTP Trieste Wannier Developers Meeting and Wannier Summer
School, which was funded by CECAM, ICTP, IYBSSD, MaX,
NCCR MARVEL, Psi-k, SISSA, and by the University of
Trieste, where the present work was conceived and organized.

A.M. acknowledges support from the ICSC -- Centro Nazionale di Ricerca
in High Performance Computing, Big Data and Quantum Computing, funded
by European Union - NextGenerationEU (CUP Grant. No. J93C22000540006, PNRR Investimento M4.C2.1.4).
The views and opinions expressed are solely those of the authors and do not necessarily reflect those of the European Union, nor can the European Union be held responsible for them.
S.B. acknowledges the Flatiron Institute, a
division of the Simons Foundation.  E.R.M. acknowledges support from
the National Science Foundation under Grant No. DMR-2035518 and Grant
No. OAC-2103991.  N.M. acknowledges early support by the US National
Science Foundation under Grant No. ASC-96-25885.  N.M., G.P. and
J.Q. acknowledge support from the NCCR MARVEL (a National Centre of
Competence in Research, funded by the Swiss National Science
Foundation, grant No. 205602).  A.A.M. acknowledges the European
Science Foundation INTELBIOMAT programme, and the Thomas Young Centre
(grant TYC-101) for support.  The work of I.S. and S.S.T. was
supported by Grant No. PID2021-129035NB-I00 funded by
MCIN/AEI/10.13039/501100011033, and by the IKUR
  Strategy under the collaboration agreement between the Ikerbasque
  Foundation and the Material Physics Center on behalf of the
  Department of Education of the Basque Government.
G.P. acknowledges support from the Swiss National Science Foundation
(SNSF) Project Funding (grant 200021E\_206190 ``FISH4DIET''), and from
the Open Research Data Program of the ETH Board (project ``PREMISE'':
Open and Reproducible Materials Science Research).

\bibliography{biblio}

%%\printglossaries
\end{document}